\documentclass[twocolumn]{aastex63}

\newcommand{\Teff}{\ensuremath{T_\mathrm{eff}}\xspace}
\newcommand{\logg}{\ensuremath{\log{g}}\xspace}
\newcommand{\vsini}{\ensuremath{v \sin i}\xspace}

\usepackage{natbib}
\usepackage{CJK}
\usepackage{amsmath}
\usepackage{natbib}
\usepackage{graphicx}
\usepackage{amsmath}
\usepackage{booktabs}
\usepackage{longtable}
\usepackage{xspace}
\usepackage{hyperref}
\usepackage[T1]{fontenc}
\usepackage{lipsum}
\usepackage{comment}
\usepackage{xcolor}
\usepackage{chngcntr}
\usepackage{url}

\newcommand{\TESS}{\textit{TESS }}
\newcommand{\HIP}{\textit{Hipparcos }}
\newcommand{\Gaia}{\textit{Gaia }}

\usepackage{lineno}
\newcommand{\caltech}{Department of Astronomy, California Institute of Technology, Pasadena, CA 91125, USA}

\newcommand{\ucsc}{Department of Astronomy \& Astrophysics, University of California, Santa Cruz, CA 95064, USA}
\newcommand{\keck}{W. M. Keck Observatory, 65-1120 Mamalahoa Hwy, Kamuela, HI, USA}
\newcommand{\ucla}{Department of Physics \& Astronomy, 430 Portola Plaza, University of California, Los Angeles, CA 90095, USA}
\newcommand{\jpl}{Jet Propulsion Laboratory, California Institute of Technology, 4800 Oak Grove Dr.,Pasadena, CA 91109, USA}
\newcommand{\ucsd}{Department of Astronomy \& Astrophysics,  University of California, San Diego, La Jolla, CA 92093, USA}

\newcommand{\osu}{Department of Astronomy, The Ohio State University, 100 W 18th Ave, Columbus, OH 43210 USA}

\newcommand{\arizona}{James C. Wyant College of Optical Sciences, University of Arizona, Meinel Building 1630 E. University Blvd., Tucson, AZ 85721}

\shorttitle{Architectures of S-type Planets II} 
\shortauthors{Zhang et al.}

\graphicspath{{./}{figures/}}

\begin{document}
\begin{CJK*}{UTF8}{gbsn}


\title{Dynamical Architectures of S-type Transiting Planets in Binaries II: A Dichotomy in Orbital Alignment of Small Planets in Close Binary Systems}


\correspondingauthor{Jingwen Zhang}
\email{jwzhang@ucsb.edu}


\author[0000-0002-2696-2406]{Jingwen Zhang (张婧雯)}
\affiliation{Institute for Astronomy, University of Hawaiʻi at M\=anoa, 2680 Woodlawn Drive, Honolulu, HI 96822, USA}
\affiliation{Department of Physics, University of California, Santa Barbara, CA 93106, USA}

\author[0000-0001-8832-4488]{Daniel Huber}
\affiliation{Institute for Astronomy, University of Hawaiʻi at M\=anoa, 2680 Woodlawn Drive, Honolulu, HI 96822, USA}

\author[0000-0003-1341-5531]{Michael Bottom}
\affiliation{Institute for Astronomy, University of Hawaiʻi at M\=anoa, 640 N. Aohoku Pl, Hilo, HI 96720, USA}

\author[0000-0002-3725-3058]{Lauren M. Weiss}
\affiliation{Department of Physics and Astronomy, University of Notre Dame, Notre Dame, IN 46556, USA}

\author[0000-0002-6618-1137]{Jerry W. Xuan}
\affiliation{\caltech}

\author[0000-0001-9811-568X]{Adam L. Kraus}
\affiliation{Department of Astronomy, University of Texas at Austin, 2515 Speedway Ave, Austin, TX 78712, USA}

\author[0000-0002-5370-7494]{Chih-Chun Hsu}
\affil{Center for Interdisciplinary Exploration and Research in Astrophysics (CIERA), Northwestern University,
1800 Sherman Ave, Evanston, IL, 60201, USA}

\author[0000-0003-0774-6502]{Jason J. Wang
\begin{CJK*}{UTF8}{gbsn}(王劲飞)\end{CJK*}}
\affil{Center for Interdisciplinary Exploration and Research in Astrophysics (CIERA), Northwestern University,
1800 Sherman Ave, Evanston, IL, 60201, USA}
\affil{Department of Physics and Astronomy, Northwestern University, 2145 Sheridan Rd, Evanston, IL 60208, USA}

\author[0000-0002-8958-0683]{Fei Dai}
\affiliation{Institute for Astronomy, University of Hawaiʻi at M\=anoa, 2680 Woodlawn Drive, Honolulu, HI 96822, USA}

\author[0000-0001-9708-8667]{Katelyn Horstman}
\affiliation{\caltech}


\author[0000-0002-6525-7013]{Ashley Baker}
\affiliation{\caltech}

\author{Randall Bartos} 
\affiliation{\jpl}

\author[0000-0003-4737-5486]{Benjamin Calvin}
\affiliation{\caltech}
\affiliation{\ucla}

\author{Sylvain Cetre} 
\affiliation{\keck}

\author[0000-0002-2361-5812]{Catherine A. Clark}
\affil{NASA Exoplanet Science Institute-Caltech/IPAC, Pasadena, CA 91125, USA}

\author[0000-0002-5741-3047]{David R. Ciardi}
\affil{NASA Exoplanet Science Institute-Caltech/IPAC, Pasadena, CA 91125, USA}

\author[0000-0001-8953-1008]{Jacques-Robert Delorme}
\affiliation{\keck}

\author{Gregory W. Doppmann} 
\affiliation{\keck}

\author[0000-0002-1583-2040]{Daniel Echeverri}
\affiliation{\caltech}

\author[0000-0002-1392-0768]{Luke Finnerty}
\affiliation{\ucla}

\author[0000-0002-0176-8973]{Michael P. Fitzgerald}
\affiliation{\ucla}

\author[0000-0002-2532-2853]{Steve~B.~Howell}
\affil{NASA Ames Research Center, Moffett Field, CA 94035, USA}

\author[0000-0002-0531-1073]{Howard Isaacson}
\affiliation{Department of Astronomy, 501 Campbell Hall, University of California, Berkeley, CA 94720, USA}

\author[0000-0001-5213-6207]{Nemanja Jovanovic}
\affiliation{\caltech}

\author[0000-0002-9903-9911]{Kathryn V. Lester}
\affiliation{Mount Holyoke College, South Hadley MA 01075, USA}

\author[0000-0002-4934-3042]{Joshua Liberman}
\affiliation{\arizona}
\affiliation{Steward Observatory, University of Arizona, 933 N Cherry Ave, Tucson, AZ, USA 85719}

\author[0000-0002-2019-4995]{Ronald A. L\'opez}
\affiliation{Department of Physics, University of California, Santa Barbara, CA 93106, USA}

\author[0000-0002-8895-4735]{Dimitri Mawet}
\affiliation{\caltech}
\affiliation{\jpl}

\author[0000-0003-3165-0922]{Evan Morris}
\affiliation{\ucsc}

\author{Jacklyn Pezzato-Rovner} 
\affiliation{\caltech}

\author[0000-0003-2233-4821]{Jean-Baptiste Ruffio}
\affiliation{\ucsd}

\author[0000-0003-1399-3593]{Ben Sappey}
\affiliation{\ucsd}

\author{Tobias Schofield}
\affiliation{\caltech}

\author[0000-0001-6098-3924]{Andrew Skemer}
\affiliation{\ucsc}

\author[0000-0001-5299-6899]{J. Kent Wallace}
\affiliation{\jpl}

\author[0000-0002-4361-8885]{Ji Wang \begin{CJK*}{UTF8}{gbsn}(王吉)\end{CJK*}}
\affiliation{\osu}

\author[0000-0002-6171-9081]{Yinzi Xin}
\affiliation{\caltech}

\author[0000-0002-4290-6826]{Judah Van Zandt}
\affiliation{\ucla}


\begin{abstract}


Stellar multiplicity plays a crucial role in shaping planet formation and dynamical evolution. We present a survey of 54 TESS Objects of Interest (TOIs) within 300 pc that exhibit significant \textit{Hipparcos}-\textit{Gaia} astrometric accelerations. We identified 35 TOIs with stellar companions at projected separations between $0.1^{\prime\prime}$ to $2^{\prime\prime}$ (or $10-200$ AU). We also identified 12 TOIs that could host planetary-mass or brown dwarf companions, including 6 that are newly discovered. Furthermore, we perform three-dimensional orbital characterization for 12 binaries hosting confirmed planets or planet candidates, allowing us to constrain the line-of-sight mutual inclination, $\Delta I_{\mathrm{los}}$, between the planetary and binary orbits. Combining our sample with previous measurements, we apply Bayesian hierarchical analysis to a total of 26 binary systems with S-type transiting planets ($r_p<5R_{\oplus}$). Specifically, we fit the $\Delta I_{\mathrm{los}}$ distribution with both single (Rayleigh) and mixture models (two-component Rayleigh and Rayleigh-isotropic mixture). We find the mixture models are strongly favored ($\log Z\gtrsim13.9$, or $\approx$5$\sigma$), indicating the observed planet-binary $\Delta I_{\mathrm{los}}$ values likely originate from two underlying populations: one nearly aligned ($\sigma_1 = 2^{\circ}.4^{+0.7}_{-0.9}$) and one with more scattered mutual inclinations ($\sigma_2 = 23^{\circ}.6^{+8.8}_{-7.1}$). Alternatively, the misaligned systems can be equally well described by an isotropic distribution of inclinations. This observed dichotomy likely reflects different dynamical histories. Notably, the misaligned population only emerges in systems with stellar periastron distances $>40$ AU while systems with close-in or eccentric stellar companions (periastron distances $<40$ AU) preserve planet-binary alignment.

\end{abstract}

\keywords{editorials, notices --- 
miscellaneous --- catalogs --- surveys}

\section{Introduction}\label{sec:intro}

Binary systems are common in the solar neighborhood, with one-third of nearby solar-type stars having at least one companion \citep{Raghavan2010}. Understanding the effects of stellar multiplicity on planetary architecture and orbital dynamics is crucial for a comprehensive view of planet formation and evolution. Several surveys have been conducted to search for stellar companions to planet-hosting stars \citep{Kraus2016, Ziegler2020, Ziegler2021, Howell2021, Lester2021, Hirsch2021, Matson2025}. \citet{Kraus2016} used high-resolution adaptive optics (AO) imaging of 382 \textit{Kepler} Objects of Interest (KOIs) with NIRC2 on the Keck II telescope, detecting 23 stellar companions with separation $<50$ AU, which is $4.6\sigma$ lower than the predicted field binary population (58 stellar companions) from \cite{Raghavan2010}. Surveys searching for stellar companions to TESS Objects of Interest (TOIs) have also identified a deficit of planets in close binaries (e.g. \citealt{Ziegler2020, Ziegler2021, Howell2021, Lester2021}). For giant planets discovered by radial velocity observations, \citet{Hirsch2021} found that the planet occurrence rate in close binaries ($a< 100$ AU) is $0.04^{+0.04}_{-0.02}$,  significantly lower than  that of  single stars ($0.18^{+0.04}_{-0.03}$). These studies leveraging various planet detection techniques consistently conclude that there is a deficit of Neptune-sized and larger planets in close binaries, particularly those with separations below 100 AU. These findings align with theoretical expectations that close companions would suppress planet formation, through disk truncation \citep{AandL1994,Jang-Condell2015} or dynamical stirring of planetesimals \citep{Quintana2007,Silsbee2021}. They are also consistent with recent ALMA observations that protoplanetary disks in binaries have lower masses \citep{Akeson2019} and smaller radii \citep{Cox2017,Manara2019,Rota2022}.  



However, some planets survive in dynamically challenging environments for reasons that are still unclear \citep[]{Hatzes2003,Correia2008,Kane2015,Dupuy2016}. An intriguing question is whether the survival of these planets is linked to the orbital characteristics of the binary systems. A key characteristic in this context is the mutual inclination between the orbital plane of the planet(s) and that of the binary orbit. Transiting planets in binary systems offer a valuable opportunity to probe this alignment, since the planetary orbital inclinations are already constrained well from the transit geometry. If the inclination of the stellar companion can also be determined, the mutual inclination—at least along the line of sight—can be inferred. Several studies have revealed a possible preference toward alignment between planetary orbits and stellar companions in binary systems. \cite{Dupuy2022} analyzed orbital arcs in 45 binaries hosting \textit{Kepler} planet candidates. By comparing the amplitudes of the stellar companions’ orbital motions in separation and position angle, they inferred that the mutual inclinations between the planetary and binary orbits are typically low. Using single-epoch Gaia astrometry, \cite{Christian2022} constrained orbits of 67 wide binaries ($100-10^5$ AU) and found a preference for alignment when the binary semi-major axis is below 700 AU. Similarly, \cite{Lester2023} used relative astrometry from imaging to constrain orbits in 13 binaries and also found evidence for planet$-$binary alignment.

Direct measurements of the mutual inclination between planets and stellar companions are challenging, due to the need for high-resolution and multi-epoch observations to track the orbital motion of stellar companions around the primary star. At tens to hundreds of AU, stellar companions have orbital periods on the order of hundreds to thousands of years. When fitting their three-dimensional orbits using relative astrometry from imaging, the limited orbital coverage  poses a challenge and can introduce biases into the results. \cite{Rodrigo2021} investigated orbit fitting with MCMC methods in scenarios where observations cover only $\sim 0.5\%$ of the orbital periods of the companions. They found a degeneracy between eccentricity and inclination, making it difficult to distinguish between face-on eccentric orbits and edge-on circular orbits. To date, detailed and precise characterization of planet-hosting binary orbits has mostly been limited to individual systems such as Kepler-444 \citep{Dupuy2016,Zhangzj2022}, $\tau $ Boo \citep{Justesen2019} and LTT 1445 \citep{Zhang2024a}, where multiple complementary datasets (e.g. RVs, relative astrometry, absolute astrometry) are available.

Astrometry offers to provide new insights on the architectures of planets in S-type binary systems.  Exoplanets, brown dwarfs, and low-mass stars can slightly pull on their host stars, causing the primary stars to deviate from a linear motion — an effect known as astrometric acceleration \footnote{Also known as proper motion anomalies}.
\cite{Kervella2019} and \cite{Brandt2019a} showed that we can measure this effect by comparing the stellar positions and instantaneous proper motions from the \textit{Hipparcos} and \textit{Gaia} astrometric surveys. This method is especially sensitive to companions orbiting between about 1 and 100 AU. Specifically, the \textit{Hipparcos}-\textit{Gaia} Catalog of Accelerations (HGCA; \citealt{Brandt2021})  provides three proper motions: (1) the \textit{Hipparcos} proper motion $\mu_{\rm{H}}$ measured at an epoch near 1991.25; (2) the \textit{Gaia} EDR3 proper motion $\mu_{\rm{G}}$ measured at an epoch near 2016.01; (3) the long-term proper motion $\mu_{\rm{HG}}$ calculated as the difference in positions between \textit{Hipparcos} and \textit{Gaia} divided by the $\sim 25$-year baseline. 


This paper is the second in a survey of transiting planets that appear to also harbor stellar, brown dwarf, or planetary companions at large separations based on their accelerations in HGCA. In Paper I (\citealt{Zhang2024a}), we identified the sample and performed an analysis based on data from HGCA, as described in Section~\ref{sec:sample}.  In Section~\ref{sec:obs}, we present new observations that refine the disposition of the planets and their distant companions, including adaptive optics (AO) imaging, primary star radial velocities (RVs), and relative radial velocities between primary and secondary stars.  In Section~\ref{sec:surveyre}, we present the survey results for companions to TOIs in our sample, including detected stellar companions, unresolved stellar companions, and giant planet or brown dwarf companions. Section~\ref{sec:3dorbit} shows the three-dimensional (3D) orbital fitting of stellar companions in 12 systems by combining  multiple datasets. Finally, in Section~\ref{sec:mul} and \ref{sec:hba}, we present a statistical analysis of the planet-binary mutual inclination at different binary separations. In Section \ref{sec:discussion}, we compare our findings with previous studies and discuss their implications for planet formation in S-type binaries. Our conclusions are summarized in Section \ref{sec:conclusion}.
. 

\section{Sample}\label{sec:sample}





The objective of Paper I was to identify a sample of transiting planets around nearby stars that exhibited significant astrometric accelerations.  We defined our sample by cross-matching TESS
Object of Interest (TOIs, up to TOI-6682) \footnote{In Paper I, we also included Kepler Objects of Interest (KOIs) and K2 planet candidates. However, all of them—except for Kepler-444—were identified as false positives. Therefore, we only present TOIs in this work.} within 300 pc of the Sun with the HGCA catalog, selecting stars that exhibit astrometric acceleration at a significance level greater than $3\sigma$.

Following \cite{Kervella2019} and \cite{Brandt2021}, Paper I computed the astrometric acceleration by subtracting the long-term proper motion $\mu_{\rm{HG}}$ from the \textit{Hipparcos} or \textit{Gaia} proper motion.  Additionally, Paper I queried the Renormalized Unit Weight Error (RUWE) values from Gaia DR3 for the planet hosts in our sample. RUWE serves as a goodness-of-fit metric for single-star astrometric model, which fits an object's position, parallax, and proper motion \citep{Lindegren2018, Ziegler2020}.
Values above 1.4 statistically suggest a non-single source, either because the components are too close to be fully resolved by Gaia \citep{Lindegren2018}, or because Gaia reports erroneous astrometry for spatially resolved binaries. Table~\ref{tab:table1X} summaries the resulting astrometric accelerations, distances, and RUWE values.  We adopt all of the stars in Table~\ref{tab:table1X} as our sample for this work.

Paper I also described which transit-like signals are dispositioned as false positives, based on TESS Follow-Up Observing Program (TFOP) and RV measurements.  The tables throughout this paper have two regions, with the top region corresponding to planets and planet candidates, and the botton region corresponding to signals that are not from transiting planets.  Table~\ref{tab:table1Y} in the appendix further distinguishes which signals are from confirmed planets (CPs), planet candidates (PCs), false positives (FPs), and false alarms (FAs).  It is important to note that transiting planets typically have orbital periods on the order of days, whereas sub-stellar or stellar companions  have orbital periods ranging from dozens to thousands of years.


\section{Observations}\label{sec:obs}
We obtained observations to characterize the masses and orbits of the companions to the targets in our sample.  These observations included new observations using high-resolution Adaptive Optics (AO) imaging on Keck/NIRC2, archival imaging observations from several facilities, archival radial velocity (RV) observations from various instruments.  In some cases, relative RVs between the primary star and its companion were obtained using Keck/KPIC. When Gaia provided photometry and astrometry for the companions, we also incorporated relative RVs and companion proper motions from Gaia DR3 astrometry.


\subsection{Adaptive Optics Imaging}

\subsubsection{Keck/NIRC2}
We obtained AO imaging of targets in the northern hemisphere ($\rm{DEC.} > -40^\circ$) using the near-infrared imager (NIRC2) in the Kcont bandpass (2.2558 $\mu m$ -2.2854 $\mu m$) on the 10-meter Keck II telescope. Our observations were taken over 11 half or full nights between 2022 February and 2024 February. All observations were in vertical angle mode and used natural guide stars. We used an exposure time of 2 mins per frame and 4 frames for each target during each run. For each target, we observed a reference star with similar magnitude and color, located within 10 degrees of the target. Additionally, we collected dark frames and dome flats  for each observing run. 

We used the Vortex Imaging Processing (VIP) software package \citep{Gomez2017, Christiaens2023} to perform dark-subtraction, flat-fielding and bad-pixel removal. We applied the solution in \cite{Service2016} to correct for geometric distortion. For frame registration, we measured the primary star's location by fitting 2D Gaussians to its point spread function (PSF), and aligned each frame with the primary star at the center. 
\begin{figure}
    \centering
    \includegraphics[width=\linewidth]{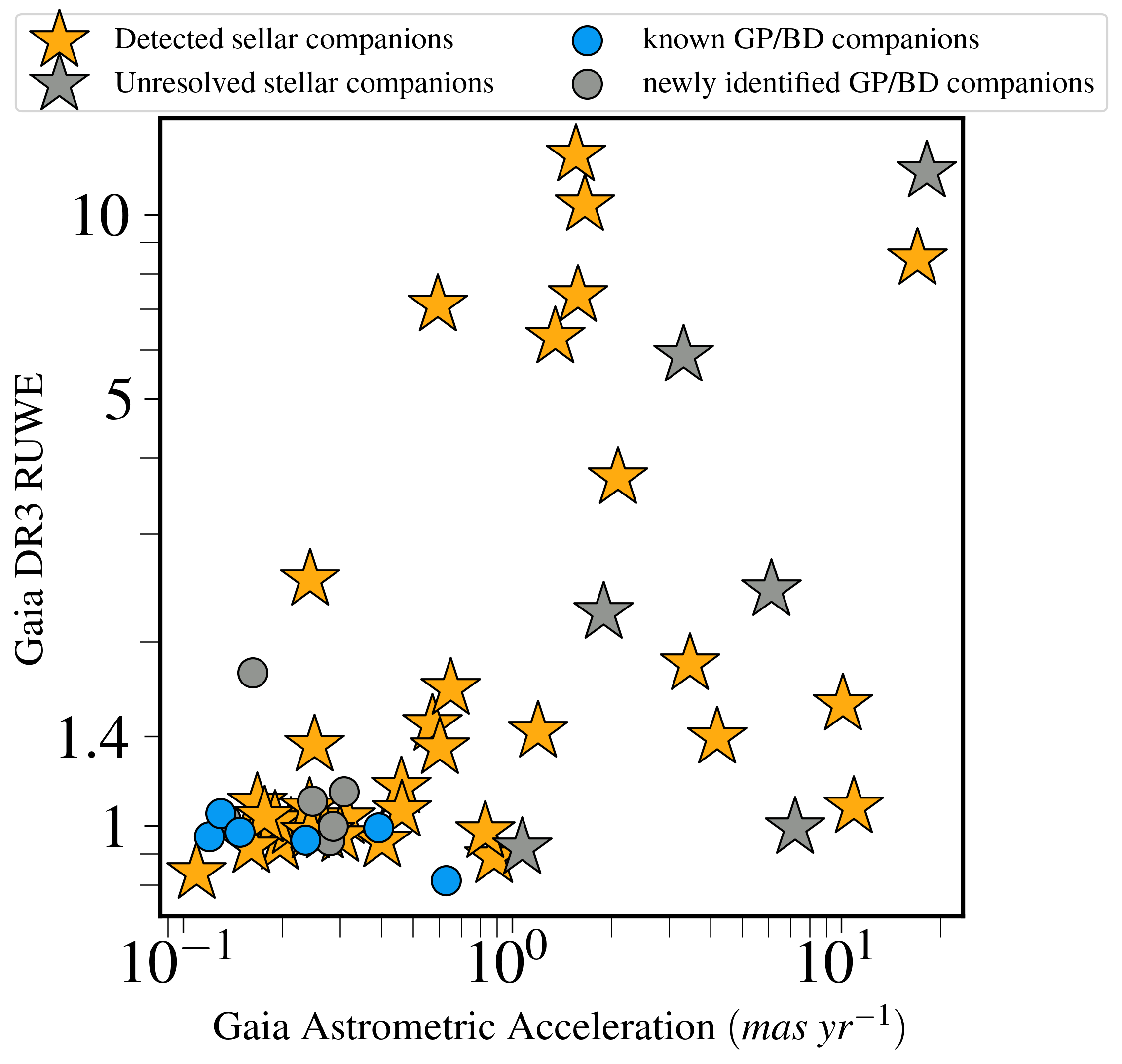}
    \caption{Gaia DR3 RUWE versus astrometric acceleration for stars in our sample. Orange stars represent systems with detected stellar companions. Gray stars indicate unresolved stellar companions ($<0.1^{\arcsec}$) inferred from astrometric acceleration. Blue circles mark systems with confirmed planetary or brown dwarf companions. Gray circles indicate newly identified planetary or brown dwarf companion candidates suggested by astrometric acceleration and AO imaging detection limit. }
    \label{fig:fig_sample}
\end{figure}

After preprocessing the data cubes, we used the \texttt{PyKLIP} package \citep{Wang2015} to forward model the stellar PSF through Karhunen-Loeve (KL) image processing, following the methods outlined in \cite{Soummer2012} and \cite{Pueyo2016}. We applied the Reference Differential Imaging (RDI) mode in \texttt{PyKLIP} to subtract the stellar PSF. This method leverages a PSF library composed of science data and reference star images. Since the observations were taken in vertical angle mode, \texttt{PyKLIP} also de-rotated each image according to its parallactic angle and stacked the individual frames into a combined image. For systems with visually identified stellar companions, we then fitted the model PSF to the data to extract the astrometry and photometry of the stellar companions using a Markov Chain Monte Carlo (MCMC) sampler. The resulting astrometry and photometry are presented in Table~\ref{tab:table2X}. For systems with no resolved stellar companions, we computed the $5\sigma$ contrast curves by injecting fake companions into the dataset at various separations and position angles.


\subsubsection{Archival relative astrometry}

For targets with available data, we used archival astrometry from previous imaging surveys, including: (1) the speckle imaging with the Southern Astrophysical Research (SOAR) telescope \citep{Ziegler2020, Ziegler2021}, (2) the speckle imaging with ‘Alopeke and Zorro speckle cameras on Gemini 8.1 m North and South telescopes \citep{Lester2021, zhou2022, Lester2023}, (3) AO imaging observation using Palomar/PHARO (PI: D. Ciardi), (4) speckle imaging using NESSI at WIYN telescope \citep{Howell2021},  and (5) catalogs of Two Micron All Sky Survey (2MASS, \citealt{Cutri2003}), \textit{Hipparcos} \citep{ESA} and \textit{Gaia} \citep{Gaia}. The details are listed in Table~\ref{tab:table2X}.

\subsection{Radial Velocities}

\subsubsection{TOI-201}

We used published RV data for TOI-201 from \citet{Hobson2021}, which were originally used to confirm the inner transiting planet. The dataset includes 53 RVs from the FEROS spectrograph on the MPG 2.2 m telescope at La Silla Observatory (November 2018–March 2020), 10 RVs from the CORALIE spectrograph on the Swiss 1.2 m Euler telescope (November 2018–March 2019), 62 RVs from the MINERVA-Australis array (January–April 2019), and 39 RVs from HARPS \citep{Mayor2003} on the ESO 3.6 m telescope at La Silla (November 2018 to February 2020).

\subsubsection{TOI-402}
We used 85 published RVs for TOI-402 taken with HARPS \citep{Mayor2003} on the ESO 3.6 m telescope in La Silla, Chile, between December 2003 and September 2017 from \cite{Dumusque2019} and \cite{Gandolfi2019}. We also include 48 RVs of TOI-402 from Planet Finder Spectrograph (PFS), located on the Magellan II Clay telescope at the Las Campanas Observatory. The observations were taken between 2019 February 12 and 2019 December 20 as part of the Magellan-TESS Survey (MTS, \citealt{Teske2021}). Furthermore, we also include 28 new RVs we collected between 2004 October and 2023 December using High Resolution Echelle Spectrometer (HIRES, \citealt{Vogt1994}) at the Keck I telescope on Maunakea. We used the standard California Planet Search (CPS) pipeline described in \cite{Howard2010} to determine RVs. Spectra were obtained with an iodine gas cell in the light path for wavelength calibration.

%
\subsection{Relative RVs between primaries and companions}

\subsubsection{Keck/KPIC}
We observed three systems using the upgraded Keck/NIRSPEC \citep{martin_overview_2018, lopez_Characterization_2020}: TOI-402 HIP 11433 on UT 2024-09-26, TOI-1684 HIP 20334 on UT 2022-07-22 and 23, and TOI-5962 HIP 112486 on UT 2024-11-22. The data were collected with the Keck Planet Imager and Characterizer (KPIC) fiber injection unit \citep[][]{delorme_Keck_2021a}, specifically the upgraded phase 2 system \citep[][]{Echeverri2022, Jovanovic2025}. The fiber injection unit is located downstream of the Keck II adaptive optics system and is used to inject light from a given target into one of the single-mode fibers connected to the NIRSPEC slit. In each observation, we collect spectra for both the primary star and the secondary companion with KPIC. The observations were conducted in $K$ band ($1.94-2.49~\mu$m) and the resolving power is $R\sim35,000$. In each observation, we also observe a telluric standard star and an RV standard star (early M giant stars) for calibration purposes (see \citealt{wang_Detection_2021}). KPIC relative RVs have been used to improve orbital and companion mass constraints for several low-mass companions \citep{Xuan2024, DoO2024, YZhang2024}.

We reduce the data using the public KPIC pipeline\footnote{\url{https://github.com/kpicteam/kpic_pipeline}} to extract 1D, wavelength-calibrated, spectra. For details, see \citet{wang_Detection_2021, Xuan2024_Survey}. In brief, the main data reduction steps are 1) instrumental background subtraction, 2) trace-finding and measuring the trace widths using the telluric star data, 3) optimal extraction \citep{Horne1986}, and 4) deriving the wavelength solution using the M giant spectra taken on the same night. For the analysis, we use three spectral orders from 2.29-2.49~$\mu$m, which have the most reliable wavelength solution \citep{Hsu2024}, and contain the strongest spectral lines from the stars. We describe the spectral analysis procedure to measure the primary and companion RVs in Appendix~\ref{sec:revel_pm}. 


\subsubsection{Gaia}
We also used the relative RVs between primaries and companions reported by Gaia DR3 for TOI-128, TOI-1099, TOI-4175.  \textit{Gaia}’s Radial Velocity Spectrometer (RVS) is a near-infrared (845-872 nm), medium-resolution ($\lambda/\Delta\lambda \sim 11{,}500$) integral-field spectrograph that disperses all incoming light within its field of view \citep{Katz2023}. The RVS measures the line-of-sight velocity component for stars down to approximately 17th magnitude. For the Gaia's resolved binaries listed above, \textit{Gaia} provides radial velocities for both components of the systems. The relative RVs used in our analysis are given in Table~\ref{tab:table_relRV} in the appendix.

\subsection{Proper motion of companions from Gaia DR3}

We also used proper motions of stellar companions from \textit{Gaia} DR3 for TOI-128, TOI-906, TOI-4175 and TOI-5962. Combined with the proper motions of the host stars, this allows us to better constrain the binary orbits by directly measuring the relative motion between the two components in the sky plane. The proper motions of companions used in our analysis are given in Table~\ref{tab:table_relpm} in the appendix.

\begin{center}
\begin{longtable*}{lccccccccccc}
\caption{Hipparcos-Gaia astrometric acceleration in declination and right ascension for stars in our sample. The $\Delta \mu_{\delta *}^{\dag}$ components have the $\cos{\delta}$ factor included. $\sigma[\Delta\mu]  $ represent the uncertainties. We assume that the uncertainties on the proper motions and $\mu_{\rm{HG}}$ are independent and add them in quadrature to calculate these uncertainties (see \citealt{Zhang2024a} for details).}\label{tab:table1X}\\
\hline
    &     &    \multicolumn{3}{c}{\textit{Gaia}}   & &  \multicolumn{3}{c}{\textit{Hipparcos}} & & &Stellar \\
\cline{3-5} \cline{7-9}
TOI&  HIP&   $\Delta\mu_{\alpha}$  & $\Delta\mu_{\delta *}$ & $\rm S/N$ & & $\Delta\mu_{\alpha}$ &  $\Delta\mu_{\delta *}$ & $\rm S/N$  & Dist. &RUWE &  Comp.\\
  &  number & $\mathrm{mas}\ \mathrm{yr}^{-1}$ & $\mathrm{mas}\ \mathrm{yr}^{-1}$&  &  & $\mathrm{mas}\ \mathrm{yr}^{-1}$ & $\mathrm{mas}\ \mathrm{yr}^{-1}$ & & pc & &detection$^{*}$   \\
\hline
\hline
\endfirsthead
\hline
    &     &    \multicolumn{3}{c}{\textit{Gaia}}   & &  \multicolumn{3}{c}{\textit{Hipparcos}} & & &Stellar\\
\cline{3-5} \cline{7-9}
TOI&  HIP&   $\Delta\mu_{\alpha}$  & $\Delta\mu_{\delta *}$ & $\rm S/N$ & & $\Delta\mu_{\alpha}$ &  $\Delta\mu_{\delta *}$ & $\rm S/N$  & Dist. &RUWE & Comp. \\
  &  number & $\mathrm{mas}\ \mathrm{yr}^{-1}$ & $\mathrm{mas}\ \mathrm{yr}^{-1}$&  &  & $\mathrm{mas}\ \mathrm{yr}^{-1}$ & $\mathrm{mas}\ \mathrm{yr}^{-1}$ & & pc & &detection$^{*}$   \\
\hline
\hline
\endhead
\hline
\endfoot
\hline
\multicolumn{12}{p{\textwidth}}{
{\footnotesize
\textsuperscript{\dag} The $\Delta \mu_{\delta *}$ components have the $\cos \delta$ factor included. $\sigma[\Delta\mu]$ represent the uncertainties.
We assume that the uncertainties on the proper motions and $\mu_{\rm HG}$ are independent and add them in quadrature (see \citealt{Zhang2024a}). \newline
\textsuperscript{*} YES: stellar companions were detected in our observations or previous surveys; NO(Unres.): no detected stellar companions, but the astrometric acceleration is consistent with the presence of a unresolved stellar companion closer than Keck resolution; NO(GP)/NO(BD): no detected stellar companion, and the observed astrometric acceleration is consistent with the presence of giant planets / brown dwarf companions. \newline
\textsuperscript{\ddag}  For stars in Gaia DR3 NSS catalog, including TOI5383 (HIP69275), the S/N of astrometric acceleration may be less significant if we used proper motion reported in NSS catalog, instead of regular DR3 catalog.
}
}
\endlastfoot
\multicolumn{12}{c}{\textit{Confirmed and Candidate Planet Host Stars}} \\
\hline
  128 &  24718    & $0.08\pm{0.04}$ & $-0.18\pm{0.04}$&4.8 & &$-0.84\pm{0.91}$ & $2.02\pm{1.06}$ & 2.1 & 68.23 & 0.94& YES\\
  141 &  111553    & $0.05\pm{0.03}$ & $-0.16\pm{0.03}$&4.92 & &$0.3\pm{0.53}$ & $-0.85\pm{0.54}$ & 1.67 & 47.77 & 1.09& YES \\
  144 &  26394    & $0.19\pm{0.07}$ & $-0.6\pm{0.08}$&7.66 & &$0.95\pm{0.4}$ & $0.58\pm{0.44}$ & 2.7 & 18.29 & 0.81& NO(GP)\\
  179 &  13754    & $-0.39\pm{0.03}$ & $-0.04\pm{0.03}$&12.06 & &$-0.45\pm{0.75}$ & $0.22\pm{0.72}$ & 0.67 & 38.63 & 0.99& NO(BD) \\
  201 &  27515    & $-0.13\pm{0.04}$ & $-0.21\pm{0.04}$&6.28 & &$0.79\pm{0.87}$ & $0.35\pm{1.0}$ & 0.96 & 112.18 & 1.1& NO(BD)   \\
  271 &  21952    & $13.18\pm{0.2}$ & $10.78\pm{0.23}$&79.79 & &$1.23\pm{0.72}$ & $3.45\pm{0.82}$ & 4.54 & 99.93 & 8.47& YES\\
  369 &  115594    & $0.12\pm{0.08}$ & $-0.22\pm{0.08}$&3.21 & &$1.34\pm{2.0}$ & $-1.41\pm{1.88}$ & 1.0 & 227.75 & 1.36& YES\\
  402 &  11433    & $-0.75\pm{0.04}$ & $-0.46\pm{0.04}$&22.46 & &$0.82\pm{1.02}$ & $0.74\pm{0.92}$ & 1.13 & 44.86 & 0.89& YES\\
  455 &  14101    & $-6.75\pm{0.86}$ & $8.54\pm{0.71}$&14.1 & &$7.34\pm{28.25}$ & $-25.74\pm{27.77}$ & 0.96 & 6.86 & 1.07& YES\\
  906 &  78301    & $0.08\pm{0.06}$ & $0.14\pm{0.05}$&3.04 & &$-1.42\pm{1.44}$ & $3.98\pm{1.49}$ & 2.84 & 135.41 & 0.92& YES\\
  930 &  16881    & $0.79\pm{0.06}$ & $0.9\pm{0.07}$&17.24 & &$0.12\pm{1.05}$ & $1.22\pm{1.14}$ & 1.08 & 294.7 & 1.43& YES\\
  1099 &  108162    & $-0.34\pm{0.04}$ & $0.31\pm{0.04}$&11.16 & &$-1.93\pm{0.97}$ & $1.04\pm{0.88}$ & 2.3 & 23.64 & 1.16& YES\\
  1131 &  81087    & $1.47\pm{0.25}$ & $0.52\pm{0.29}$&6.06 & &$-0.72\pm{1.07}$ & $0.65\pm{1.29}$ & 0.83 & 247.64 & 12.51& YES   \\
  1144 &  97657    & $-0.23\pm{0.04}$ & $-0.05\pm{0.03}$&6.52 & &$0.26\pm{0.9}$ & $-1.57\pm{0.85}$ & 1.86 & 37.84 & 0.95& NO(GP)     \\
  1151 &  96618    & $-0.1\pm{0.03}$ & $-0.26\pm{0.04}$&7.76 & &$-0.83\pm{0.46}$ & $-0.36\pm{0.52}$ & 1.9 & 136.98 & 0.95& NO(BD)    \\
  1204 &  55069    & $0.65\pm{0.04}$ & $0.51\pm{0.03}$&23.51 & &$-1.17\pm{0.75}$ & $-0.98\pm{0.71}$ & 2.09 & 106.45 & 0.98& YES\\
  1271 &  66192    & $0.0\pm{0.03}$ & $0.12\pm{0.04}$&3.47 & &$-1.63\pm{0.88}$ & $0.84\pm{0.66}$ & 2.18 & 92.04 & 0.96& NO(GP)      \\
  1339 &  99175    & $-0.12\pm{0.03}$ & $-0.05\pm{0.04}$&3.73 & &$0.95\pm{0.73}$ & $-0.67\pm{0.93}$ & 1.45 & 53.49 & 1.05&  NO(GP)     \\
  1730 &  34730    & $-0.07\pm{0.13}$ & $0.3\pm{0.09}$&3.32 & &$-3.82\pm{4.43}$ & $-5.81\pm{3.08}$ & 1.96 & 35.65 & 1.14& NO(GP)    \\
  1799 &  54491    & $0.09\pm{0.05}$ & $-0.1\pm{0.04}$&3.31 & &$2.72\pm{1.22}$ & $0.08\pm{0.87}$ & 2.23 & 62.23 & 1.03& NO(GP)    \\
  4175 &  64573    & $-0.6\pm{0.08}$ & $0.04\pm{0.06}$&7.18 & &$-4.27\pm{1.61}$ & $-1.41\pm{1.01}$ & 2.88 & 46.32 & 1.34&  YES \\
  4399 &  94235    & $0.12\pm{0.04}$ & $-0.21\pm{0.03}$&7.4 & &$1.25\pm{0.9}$ & $1.01\pm{0.8}$ & 1.87 & 58.55 & 1.07&  YES\\
  4568 &  77921    & $-1.1\pm{0.08}$ & $0.78\pm{0.13}$&13.4 & &$0.41\pm{0.71}$ & $0.54\pm{0.82}$ & 0.87 & 48.29 & 6.3& YES\\
  4603 &  26250    & $0.23\pm{0.06}$ & $0.17\pm{0.03}$&5.92 & &$1.0\pm{1.6}$ & $1.33\pm{0.72}$ & 1.49 & 224.15 & 1.0& NO(BD)    \\
  5140 &  42403    & $-0.11\pm{0.04}$ & $0.1\pm{0.03}$&3.85 & &$0.57\pm{0.83}$ & $-0.03\pm{0.59}$ & 0.69 & 35.24 & 0.98& NO(GP)    \\
  5811 &  102295    & $3.46\pm{0.1}$ & $-0.0\pm{0.08}$&34.99 & &$-3.26\pm{1.1}$ & $-1.76\pm{1.01}$ & 3.42 & 167.79 & 1.84& YES     \\
  5962 &  112486    & $0.08\pm{0.07}$ & $0.23\pm{0.06}$&3.8 & &$-4.66\pm{2.2}$ & $-7.47\pm{2.31}$ & 3.86 & 127.27 & 0.98& YES \\
\hline
\multicolumn{12}{c}{\textit{Transit False Positive Stars}} \\
\hline
  230 &  114003    & $0.16\pm{0.04}$ & $-0.02\pm{0.05}$&3.78 & &$-0.4\pm{1.16}$ & $0.54\pm{1.22}$ & 0.56 & 141.62 & 1.01& NO(BD.)     \\
  394 &  15053    & $4.05\pm{0.12}$ & $-9.59\pm{0.11}$&94.97 & &$0.66\pm{1.84}$ & $-1.47\pm{1.24}$ & 1.18 & 141.7 & 3.18& YES     \\
  510 &  33681    & $5.81\pm{0.05}$ & $1.96\pm{0.06}$&116.11 & &$2.64\pm{0.81}$ & $5.24\pm{1.07}$ & 5.74 & 92.84 & 2.43& NO(Unres.)    \\
  522 &  40694    & $-0.15\pm{0.07}$ & $-0.15\pm{0.06}$&3.36 & &$1.13\pm{0.7}$ & $-0.33\pm{0.7}$ & 1.67 & 111.73 & 0.99& YES \\
  575 &  41849    & $-0.01\pm{0.05}$ & $-0.19\pm{0.04}$&4.96 & &$-0.06\pm{0.98}$ & $-0.23\pm{0.71}$ & 0.33 & 178.49 & 1.02& YES\\
  621 &  44094    & $-0.11\pm{0.03}$ & $0.0\pm{0.03}$&3.69 & &$0.08\pm{0.66}$ & $0.44\pm{0.62}$ & 0.72 & 182.31 & 0.84& YES \\
  635 &  47990    & $-0.31\pm{0.05}$ & $-0.03\pm{0.03}$&6.22 & &$1.27\pm{1.34}$ & $-1.33\pm{0.72}$ & 1.74 & 59.04 & 1.03& YES \\
  680 &  58234    & $0.06\pm{0.11}$ & $2.09\pm{0.07}$&29.68 & &$0.96\pm{1.38}$ & $-2.72\pm{1.03}$ & 2.67 & 160.05 & 3.71& YES \\
  896 &  28122    & $1.37\pm{0.14}$ & $-3.96\pm{0.08}$&45.59 & &$-1.38\pm{4.71}$ & $3.44\pm{2.8}$ & 1.18 & 155.53 & 1.4& YES \\
  909 &  82032    & $-0.12\pm{0.02}$ & $-0.13\pm{0.03}$&5.81 & &$0.46\pm{0.46}$ & $1.28\pm{0.69}$ & 2.04 & 51.65 & 1.04& YES \\
  951 &  21000    & $-6.45\pm{0.25}$ & $3.28\pm{0.19}$&30.07 & &$52.15\pm{8.46}$ & $131.99\pm{6.39}$ & 21.16 & 202.99 & 0.99&  NO(Unres.)     \\
  1124 &  98516    & $15.88\pm{0.35}$ & $8.81\pm{0.3}$&53.18 & &$1.19\pm{1.35}$ & $6.98\pm{0.98}$ & 7.17 & 80.17 & 11.77& NO(Unres.)    \\
  1418 &  83168    & $0.3\pm{0.04}$ & $-1.03\pm{0.04}$&30.36 & &$-0.62\pm{0.46}$ & $0.14\pm{0.44}$ & 1.4 & 159.34 & 0.92& NO(Unres.)     \\
  1665 &  27844    & $0.04\pm{0.04}$ & $0.4\pm{0.03}$&13.17 & &$2.32\pm{1.39}$ & $0.4\pm{0.84}$ & 1.72 & 66.44 & 0.94& YES \\
  1684 &  20334    & $-0.75\pm{0.04}$ & $-10.08\pm{0.04}$&226.52 & &$-6.11\pm{0.77}$ & $-5.26\pm{0.9}$ & 9.75 & 87.11 & 1.58& YES \\
  1719 &  44289    & $1.61\pm{0.2}$ & $0.41\pm{0.21}$&8.31 & &$1.28\pm{1.14}$ & $-0.88\pm{0.84}$ & 1.47 & 236.54 & 10.38& YES \\
  1831 &  65205    & $-0.24\pm{0.06}$ & $0.18\pm{0.03}$&6.08 & &$-1.64\pm{1.24}$ & $0.78\pm{0.77}$ & 1.56 & 126.45 & 0.96&  YES\\
  1837 &  67650    & $1.56\pm{0.12}$ & $0.27\pm{0.13}$&12.85 & &$-1.46\pm{1.15}$ & $0.28\pm{0.7}$ & 1.31 & 152.67 & 7.37& YES\\
  1946 &  70833    & $-3.31\pm{0.18}$ & $-0.19\pm{0.17}$&18.81 & &$0.67\pm{1.0}$ & $-1.0\pm{1.05}$ & 1.16 & 250.89 & 5.88& NO(Unres.)     \\
  2017 &  74685    & $-0.4\pm{0.12}$ & $0.44\pm{0.15}$&4.37 & &$-0.38\pm{1.25}$ & $0.44\pm{1.19}$ & 0.48 & 98.57 & 7.12& YES \\
  2118 &  79105    & $0.11\pm{0.08}$ & $-2.8\pm{0.1}$&27.9 & &$0.47\pm{0.75}$ & $2.42\pm{0.91}$ & 2.73 & 188.03 & 4.37& NO(Unres.)    \\
  2299 &  93711    & $-0.45\pm{0.07}$ & $0.1\pm{0.06}$&6.87 & &$1.11\pm{1.55}$ & $-0.14\pm{1.58}$ & 0.72 & 34.27 & 1.06& YES \\
  2666 &  45621    & $-0.19\pm{0.06}$ & $0.54\pm{0.04}$&12.83 & &$1.57\pm{0.89}$ & $-1.77\pm{0.81}$ & 2.78 & 32.28 & 1.46& YES \\
  4314 &  22084    & $-1.25\pm{0.07}$ & $-1.42\pm{0.06}$&29.95 & &$1.81\pm{0.92}$ & $2.23\pm{0.7}$ & 3.62 & 153.05 & 2.23& NO(Unres.)    \\
  5079 &  24007    & $0.25\pm{0.08}$ & $-0.1\pm{0.05}$&3.47 & &$-3.6\pm{3.18}$ & $4.73\pm{1.33}$ & 2.71 & 76.93 & 0.98& YES \\
  5521 &  57386    & $-0.31\pm{0.06}$ & $-0.57\pm{0.06}$&10.78 & &$-1.03\pm{1.63}$ & $-1.33\pm{1.87}$ & 0.94 & 136.95 & 1.68& YES\\
  5383$^{\ddag}$ &  69275    & $-0.04\pm{0.07}$ & $0.24\pm{0.07}$&3.73 & &$0.47\pm{0.57}$ & $0.93\pm{0.61}$ & 1.74 & 102.09 & 2.53& YES\\
\hline
\end{longtable*}
\end{center}

\section{Survey Results}\label{sec:surveyre}

Figure~\ref{fig:fig_sample} provides an overview of our sample by showing Gaia RUWE versus astrometric accelerations. We highlight three types of new detections. First, we report the detection of 22 stellar companions in 20 TOIs using high-resolution imaging with Keck/NIRC2. In addition, 15 additional TOIs in our sample have known stellar companions identified by previous surveys \citep{Ziegler2020, Ziegler2021, Lester2021}. This brings the total to 35  resolved binaries (or triples) in our sample. Second, our analysis reveals 7  TOIs hosting unresolved stellar companions, as indicated by their \textit{Hipparcos}-\textit{Gaia} astrometric acceleration. These companions are inferred to lie within $<0.1^{\arcsec}$, making them too close to be resolved in imaging. Third, we present evidence for giant planet or brown dwarf companions in the remaining 12 TOIs. A detailed breakdown of each category is provided below. 

\subsection{Resolved Stellar Companions}

\begin{figure*}
    \centering
    \includegraphics[width=\linewidth]{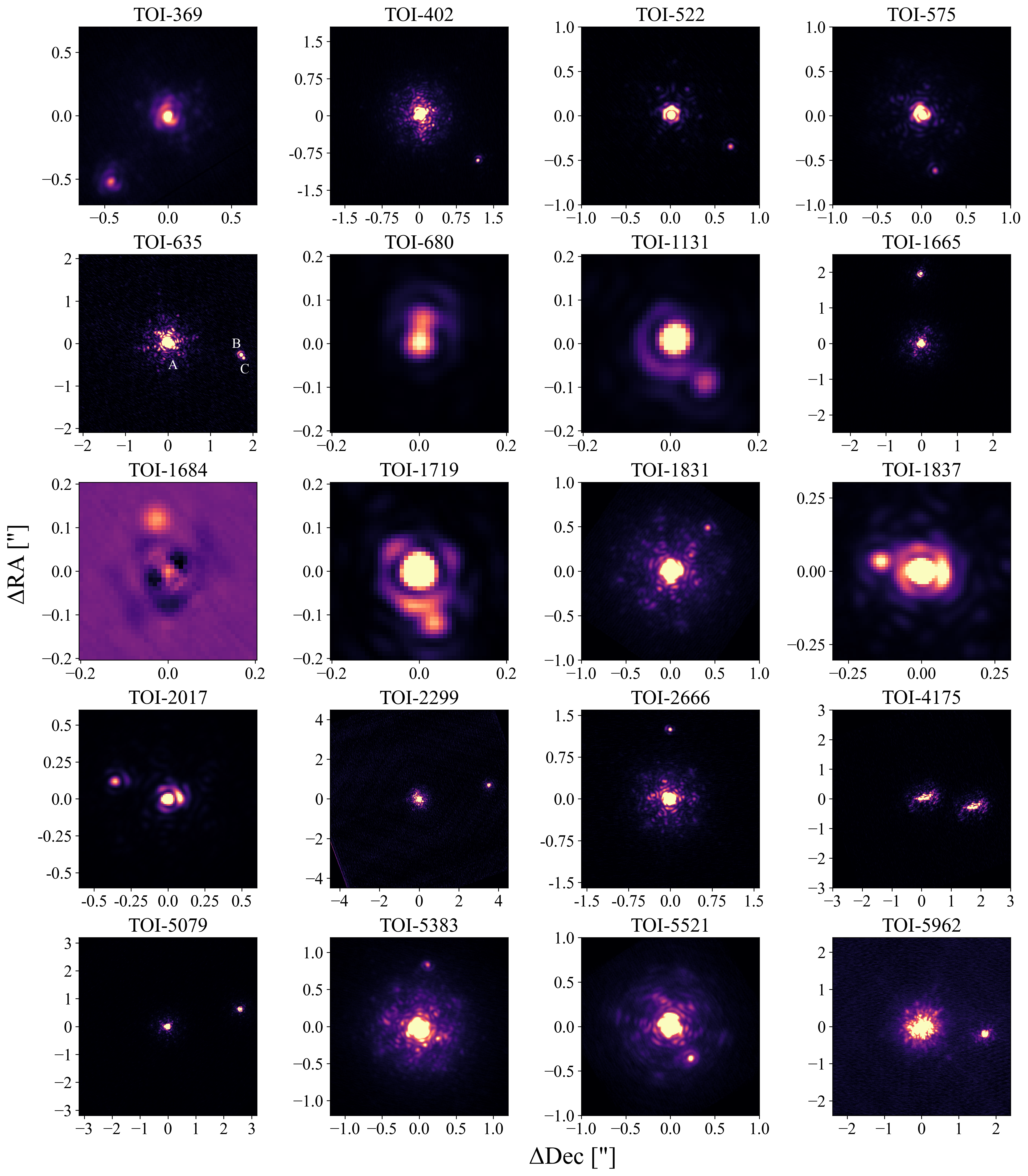}
    \caption{Keck/NIRC2 AO images of TOIs with resolved stellar companions detected from our survey. Images are present in linear scale. For TOI-1684, we present the AO image after PSF subtraction to enhance visibility. The orientation is the same in all images, with North pointed up and East to the left. }
    \label{fig:gallery}
\end{figure*}

\begin{figure*}[!t]
    \centering
    \includegraphics[width=\linewidth]{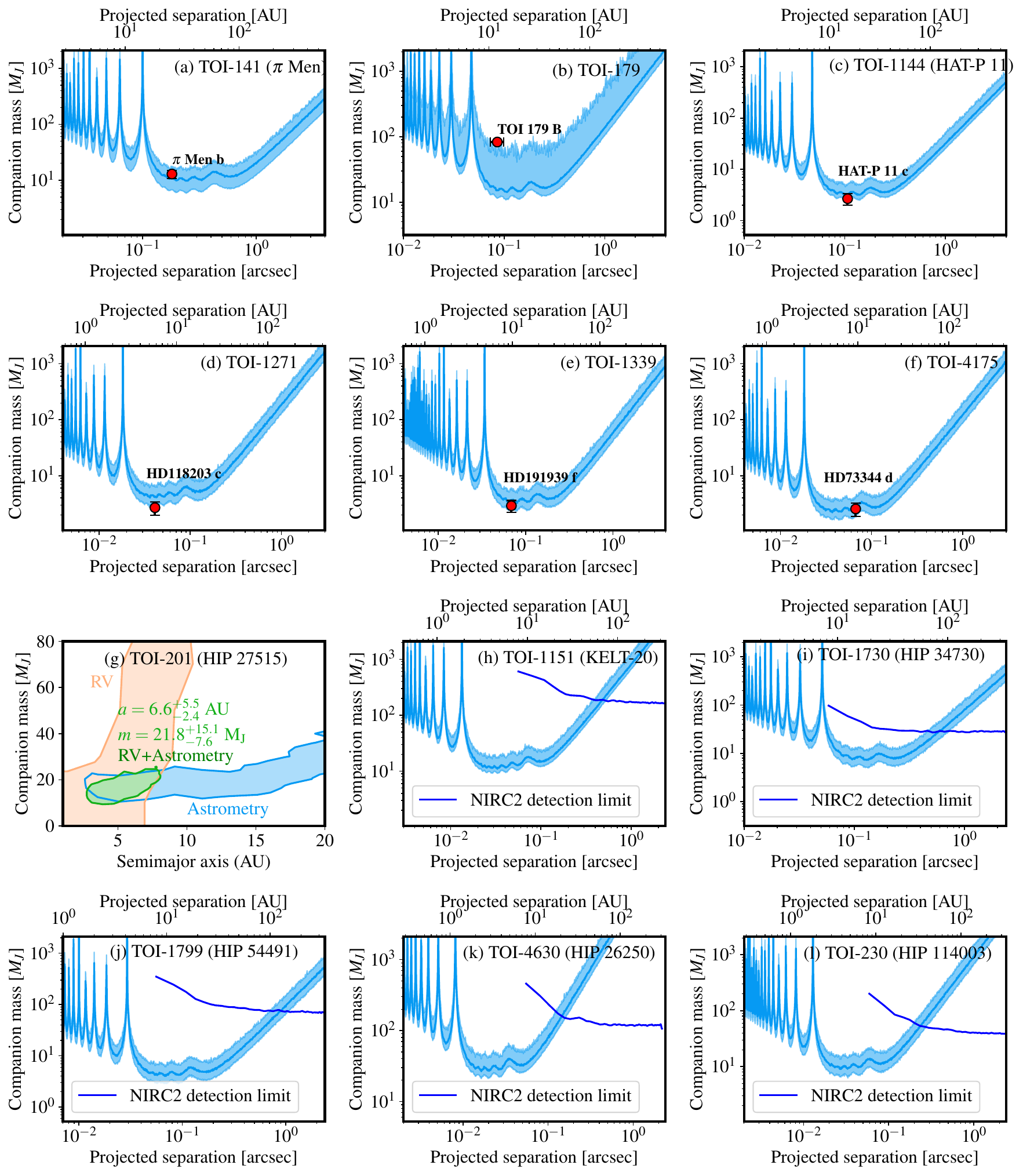}
    \caption{Constraints on the masses and semi-major axes of outer giant planet or brown dwarf companions to 12 TOIs, derived from \textit{Hipparcos}-\textit{Gaia} astrometric acceleration. Panels (a)$-$(f) show six systems with previously known companions identified via RVs or direct imaging; their properties are consistent with the observed astrometric accelerations. Panels (g)$-$(l) present six newly identified sub-stellar companions. For TOI-201, constraints come from both the RV trend and astrometric acceleration. For the remaining five TOIs, constraints are based on the combination of astrometric acceleration and AO imaging detection limits.}
    \label{fig:figure201}
\end{figure*}

Our Keck/NIRC2 observations have revealed 22 stellar companions in 20 TOIs.
Table~\ref{tab:table2X} summarizes the properties of stellar companions such as separation, position angle and contrast, to TOIs in our sample. 
Figure~\ref{fig:gallery} presents a gallery showcasing the detected stellar companions.
The sample includes two triple systems: TOI-680, a third star orbits a close inner binary \footnote{The third stars is at $\sim11$ arcsec, which is not shown in Figure~\ref{fig:gallery}}; For TOI-635, the previously known companion reported by \cite{Ziegler2020} is resolved as a close binary pair. In addition, for TOIs with declinations below $-40^\circ$—and thus not accessible to Keck—we examined archival data from previous surveys \citep{Ziegler2020, Ziegler2021, Lester2021} and identified 14 systems with known stellar companions. Most of the systems have projected separations within $2^{\arcsec}$.

Among the resolved stars, HIP 21952, HIP 33681, HIP 98516, HIP 20334, HIP 70833, HIP 79105, and HIP 22084 are listed in the Gaia Non-Single Star (NSS) acceleration catalog\citep{NSS2023}, showing reported second-derivative (Acceleration-7) or third-derivative (Acceleration-9) solutions consistent with the presence of stellar companions \citep{Halbwachs2023}. HIP 102295 and HIP 21000 appear in the SB1 catalog, though no orbital periods or mass ratios are provided. For HIP 108162, Gaia DR3 resolves the system into two components: A (Gaia DR3 6356417496318028800, RUWE = 1.15) and B (Gaia DR3 6356417496318028928, RUWE = 37.8), with a projected separation of $7.6\arcsec$. We identify HIP 108162B as the source associated with the NSS two-body orbital solution (period = 1046 days, mass ratio $\approx 0.2$), suggesting that the system is likely a hierarchical triple in which the primary star A is orbited by an unresolved binary. HIP 69275 (Gaia DR3 1667489892685370624) is included in the Gaia DR3 NSS catalog with a combined SB1 and astrometric solution (period = 195.8 days, mass ratio = 0.16). Because this detection is supported by both radial velocities and astrometry, we consider it robust. The reported transiting planet candidate for this star is therefore most likely a false positive, since the survival of a planet in such a tight binary is highly unlikely.

\subsection{Unresolved Stellar Companions}

We follow \citealt{Kervella2019} to use Hipparcos-Gaia astrometric accelerations to identify seven TOIs (TOI-501, TOI-951, TOI-1418, TOI-1924, TOI-1946, TOI-2118, and TOI-4314) with stellar companions at separations smaller than $0.1^{\arcsec}$. This is consistent with the non-detections in our Keck/NIRC2 observations and previous high-resolution imaging surveys \citep{Ziegler2020, Ziegler2021, Lester2021}, as the companions lie below the resolution limit. Furthermore, although TOI-394, TOI-1131 and TOI-5811 each have known stellar companions, their astrometric accelerations indicate the presence of additional companions at separations below $0.1^{\arcsec}$. Constraints on companion masses and separations for these systems are provided in the Figure~\ref{fig:unres} in the appendix. We acknowledge that some of these systems may exhibit spurious HGCA signals rather than hosting very close-in stellar companions. Misclassifications of spurious HGCA signals (at most 7) would contribute negligibly to our sample (35 stars with companions, either resolved or unresolved) and our statistical analysis only uses the resolved binaries, so we leave the detailed validation of the unresolved binary companions to future work.

\subsection{ Giant Planet or Brown Dwarf Companions}

In addition to stellar companions, our sample includes 12 TOIs whose astrometric accelerations are caused by long-period giant planets and brown dwarfs, typically at a few AU. Among them, six systems have known giant planets or brown dwarfs, including TOI-144 b ($\pi$ Men b, \citealt{Xuan2020,DR2020,Damasso2020}), TOI-179 B \citep{Desidera2023}, TOI-1144 c (HAT-P-11 c, \citealt{Yee2018, Xuan2020}), TOI-1339 f (HD 191939 f, \citealt{Lubin2022}), TOI-1271 c (HD 118203 c, and \citealt{Zhang2024c, Maciejewski2024}), TOI-5140 d (HD 73344 d, \citealt{Zhang2024d}).

Furthermore, we identified six additional systems—TOI-201, TOI-230, TOI-1151 (KELT-20), TOI-1730, TOI-1799, and TOI-4063—that likely host giant planet or brown dwarf companions. While our AO imaging observations could not directly detect these sub-stellar companions, they provide upper limits on their masses and separations. By combining these limits with radial velocities or astrometric acceleration (following the methods of \citealt{Kervella2019, Lubin2022}), we place tight constraints on the allowed companion parameter space (see Figure~\ref{fig:figure201}). Notably, TOI-201 \citep{Hobson2021}, TOI-1151 \citep{Lund2017}, and TOI-4603 \citep{Khandelwal2023} each host a known hot Jupiter, making the characterization of their outer companions especially valuable for probing hot Jupiter migration histories. For TOI-201, using published RVs from 2018 to 2020, we combined the RV trend with astrometric acceleration to constrain the mass of the companion to $21.8^{+15.1}_{-7.6}\ M_J$ and the semi-major axis to $6.6^{+5.5}_{-2.4}$ AU (1$\sigma$). For the other five systems, AO imaging rules out the presence of stellar companions and instead indicates  the existence of planetary-mass or brown dwarf companions.

\section{Orbits of stellar companions}\label{sec:3dorbit}

We performed three-dimensional orbital fitting for a subset of systems with resolved stellar companions, focusing on those where sufficient astrometric information is available. Our selection criteria for systems to perform orbital fitting are: (1) the binary with planets or planet candidate (excluding known false positives), (2) the systems have a resolved stellar companion with astrometry spanning more than one epoch, and (3) there are only one stellar companion \footnote{TOI-455 and TOI-1099 are exceptions, as it is the primary star orbited by a close binary pair; for the purposes of orbital fitting, we treat the companion pair as a single object}. This leaves a sample of 12 TOIs for which we derive well-constrained stellar companion orbits.

Our analysis follows Paper I, in which we presented a detailed characterization of the planet-hosting triple system TOI-455 (LTT 1445), where the companion pair orbits the primary in an edge-on configuration, and all three stars share a nearly coplanar orbital plane. We do not repeat the orbital fitting of TOI-455 but include the system in the statistical analyses in Section~\ref{sec:mul}.

\begin{center}
\begin{longtable*}{lccccccccc}
\caption{Summary of relative astrometry for stellar companions}\label{tab:table2X}\\
\hline
TOI& HIP  & Obs. & Comp. & $\rho$ &  $\theta$ & $\Delta \rm m$ & Inst. & Filter &Ref.  \\
 &  Number &  Time &  & $(\arcsec) $ & (degree) & (mag) & & &  \\
\hline
\hline
\endfirsthead
\hline
TOI & HIP  & Obs. & Comp. & $\rho$ &  $\theta$ & $\Delta \rm m$ & Inst. & Filter &Ref.  \\
 &  Number &  Time &  & $(\arcsec) $ & (degree) & (mag) & & &  \\
\hline
\hline
\endhead
\hline
\endfoot
\multicolumn{10}{p{\textwidth}}{
{\footnotesize
\textsuperscript{1} \cite{ESA}. For HIP 108162 and HIP 78301, \textit{Hipparcos} reported ambiguous double-star detections that are inconsistent with measurements from other instruments. We therefore exclude them from the tables.} \newline
\textsuperscript{2} \cite{Gaia}.For binaries resolved by \textit{Gaia}, we inflated the separation uncertainties to 0.1 mas to prevent the single Gaia measurement from dominating the orbital fit. \newline
\textsuperscript{3}{\cite{Ziegler2020}}, \textsuperscript{4}{\cite{Lester2021}}, \textsuperscript{5}{\cite{Lester2023}}, \textsuperscript{6}{\cite{Desidera2023}}, \textsuperscript{7}{\cite{Cutri2003}}, \textsuperscript{8}{TFOP}, \textsuperscript{9}{  \cite{Howell2021}}, \textsuperscript{10}{  \cite{Dieterich2012}}, \textsuperscript{11}{  \cite{Rodriguez2015}}, \textsuperscript{12}{  \cite{zhou2022}}
}
\endlastfoot
\multicolumn{10}{c}{\textit{Confirmed and Candidate Planet Host Stars}} \\
\hline
 128 & 24718 & 1991.51 &A-B& $2.154\pm{0.017}$  & $ 153.1\pm{0.2}$&  2.9 & \textit{Hipparcos}  & Hp  & 1\\
 128 & 24718 & 2016.00 &A-B& $2.2014\pm{0.0001}$  & $ 153.00\pm{0.01}$& 2.6 & \textit{Gaia} DR3  & G  & 2 \\
 128 & 24718 & 2018.81 &A-B& $2.220\pm{0.005}$  & $ 153.8\pm{0.5}$& 2.4  & SOAR  & I  & 3 \\
 141 & 111553 & 2018.73 &A-B& $0.443\pm{0.005}$  & $ 239.5\pm{0.5}$& 4.9 & SOAR  & I  & 3 \\
 141 & 111553 & 2018.73 &A-C& $ 1.200\pm{0.005}$  & $ 305.2\pm{0.5}$& 5.4 & SOAR  & I  & 3 \\
 179 & 13754 &  2019.77  & A-B & $0.0845\pm{0.0036}$ &   $212.0\pm{1.5}$  & 6.4 & SPHERE &J band & 6 \\     
 271 & 21952& 2019.70 &A-B& $ 0.133\pm{0.005}$  & $ 231.1\pm{0.5}$& 5.0 & Gemini/Zorro  & 832 nm & 5\\
 271 & 21952& 2021.71 &A-B& $ 0.153\pm{0.005}$  & $ 225.6\pm{0.5}$& 5.0 & Gemini/Zorro  & 832 nm & 5\\
 369  & 115594 & 1991.50 &A-B& $0.75\pm{0.01}$  & $ 141.9\pm{0.6}$& 1.4 & \textit{Hipparcos}  & Hp  & 1  \\
 369  & 115594 & 2016.00 &A-B& $0.6986\pm{0.0001}$  & $ 139.93\pm{0.01}$& 1.3 & \textit{Gaia} DR3  & G  & 2  \\
 369  & 115594 & 2019.38 &A-B& $0.695\pm{0.005}$  & $139.3\pm{0.5}$& 1.3 & SOAR  & I  & 3  \\
 369  & 115594 & 2022.46 &A-B& $0.687\pm{0.005}$  & $139.1\pm{0.5}$& 1.3 & Keck/NIRC2 & Kcont  & This work  \\
 369  & 115594 & 2023.57 &A-B& $0.685\pm{0.005}$  & $138.9\pm{0.5}$& 1.1 & Keck/NIRC2 & Kcont  & This work \\
 402  & 11433 &2019.53 &A-B&$1.439\pm{0.005}$ & $233.4\pm{0.5}$& 5.2 & SOAR & I  & 3 \\
 402  & 11433 &2019.78 &A-B&$1.467\pm{0.005}$ & $234.2\pm{0.5}$& 5.3 & SOAR & I  & 3 \\
 402  & 11433 & 2021.82 &A-B&$1.466\pm{0.005}$ & $233.1\pm{0.5}$& 3.6 & Keck/NIRC2 &Kcont  & This work \\
 402  & 11433 &2022.11 &A-B&$1.471\pm{0.005}$ & $232.85\pm{0.5}$& 3.8 & Keck/NIRC2 & Kcont  & This work \\
 402  & 11433 & 2023.57 &A-B& $1.489\pm{0.005}$& $232.42\pm{0.5}$& 3.8 & Keck/NIRC2 & Kcont  & This work\\
455 & 14101 & 2003.46  & A-BC & $7.13\pm{0.005}$ & $314.74\pm{0.5}$ & - & HST/NICMOS & - &10\\
455 & 14101 & 2010.59  & A-BC & $7.02\pm{0.005}$ & $314.91\pm{0.5}$ & - & Lick/IRCAL& - &11\\
906  & 78301 & 2016.00 &A-B& $1.245\pm{0.0001}$  & $ 50.47\pm{0.01}$& 3.2 & \textit{Gaia} DR3  & G  & 2  \\
 906 & 78301 & 2019.61 &A-B& $1.246\pm{0.005}$  & $ 50.8\pm{0.5}$& 2.8  & SOAR  & I  & 3 \\
 930 & 16881 & 1991.50 &A-B& $0.7032\pm{0.02}$  & $ 37.9\pm{0.5}$& 3.0  & \textit{Hipparcos}  & Hp & 1 \\
 930 & 16881 & 2019.61 &A-B& $0.692\pm{0.005}$  & $ 32.6\pm{0.5}$& 2.5  & SOAR  & I  & 3 \\
 930 & 16881 & 2020.97 &A-B& $0.695\pm{0.005}$  & $ 33\pm{0.5}$& 3.1  & Gemini/Zorro & 832 nm  & 4 \\
 1099 & 108162  & 1999.926 &A-B& $7.71\pm{0.08}$  & $ 308.9\pm{0.1}$& 1.4  & 2MASS  & $K_s$ & 7 \\
 1099 & 108162  & 2016.00 &A-B& $7.5977\pm{0.0001}$  & $ 310.51\pm{0.01}$& 2.9  & \textit{Gaia} DR3  & G  & 2 \\
 1131  & 81087 &2022.46 &A-B& $0.1183\pm{0.0050}$&$212.6\pm{2.5}$ & 2.0 & Keck/NIRC2 & Kcont  & This work \\
 1131  & 81087 &2023.11 &A-B& $0.1186\pm{0.0050}$&$214.4\pm{2.5}$ & 1.6 & Keck/NIRC2 & Kcont  & This work \\
 1131  & 81087 &2023.48 &A-B& $0.1206\pm{0.0050}$&$214.4\pm{2.5}$ & 1.8 & Keck/NIRC2 & Kcont  & This work \\ 
 1204  & 55069 & 2020.02 &A-B& $0.344\pm{0.010}$&$38.8\pm{0.5}$ & 5.0 & Gemini/Zorro & 832 nm  & 4\\
 1204  & 55069 & 2023.50 &A-B& $0.353\pm{0.010}$&$30.7\pm{0.5}$ & 5.9 & Gemini/Zorro & 832 nm  & 8\\
1204  & 55069 & 2025.02 &A-B& $0.337\pm{0.010}$&$32.4\pm{0.5}$ & 5.9 & Gemini/Zorro & 832 nm  & 8\\
 4175  & 64573 & 1991.50 &A-B& $1.327\pm{0.005}$&$260.65\pm{0.1}$ & 0.2 & \textit{Hipparcos}  & Hp  & 5 \\
 4175  & 64573 & 2016.00 &A-B& $1.6103\pm{0.0001}$&$259.77\pm{0.01}$ & & \textit{Gaia} DR3  & G  & 2 \\
 4175  & 64573 & 2021.53  &A-B& $1.678\pm{0.005}$&$259.6\pm{0.5}$ & & Gemini/Zorro & I: (832) nm  & 4\\
 4175  & 64573 & 2023.00 &A-B& $0.172\pm{0.005}$&$262.9\pm{0.5}$ & 0.5& Keck/NIRC2 & Kcont  & This work \\
 4399  & 94235 & 2010.57 &A-B& $0.506\pm{0.007}$ & $150.6\pm{0.8}$ & 3.8 & VLT-NaCo & Hband & 12 \\
 4399  & 94235 & 2021.56 &A-B& $0.596\pm{0.005}$ & $162.9\pm{0.5}$ & 5.8 &Gemini/Zorro & 832 nm & 12 \\
 4399  & 94235 & 2021.81 &A-B& $0.600\pm{0.008}$ & $161.7\pm{0.8}$ & 5.3 & Gemini/Zorro & 832 nm & 12 \\
4568 &  77921 &  1991.50 &  A-B &  $0.213\pm{0.005}$ &  $288.5\pm{1.0}$&  0.14 &  \textit{Hipparcos}  &  Hp  &  5\\
 4568 & 77921 & 2023.17 & A-B & $0.348\pm{0.005}$ & $299.2\pm{0.5}$ & 0.3 & Gemini/Zorro & 832 nm & 8\\
  4568 & 77921 & 2023.17 & A-C & $1.199\pm{0.005}$ & $50.1\pm{0.5}$ & 7.8 & Gemini/Zorro & 832 nm & 8\\
 5811 & 102295 &  1991.50  &  A-B &  $7.434\pm{0.002}$ &  $306.5\pm{0.1}$ &  0.9 &  \textit{Hipparcos}  &  Hp &  5\\
5811 & 102295 & 2023.43 & A-B & $7.445\pm{0.053}$ & $307.2\pm{0.03}$ & 0.2 & Palomar & Brgamma & 8\\
 5962 & 112486 & 2016.00 &A-B&  $1.7041\pm{0.0001}$&$262.83\pm{0.01}$ & 3.9 & \textit{Gaia} DR3 & G  & 2 \\
 5962 & 112486 & 2023.49 &A-B&  $1.709\pm{0.005}$&$263.2\pm{0.5}$ & 2.5 & Keck/NIRC2 & Kcont  & This work \\
 5962 & 112486 & 2024.97 &A-B&  $1.711\pm{0.005}$&$263.2\pm{0.5}$ & 2.6 & Keck/NIRC2 & Kcont  & This work \\
\hline
\multicolumn{10}{c}{\textit{Transit False Positive Stars}} \\
\hline
 394 & 15053 &  1991.50 &  A-B &  $3.248\pm{0.002}$ &  $319.5\pm{0.2}$ &  3.8&  \textit{Hipparcos} &  Hp  &  5\\
 522 & 40694 & 2023.57 &A-B&$0.757 \pm{0.005}$ & $241.9\pm{0.5}$&  3.2 & Keck/NIRC2 & Kcont  & This work \\
 575  & 41849 & 2023.00 &A-B& $0.643 \pm{0.005}$& $192.9\pm{0.5}$& 3.0 & Keck/NIRC2 & Kcont  & This work \\
621 & 44094 & 2019.38 &A-B& $2.066\pm{0.005}$  & $ 19.2\pm{0.5}$& 6.5  & SOAR  & I  & 3 \\
 635  & 47990 & 2022.11 & A-B&$1.7403 \pm{0.0050}$ &$260.9\pm{0.2}$ & 3.5 & Keck/NIRC2 & Kcont  & This work \\
 635  & 47990 & 2023.00 & B-C& $0.0995 \pm{0.0050}$& $216.6\pm{2.0}$& 1.4 & Keck/NIRC2 & Kcont  & This work \\
 680  & 58234 & 2022.11 &A-B&$0.0446 \pm{0.0050}$ &$352.7\pm{1.0}$ & 0.5 & Keck/NIRC2 & Kcont  & This work \\
 680 & 58234 & 2022.11 &A-C& $0.7968\pm{0.0050}$&$332.1\pm{0.4}$ &2.4 & Keck/NIRC2 & Kcont  & This work \\
 896  & 28122 & 2020.91 & A-B &  $0.062\pm{0.005}$ & $126.2\pm{0.5}$ & 3.5 & Gemini/Zorro&  I: (832) nm  & 4 \\
 909 & 82032 & 2021.53 &A-B& $1.347\pm{0.005}$  & $ 268.3\pm{0.5}$& 4.3  & SOAR  & I  & 3 \\
 1665 & 27844 &  1991.50  &  A-B &  $1.712\pm{0.001}$ &  $355.2\pm{0.1}$ &  1.9 &  \textit{Hipparcos}  &  Hp  &  5\\
 1665 & 27844 & 2023.00 &A-B&$ 1.953\pm{0.005}$&$0.79\pm{0.5}$ & 1.7 & Keck/NIRC2 & Kcont  & This work \\
 1684  & 20334 & 2022.10 &A-B&$0.1203\pm{0.005}$& $5.45\pm{1.00}$ &  4.5 & Keck/NIRC2 & Kcont  & This work \\
 1719  & 44289 & 2022.10 &A-B&  $0.1199\pm{0.0050}$& $195.94\pm{1.00}$ & 2.0 & Keck/NIRC2 & Kcont  & This work \\
 1719 & 44289 & 2023.00&A-B&  $0.1216\pm{0.0050}$& $196.25\pm{1.00}$ & 2.1 & Keck/NIRC2 & Kcont  & This work \\
 1719  & 44289 & 2024.96 &A-B&$0.1269\pm{0.0050}$ &$198.60\pm{1.00}$ & 2.1 & Keck/NIRC2 & Kcont  & This work \\
 1831  & 65205 & 2023.00 &A-B& $0.645\pm{0.005}$&$319.2\pm{0.5}$ & 4.3 & Keck/NIRC2 & Kcont  & This work \\
 1831  & 65205 & 2024.97 &A-B& $0.635\pm{0.005}$&$319.7\pm{0.5}$ & 4.1 & Keck/NIRC2 & Kcont  & This work \\
 1837 & 67650 & 2023.00 &A-B& $0.144\pm{0.005}$&$75.9\pm{1.0}$ & 2.5 & Keck/NIRC2 & Kcont  & This work \\
 2017  & 74685 & 2023.00  &A-B& $0.377\pm{0.005}$&$71.5\pm{0.5}$ & 1.9 & Keck/NIRC2 & Kcont  & This work  \\
 2299 & 93711 &  1991.50  &  A-B &  $13.616\pm{0.002}$ &  $279.6\pm{0.1}$ &  3.4&  \textit{Hipparcos} &  Hp &  5\\
 2299  & 93711 & 2022.46 &A-B& $3.599\pm{0.005}$&$281.3\pm{0.5}$ & 2.1 & Keck/NIRC2 & Kcont  & This work \\
 2299 & 93711 & 2023.48 &A-B& $ 3.595\pm{0.005}$&$281.3\pm{0.5}$ & 2.3 & Keck/NIRC2 & Kcont  & This work \\
 2666  & 45621 & 2023.00 &A-B& $1.249\pm{0.005}$&$0.16\pm{0.5}$& 4.2 & Keck/NIRC2 & Kcont  & This work \\
 5079  & 24007 & 2022.11&A-B& $2.686\pm{0.005}$&$283.5\pm{0.5}$ & 1.4 & Keck/NIRC2 & Kcont  & This work \\
 5079  & 24007 & 2023.00&A-B& $2.6878\pm{0.005}$&$283.5\pm{0.5}$ & 1.1& Keck/NIRC2 & Kcont  & This work\\
5521 & 57386 & 2023.00 &A-B& $0.425\pm{0.005}$&$213.1\pm{0.5}$ &4.4 & Keck/NIRC2 & Kcont  & This work \\
 5521 & 57386 & 2024.96 &A-B& $0.416\pm{0.005}$&$212.7\pm{0.5}$ &4.4 & Keck/NIRC2 & Kcont  & This work \\
5383 & 69275 & 2024.31 & A-B & $0.833\pm{0.005} $ & $353.3\pm{0.5}$ & 4.2 & Palomar/PHARO & Kcont & 8  \\
5383 & 69275 & 2022.29 & A-B & $0.839\pm{0.005} $ & $352.6\pm{0.5}$ & 6.6 & WIYN/NESSI & 832 nm & 9  \\
 5383 & 69275 & 2023.00 & A-B & $0.8438\pm{0.005} $ & $ 351.92\pm{0.5}$ & 4.4 & Keck/NIRC2 & Kcont & This work  \\
 5383 & 69275 & 2024.96 & A-B & $0.8342\pm{0.005} $ & $ 352.63\pm{0.5}$ & 4.2 & Keck/NIRC2 & Kcont & This work  \\
\hline
\end{longtable*}
\end{center}

\subsection{Binary Masses from Isochrone Model}\label{sec:mass}

We used Python package \textit{isoclassify} \citep{huber2017, berger2020} to derive stellar parameters for the primary and secondary stars in the binary systems for which we performed orbital fitting. 
First, we input the apparent magnitudes $m_{A}$ of the host stars and the contrast ($\Delta m$, in magnitude) between the primary and secondary stars. The above two parameters have to been in the same photometric band. Since the binaries in our sample were resolved using different instruments and in various photometric bands, we adopt the $m_{A}$
values corresponding to the band in which each companion was observed. For systems resolved in the K-band with Keck/NIRC2 from our observations, we adopt the 2MASS $K_{s}$ magnitude \citep{Cutri2003} for the host stars. For systems resolved by Gaia, we used magnitudes in Gaia $G$ band for the two components.  We also input the parallax for the each system from \textit{Gaia} DR3 \citep{Gaia}. Furthermore, we adopted the spectroscopic effective temperature $\rm{T_{eff}}$, metallically $\rm [Fe/H]$ of the host stars from high-resolution spectra collected by HIRES \citep{Petigura2013} or HARPS \citep{Perdelwitz2024} when they are available. We adopt  uncertainties of 100 K in $\rm{T_{eff}}$ and 0.05 in $\rm [Fe/H]$ for all of our stars. The masses, radius and effective temperature $\rm{T_{eff}}$, and surface gravity $log g$ of primary and secondary stars are presented in Table~\ref{tab:table_maas} in the appendix. We inflate the uncertainties in radius and mass to $5\%$, and in \( T_{\mathrm{eff}} \) to $2\%$, to account for systematic errors associated with isochrone models \citep{Tayar2022}. The derived primary and secondary masses are used as priors in our orbital fits.


\subsection{Three-dimensional Orbital Fitting}

We used \texttt{orvara} \citep{Brandt2021b}, which performs a parallel-tempering Markov Chain Monte Carlo (MCMC) fitting. We used six orbital parameters to define the orbit of stellar companions, including semi-major axis ($a$), inclination ($i$), longitude of the ascending node ($\Omega$), mean longitude at a reference epoch ($t_{\rm ref}$) of 2455197.5 JD ($\lambda_{ref}$), the eccentricity ($e$) and the argument of periastron ($\omega$) in the form of $e\sin \omega$ and $e\cos \omega$. Two additional parameters for the masses of the host star ($M_{\rm{A}}$) and  companion ($M_{\rm{B}}$).  We assume Gaussian priors for $M_{*}$ and $M_{\rm{B}}$ based on the masses from isochrone models in section~\ref{sec:mass}. We adopted a uniform prior between 0 and 1  on the eccentricity. We also include parallax $(\varpi)$ as a fitted parameters, adopting a Gaussian prior based on the \textit{Gaia} DR3 value and its reported uncertainty.

The likelihood is computed by comparing multiple observational constraints: relative astrometry (separation and position angle), absolute astrometry (\textit{Hipparcos}-\textit{Gaia} astrometric acceleration), and when available, radial velocities (RVs) of the host star, relative radial velocities between the host and companion, the companion's proper motion \citep{Brandt2021b}:

\begin{align}
\ln \mathcal{L} = -\frac{1}{2} (&\chi^2_{\mathrm{rel\ ast}} + \chi^2_{\mathrm{abs\ ast}} + \nonumber \\
\chi^2_{\mathrm{RV}} &+ \chi^2_{\mathrm{rel\ RV}} + \chi^2_{\mathrm{comp.\ pm}})
\end{align}

Relative astrometry and \textit{Hipparcos}-\textit{Gaia} astrometric acceleration are available for all systems.  Relative RVs are used for TOI-402, TOI-5962, TOI-128, TOI-1099, and TOI-4175. Companion proper motions are available for TOI-128, TOI-906, TOI-4175, and TOI-5962. Host star RVs are only included for TOI-402. In the fitting of TOI-402,  we included zero-point of the system velocity $\gamma$ and the intrinsic jitters $\sigma_{jit}$ for each set of the instrument. 

\begin{figure*}
    \centering
    \includegraphics[width=\linewidth]{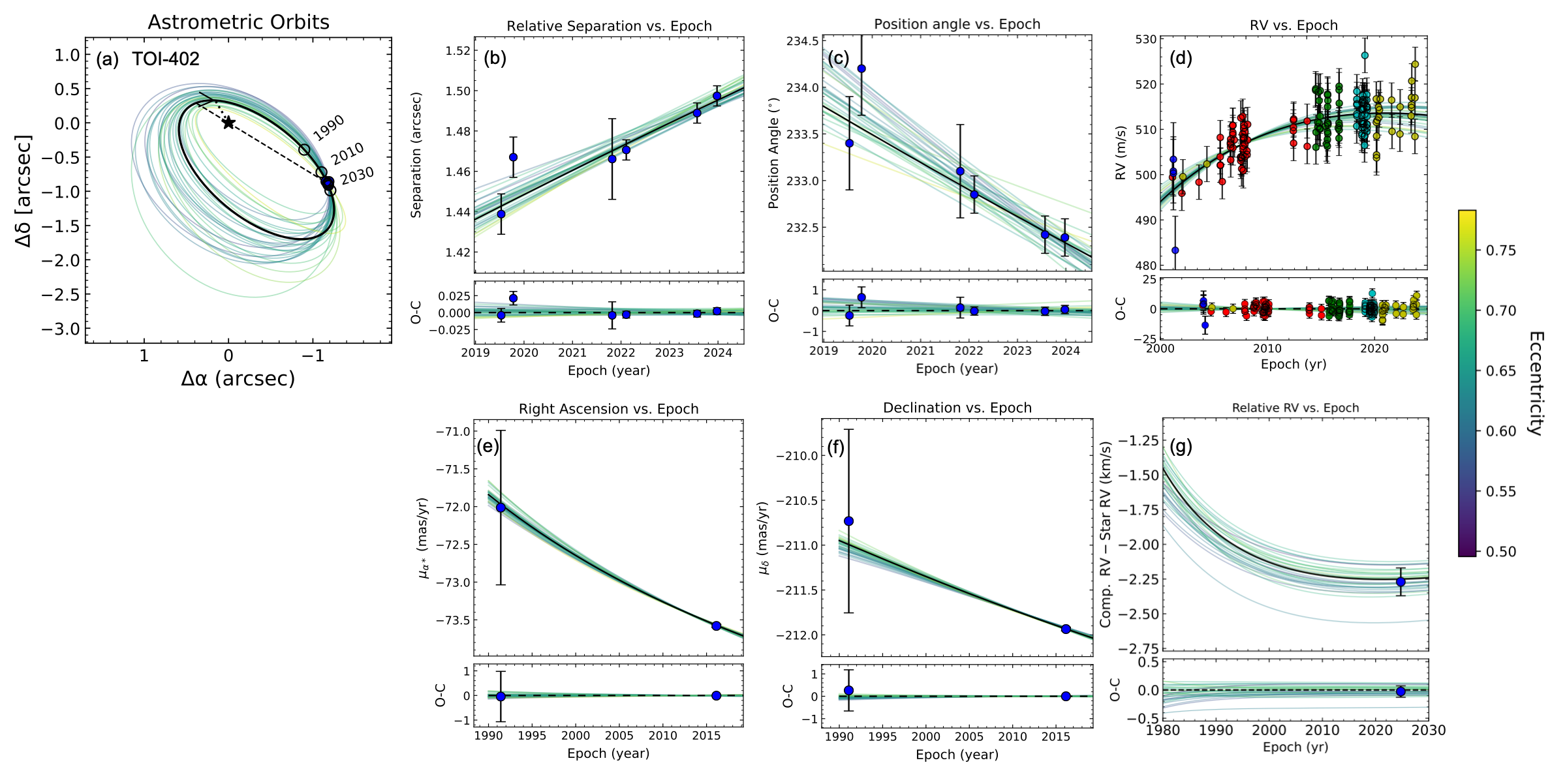}
    \caption{Orbital characterization of TOI-402 B  around A using multiple datasets.  (a): astrometric orbits of TOI-402 B  around A. The blue filled circles are observed relative astrometry used in our analysis. The empty circles are predicted locations of the stellar companion based on the best-fit solution at epoch 1990, 2010 and 2030.   (b)-(c): Observed and fitted separation and  position angle of TOI-402 B relative to TOI-402 A. (d): Observed and fitted RVs of TOI-402 A from HARPS, HIRES,and PFS. (e)-(f): Observed and fitted Hipparcos and Gaia proper motion of TOI-402 A in right ascension and declination. (g) Observed and fitted relative RV between primary star and the companion taken by KPIC. In all of above panels, the thicker black lines represent the best-fit orbit in the MCMC chain. The remaining 50 lines are random samples from the posterior chain, color-coded by eccentricity. The complete set of orbital characterization plots (12 figures) is available as a figure set in the online Journal.} 
    \label{fig:figure402}
\end{figure*}

We used 40 different temperatures and 100 walkers to sample the parameter space over $1.2\times 10^{5}$ steps. Our results are taken from the `coldest' chain, which corresponds to the original, unmodified likelihood function.  We saved every 50th step of our chains and discarded the first $50\%$ of the chain as the burn-in portion. Figure \ref{fig:figure402} and  shows the best-fit orbit (black lines) from our MCMC chains for TOI-402, including fitted astrometric orbit, RVs and Hipparcos-Gaia proper motions. The data and best-fit model for other systems are presented in Appendix.  Table~\ref{tab:table_re} presents the orbital fitting results for all the systems. 
For systems  dispositioned as ``virtual binaries'' in  \textit{Hipparcos} catalog (sources for which a single-star solution was not adequate that were improved by the inclusion of a non-linear acceleration or a binary companion: TOI-128, TOI-369, TOI-906, TOI-930, TOI-1099 and TOI-4175, \citealt{ESA1997}), the reported proper motions may be less reliable because the catalog generally constrained the same proper motion for the primary and secondary. 
To test the impact of this assumption, we re-ran the orbital fits after inflating the Hipparcos proper-motion uncertainties by a factor of 10 and obtained results consistent with those derived from the original, non-inflated uncertainties.

\subsection{Results for Individual Systems}

\subsubsection{TOI-128}
TOI-128 A is a bright, sunlike star ($\rm M_{A}={1.08}_{-0.05}^{+0.05}\ M_{\odot}$, $G\sim8.4$) with a distance of $\sim68.2$ pc. TOI-128 A has a stellar companion at $\sim2.2^{\arcsec}$ with a magnitude difference of $\sim 2.6$ in Gaia G band. The \textit{TESS} mission identified a planet candidate, TOI-128.01, with a radius of $\sim 2.3\,R_{\oplus}$ and an orbital period of 4.94 days \citep{TOI_ref}.
Assuming the magnitude difference in the TESS band is similar to that in the Gaia G band, the flux dilution from the companion will not significantly bias the measurement of the planet's radius. Ground-based follow-up observations confirmed that the transit occurs on the primary star, supporting its classification as a planet candidate. Additionally, three CORALIE radial velocity measurements taken between Aug. 09, 2018 and Oct. 25, 2018 rule out the possibility of an eclipsing binary \footnote{$\rm https://dace.unige.ch/radialVelocities/?pattern=TOI\%20128$}.  

This binary has been resolved by both \textit{Hipparcos} and \textit{Gaia}, providing measurements of separation and position angle relative to the primary star at the two epochs. We also include relative astrometry from observations by \cite{Ziegler2020} obtained in 2018, extending the observational baseline to over 27 years. Furthermore, \textit{Gaia} DR3 provides radial velocities and proper motions for both stars, which provides the information about the relative motion between two stars. 

Our results show that TOI-128 B is a K dwarf ($M_{B}={0.71}_{-0.09}^{+0.09}\ M_{\odot}$), with a semi-major axis of ${113}_{-19}^{+27}\ $ AU and eccentricity of ${0.55}_{-0.30}^{+0.26}$. The orbital inclination is ${77.8}_{-6.4}^{+2.7}$ deg. The $\chi^{2}$ of the best-fit solution compared to the astrometric measurements indicates a good fit and  well-estimated uncertainties,
with values of 6.2 and 2.1 for separations and PAs (three measurements each) and 3.49 for the four astrometric accelerations. 

\subsubsection{TOI-271}

TOI-271~A is a G0V star ($\rm M_{A}={0.97}_{-0.06}^{+0.06} \ M_{\odot}$, $G\sim 8.8$, distance $\sim 100$~pc) with a faint, close companion located approximately \(0.13^{\prime\prime}\) away, exhibiting a magnitude difference of 5 in the 832~nm filter from Gemini/Zorro observations. The \textit{TESS} mission identified a small transiting planet candidate, TOI-271.01, with a radius of \(\sim 2.8\,R_{\oplus}\) and an orbital period of 2.47~days. Flux dilution from the nearby companion in the TESS bandpass is minor compared to the transit depth and does not significantly bias the inferred planetary radius. Ground-based follow-up observations confirmed a full transit on the target star and excluded all Gaia-detected stars beyond \(0.5^{\prime\prime}\). However, these observations did not resolve the close companion at \(0.13^{\prime\prime}\), leaving ambiguity as to whether the transit occurs on the primary or the companion star.

TOI-271 lies in the southern sky ($\delta \sim -50^{\circ}$) and is not accessible from Keck. Therefore, we adopt two epochs of relative astrometry from \citet{Lester2023}, taken in September 2019 and September 2021 using Zorro speckle camera on the Gemini South telescope. Using the two measurements, \citet{Lester2023} derived a highly eccentric and nearly edge-on orbit for the stellar companion ($a_B=11.0^{+4.3}_{-3.4}$ AU, $e_B=0.95^{+0.05}_{-0.08}$, $I_B=98.5^{+11}_{-6.2}$ deg).

In our analysis, we incorporate not only the two epochs of relative astrometry but also the \textit{Hipparcos}-\textit{Gaia} astrometric acceleration, providing a 25-year baseline. Our orbital fit yields a less eccentric and more inclined solution: $a_B={12.52}_{-0.56}^{+0.80}$ AU, $e_B={0.80}_{-0.05}^{+0.06}$, $I_B={111.3}_{-7.7}^{+10}$ deg. Our results are consistent with those of \citet{Lester2023} within $1\sigma$.

\subsubsection{TOI-369}
TOI-369 has a F7V primary star ($\rm M_{A}={1.38}_{-0.07}^{+0.07} \ M_{\odot}$, $G\sim 9.5$, distance $\sim 227.8$~pc) and a G type stellar companion at a separation of $\sim0.7^{\arcsec}$ and with a magnitude difference of 1.3 in Gaia G band. \textit{TESS} observations reveal a transiting planet candidate at orbital period of 5.4 days with a planet radius of $\sim 10.9 R_{\oplus}$. If we consider the  flux dilution from the companion star, the revised radius is $\sim 12.4 R_{\oplus}$, indicating the planet radius is inflated.  
We observed the radial velocities of the star using Keck Planet Finder between August 2023 and August 2024. In addition, we observed the RM observations of the planet on July 25 2024 using KPF. These observations confirm the planetary nature of the transit, and the detailed results will be presented in a separate paper (Zhang et al. in prep).

The binary system was resolved in the \textit{Gaia} DR3 observations, providing astrometric data around the 2016 epoch. We also included an additional astrometric measurement published by \cite{Ziegler2020}, taken in May 2019, along with three Keck/NIRC2 measurements from our own survey. 

For TOI-369 system, we obtain a companion mass of $\rm{M_{B}}=1.13_{-0.06}^{+0.06}\ M_{\odot}$ for the stellar companion. The stellar companion has a semi-major axis of $a_{B}={134}_{-26}^{+53}$ AU, with eccentricity of $e_{B}={0.67}_{-0.37}^{+0.20}$ and an inclination of $I_{B}={122}_{-12}^{+26}$ deg. TOI-369 hosts a hot Jupiter candidate, and our analysis indicates a misalignment between the transiting planet and the outer stellar companion. This could suggest the occurrence of the von Zeipel-Kozai-Lodov cycle leading to the migration of the hot Jupiter driven by the misaligned stellar companion. 



\begin{center}
\begin{longtable*}{llccccc}
\caption{Orbital Fitting Results}\label{tab:table_re}\\
\hline

Name & HIP & $M_{A}$ &  $M_{B}$ & $a_{B}$ & $I_{B}$ & e   \\
 &  Number &  $\rm M_{\odot}$ & $\rm M_{\odot}$& AU & deg &  \\
\hline
\hline
\endfirsthead
\hline
Name & HIP & $M_{A}$ &  $M_{B}$ & $a_{B}$ & $I_{B}$ & e   \\
 &  Number &  $M_{\odot}$ & $M_{\odot}$& AU & degree &  \\
\hline
\hline
\endhead
\hline
\endfoot
\multicolumn{7}{p{0.6\linewidth}}{
{\footnotesize
\textsuperscript{*} Note: The masses listed for both primary and secondary  are essentially the priors from stellar isochrone fitting. \newline
\textsuperscript{\dag} For TOI-1099, our isochrone model treats the companion component as a single star and does not account for the fact that the companion is an unresolved binary pair. To reflect the potential bias introduced by this simplification, we add an additional $10\%$ uncertainty to the mass estimate.
}
}
\endlastfoot
TOI-128 & 24718 &  ${1.08}_{-0.05}^{+0.05}$ &  ${0.71}_{-0.09}^{+0.09}$ &   ${113}_{-19}^{+27}$ &  ${77.8}_{-6.4}^{+2.7}$ & ${0.55}_{-0.30}^{+0.26}$  \\
TOI-271 & 21952 &  ${0.97}_{-0.06}^{+0.06}$ &  ${0.56}_{-0.03}^{+0.03}$ &    ${12.5}_{-0.6}^{+0.80}$ &   ${111.3}_{-7.7}^{+10}$ &  ${0.80}_{-0.05}^{+0.06}$  \\
TOI-369 & 115594 & ${1.38}_{-0.07}^{+0.07}$ & $1.13_{-0.06}^{+0.06}$&  ${134}_{-26}^{+53}$ &  ${122}_{-12}^{+26}$ & ${0.67}_{-0.37}^{+0.20}$  \\
TOI-402 & 11433 & $0.88_{-0.05}^{+0.05}$ &$0.26_{-0.02}^{+0.02}$ & $59.9_{-6.7}^{+9.4}$ & $126.9_{-6.7}^{+5.8}$ &  ${0.66}_{-0.06}^{+0.06}$ \\
TOI-455 & 14101 &  ${0.25}_{-0.01}^{+0.01}$ &$0.39^{+0.01}_{-0.01}$ & ${58}_{-10}^{+16}$ & ${88.5}_{-1.3}^{+1.3}$ & ${0.38}_{-0.04}^{+0.08}$  \\
TOI-906 & 78301 &  ${1.14}_{-0.06}^{+0.06}$ &$0.75^{+0.09}_{-0.09}$ & ${154}_{-33}^{+37}$ &  ${43.5}_{-16}^{+8.5}$ &  ${0.31}_{-0.21}^{+0.20}$  \\
TOI-930 & 16881 &  ${1.86}_{-0.05}^{+0.05}$ &$1.21^{+0.09}_{-0.09}$ &  ${195}_{-34}^{+37}$ &  ${152}_{-13}^{+15}$ &  ${0.23}_{-0.15}^{+0.19}$  \\
TOI-1099$^{\dag}$ & 108162 &  ${0.81}_{-0.12}^{+0.12}$ & $0.63^{+0.10}_{-0.10}$ &  ${151}_{-10}^{+10}$ &    ${46.6}_{-3.8}^{+2.9}$ &  ${0.33}_{-0.08}^{+0.08}$\\
TOI-1204 & 55069 &  ${1.18}_{-0.10}^{+0.10}$ &  ${0.43}_{-0.05}^{+0.05}$ &${37.3}_{-6.2}^{+11}$ & ${136.2}_{-9.4}^{+11}$ & ${0.39}_{-0.21}^{+0.14}$ \\
TOI-4175 & 64573 &     ${0.86}_{-0.05}^{+0.05}$ &  $0.81^{+0.05}_{-0.05}$ &   ${84}_{-10}^{+14}$ &       ${92.98}_{-0.35}^{+0.69}$ &   ${0.33}_{-0.16}^{+0.23}$  \\
TOI-4399 & 94235 & ${1.04}_{-0.05}^{+0.05}$&   ${0.28}_{-0.05}^{+0.05}$      &${61}_{-11}^{+12}$ &    ${68.9}_{-4.2}^{+2.7}$ & ${0.21}_{-0.14}^{+0.24}$  \\
TOI-5962 & 112486 &  ${0.65}_{-0.03}^{+0.03}$ & $0.26_{-0.01}^{+0.01}$ & ${153}_{-23}^{+59}$ & ${59.9}_{-13}^{+8.6}$ & ${0.72}_{-0.15}^{+0.13}$ \\
\hline
\end{longtable*}
\end{center}

\subsubsection{TOI-402}

TOI-402 is a nearby binary at a distance of 44.81 pc. The primary star is a K dwarf ($\rm M_{A}=0.88^{+0.05}_{-0.05} \ M_{\odot}$, $G\sim8.9$) orbiting by a M dwarf companion at a separation of $\sim 1.5 ^{\arcsec}$ with a magnitude difference of 5.2 in I band (824 nm)  filter of SOAR observation \citep{Ziegler2020}. TOI-402 A has two confirmed transiting sub-Neptunes \citep{Dumusque2019, Gandolfi2019,ros2024}: TOI-402 Ab ($P_{b}=4.76$ days, $r_{b}=1.64\pm{0.06}\ R_{\oplus}$, $m_{b}=7.51\pm{1.10}\ M_{\oplus}$), TOI-402 Ac ($P_{c}=17.17$ days, $r_{c}=2.39\pm{0.12}\ R_{\oplus}$, $m_{c}=8.11\pm{1.8}\ M_{\oplus}$). The flux dilution of the faint companion does not significantly bias the measurements of the planets' radii.

We include two relative astrometric measurements from SOAR speckle imaging \citep{Ziegler2020}, along with three Keck/NIRC2 observations obtained as part of our survey. Since TOI-402 has been an active target for RV  monitoring, we were able to compile long-baseline RV measurements from multiple high-precision spectrographs of the system spanning from 2000 to 2024.  In our RV modeling, we ignore the two known transiting planets in the system, as their expected RV amplitudes are smaller than the signal induced by the stellar companion. Furthermore, we included the relative RVs between primary and secondary stars obtained using KPIC on Sep. 26 2024. The relative RV provides direct information on the stars' mutual motion and significantly enhance the orbital constraints of the stellar companion, particularly on the eccentricity.

Figure~\ref{fig:figure402} presents the data and best-fit model for the TOI-402 system. We obtain a  mass for the stellar companion of ${M_{\rm B}}=0.26^{+0.02}_{-0.02}\ M_{\odot}$. Our best-fit model shows that the the companion orbits around primary A in a moderate eccentric orbit ($\rm{a_{B}=59.9^{+9.4}_{-6.7}}$ AU, $\rm{e_{B}={0.66}_{-0.06}^{+0.06}}$) with an inclination of  $\rm{I_{B}=126.9^{+5.8}_{-6.7}}$ deg. The periastron distance of the stellar companion is $20.3^{+4.7}_{-4.5}$ AU. The well-constrained orbit of TOI-402 B shows that the two transiting planets in the system are misaligned with respect to the stellar companion. Such a misaligned companion can exert a gravitational torque on the inner planetary system, potentially tilting their orbits \citep{Huber2013, Zhang2021}.

\subsubsection{TOI-906}

TOI-906 is a binary system consisting of an early F-type primary star ($M_A={1.14}_{-0.06}^{+0.06}~M_{\odot}$, $G\sim8.8$, distance $\sim135.5$~pc) and a K-dwarf companion ($M_B=0.75^{+0.09}_{-0.09}~M_{\odot}$) separated by about $1.2^{\arcsec}$. \textit{TESS} identified a planet candidate TOI-906.01 with a radius of $5.3R_{\oplus}$ at an orbital period of 1.665 days. Since the magnitude difference between the two components is \(3.2\) in the Gaia \(G\)-band (assuming a similar difference in the TESS bandpass), flux dilution from the secondary star is small relative to the transit depth and therefore does not significantly bias the inferred planet radius.

Ground-based observations detected the transit on target. The binary was resolved by \textit{Gaia}, which provides relative astrometry and proper motion measurements for both stars in DR3. We also include a relative astrometric measurement from SOAR speckle imaging \citep{Ziegler2020}. Our orbital fit yields a moderately eccentric, nearly face-on orbit with a semi-major axis of \(\rm{a_B = {154}_{-33}^{+37}}\)~AU, eccentricity \(\rm{e_B ={0.31}_{-0.21}^{+0.20}}\), and inclination \(\rm{I_B = {43.5}_{-16}^{+8.5}}\) deg.

\subsubsection{TOI-930}

TOI-930A is a subgiant star with a mass of ($M_A = {1.86}_{-0.05}^{+0.05}\,M_{\odot}$, $G\ \sim8.5$. distance $\sim295$ pc) and a radius of $R_A = 3.98^{+0.16}_{-0.12}\,R_{\odot}$, hosting a transiting planet candidate with a radius of \(\sim6.56\,R_{\oplus}\) and an orbital period of 4.9 days. Its companion, TOI-930B, is located approximately \(0.7^{\arcsec}\) away with a magnitude difference of about 3.1 in the 832 nm filter from Gemini/Zorro observations. The faint companion's flux has a negligible impact on the planet radius measurements. The binary was resolved by \textit{Hipparcos}, which provides relative astrometry from around 1991. We also include more recent relative astrometric measurements from SOAR speckle imaging \citep{Ziegler2020} on Aug. 2019 and Gemini observations on Dec. 2020 \citep{Lester2021}. We obtained a semi-major axis of ${195}_{-34}^{+37}$ AU, eccentricity of ${0.23}_{-0.15}^{+0.19}$ and orbital inclination of ${152}_{-13}^{+15}$ deg.


\subsubsection{TOI-1099}

TOI-1099 is a young visual binary system \citep{Barros2023} consisting of a K2V primary star (\(M_A = {0.81}_{-0.12}^{+0.12}\,M_{\odot}\), \(\sim0.56\) Gyr, $G\sim8.0$, distance $\sim23.6$ pc, Gaia DR3 6356417496318028800) and a K-dwarf companion (\(M_B = 0.63^{+0.10}_{-0.10}\,M_{\odot}\), Gaia DR3 6356417496318028928) . The primary, TOI-1099A, hosts a confirmed mini-Neptune with a radius of \(\sim2.25\,R_{\oplus}\) and a mass of \(\sim6.1\,M_{\oplus}\), orbiting with a period of 6.4 days \citep{Barros2023}. The binary is a known wide (visual) pair with a projected separation of approximately \(7.7^{\arcsec}\). Gaia DR3 identifies HIP TOI-1099B as the source associated with the NSS two-body orbital solution (period = 1046 days, mass ratio ≈ 0.2), implying that the system is a hierarchical triple in which the primary star A is orbited by an unresolved binary. For the purposes of orbital fitting, we treat the companion pair as a single object. We adopt relative astrometric measurements from the 2MASS survey and \textit{Gaia}, along with proper motion data for the companion from \textit{Gaia}. 

Our fitting reveals that the binary orbit has a semi-major axis of $151^{+10}_{-10}$ AU, an moderate eccentricity of $0.33^{+0.08}_{-0.07}$ and orbital inclination of $46.6^{+2.9}_{-3.8}$ deg. TOI-1099 stands for another system with the transiting small planet to be misaligned with the stellar companions.

\subsubsection{TOI-1204}
TOI-1204 A is G0V star ($M_{A}=1.19^{+0.09}_{-0.03}\ M_{\odot}$, $G\sim8.4$, distance $\sim106.5$ pc) with a stellar companion at around $0.3^{\arcsec}$ and a magnitude difference of 5.5 in 832 nm filter of Gemini/Zorro observations. The \textit{TESS} photometry yield a small transiting planet with a radius of $2R_{\oplus}$ and orbital period of 1.38 days. The flux from the faint companion does not significantly affect the radius measurements. 

We adopted three relative astrometric measurements from speckle imaging using Gemini South telescope \citep{Lester2021}. In our analysis, we incorporate not only the astrometric measurements from speckle imaging but also the \textit{Hipparcos}-\textit{Gaia} astrometric acceleration, providing a 25-year baseline. We obtained a semi-major axis of $a_B={37.3}_{-6.2}^{+11}$ AU, eccentricity of $e_B={0.39}_{-0.21}^{+0.14}$ and orbital inclination of $I_B={136.2}_{-9.4}^{+11}$ deg.

\subsubsection{TOI-4175}

TOI-4175 (HIP 64573) is a nearly equal-brightness binary system with a separation of $1.7^{\arcsec}$ at a distance of $\sim45.9$ pc. The primary star ($M_{A}={0.86}_{-0.05}^{+0.05}\ M_{\odot}$, Gaia DR3 3511580658668406528) has a Gaia G magnitudes of $\sim8.6$ while that of the secondary star ($M_{B}=0.81^{+0.05}_{-0.05}\ M_{\odot}$, Gaia DR3 3511580658668659328) is $\sim8.6$.  \textit{TESS} identified a transiting planet candidate with a radius of $2.9 R_{\oplus}$ and orbital period of 2.2 days around northeastern component of the system. Since the two stars are similarly bright, the flux dilution could lead to a significant underestimate of the planet candidate's radius. Taking this into account, the planet's radius could be as large as $3.9 R_{\oplus}$. 

The binary was resolved by \textit{Hipparcos} and \textit{Gaia}. We adopted relative astrometric measurements from \textit{Hipparcos}, \textit{Gaia} and  specking imaging with Gemini telescope, and AO imaging using Keck/NIRC2. We also include relative RVs and proper motion of the companions provided by \textit{Gaia} DR3. Our fitting results show that the binary orbit is fairly edge-on, with a semi-major axis of $a_B={84}_{-10}^{+14}$ AU, eccentricity of $e_B={0.33}_{-0.16}^{+0.23}$ and orbital inclination of $I_B = {92.98}_{-0.35}^{+0.69}$ deg.

 \begin{figure*}
    \centering
    \includegraphics[width=\linewidth]{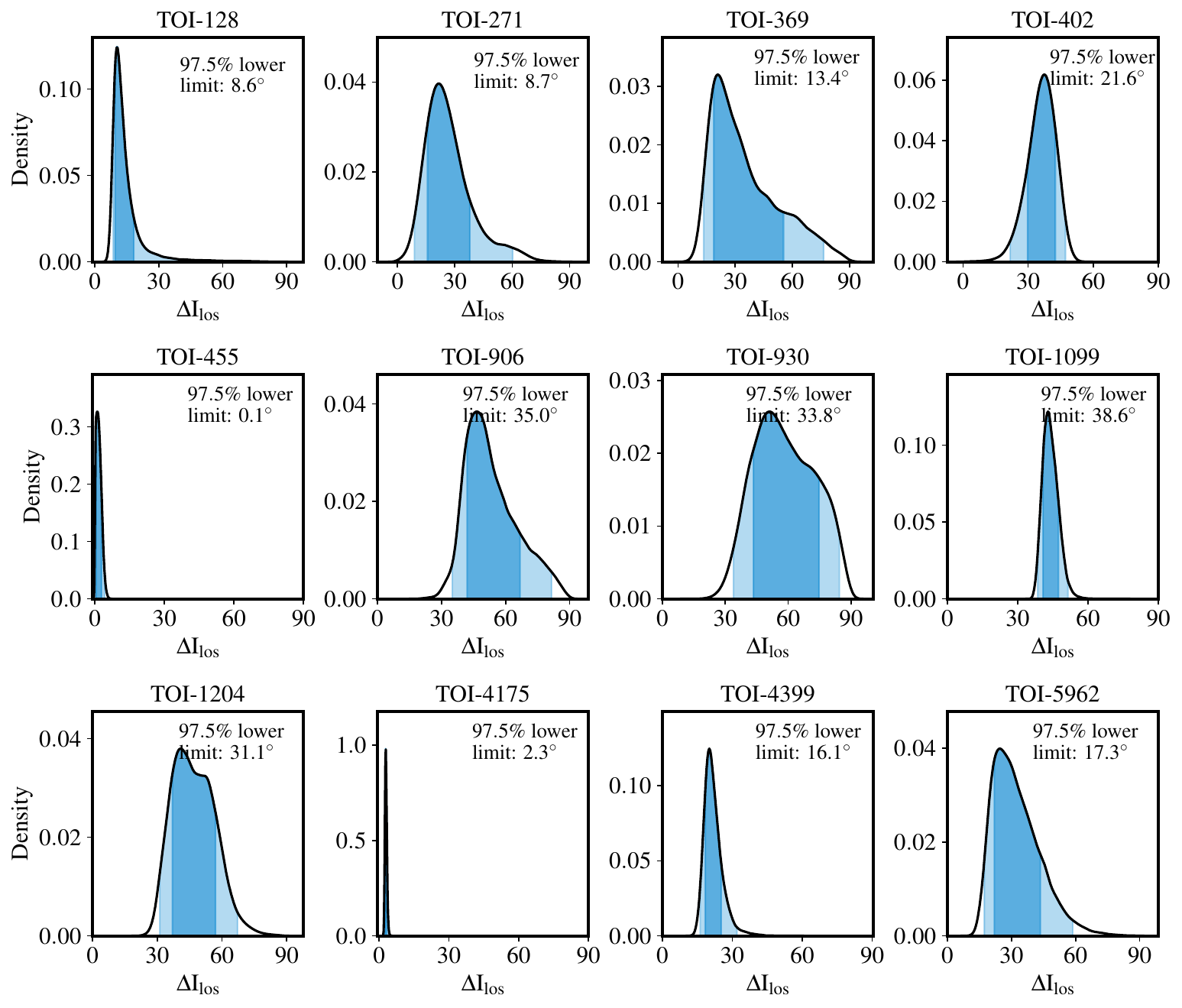}
    \caption{Posterior distributions of the line-of-sight mutual inclination $\Delta I_{\mathrm{los}}$ between transiting planets and stellar companions for \textcolor{blue}{12} TOIs with orbital fits from our survey. Shaded regions show $68\%$ and $95\%$ credible intervals. }\label{fig:dI_pdf}
\end{figure*}

\subsubsection{TOI-4399}
TOI-4399 (HIP 94235) A is a young star ($M_A={1.04}_{-0.05}^{+0.05}\ M_{\odot}$,$\sim 120$ Myr, $G\sim8.2$, distance $\sim58.5$ pc) identified as part of the AB Doradus moving group \citep{zhou2022}. It hosts a transiting mini-Neptune ($r_p=3.00^{+0.32}_{-0.28}\ R_{\oplus}$, $P=7.7$ days) discovered by TESS \citep{zhou2022}. It has a stellar companion at around $0.6^{\arcsec}$ with a magnitude difference of $\sim 5.3$ in 832 nm filter of Gemini/Zorro observations. The faint companion does not significantly bias the radius measurement of the planet. 

We used three archival relative astrometry points and \textit{Hipparcos}-\textit{Gaia} astrometric acceleration to fit the binary orbit. We obtained the stellar companion TOI-4399 B has a mass of $0.28^{+0.05}_{-0.05}\ \rm{M_{\odot}}$, semi-major axis of $61_{-11}^{+12}$ AU and orbital inclination of $68.9^{+2.7}_{-4.2}$ deg. Our results are consistent with those from \citep{zhou2022} within $1\sigma$.



\subsubsection{TOI-5962}\label{sec:5962}

TOI-5962 is a binary system, consisting of a primary star ($M_A={0.65}_{-0.03}^{+0.03}\ M_{\odot}$, $G\sim11.4$, distance $\sim127.4$ pc) with a companion at $1.7^{\arcsec}$  with a magnitude difference of 3.9 in Gaia G band . TOI-5962.01 is transiting planet candidate with radius of $4.9R_{\oplus}$ and period of 1.92 days. TOI-5962.01 was initially identified as a nearby eclipsing binary by the TFOP team, as ground-based observations could not rule out that the transit occurs on the companion star. However, we still consider this an interesting case, as our analysis indicates that the two components form a bound pair. This means the candidate could still represent a genuine planet in a binary system, even if it orbits the secondary star. If the transit around the primary star, the flux dilution will not significant bias the planet radius from the faint companion. If the planet instead orbits the secondary star, its radius must be corrected for flux dilution from the brighter primary and recalculated using the secondary star’s radius. After applying these corrections, the true planet radius is $\sim 8.3 \ R_{\oplus}$. 



The binary was resolved by \textit{Gaia}. We used relative astrometric measurements from \textit{Gaia} DR3, along with two observations from Keck/NIRC2. The proper motions of the components, as provided by \textit{Gaia} DR3, were also included. Additionally, we incorporated relative radial velocity measurements obtained with KPIC. We obtained a binary orbit with semi-major axis of $a_B={153}_{-23}^{+59}$AU, eccentricity of $e_B={0.72}_{-0.15}^{+0.13}$ and orbital inclination of $I_B={59.9}_{-13}^{+8.6}$ deg.

\section{Planet-Binary Mutual Inclination }\label{sec:mul}


The mutual inclination between the planets and stellar companions can be obtained by \citep{Fabrycky2009}:

\begin{equation}
    \cos{\Delta I} = \cos{I_{p}}\cos{I_*} + \sin{I_p}\sin{I_*}\cos{(\Omega_p - \Omega_*)}
    \label{eq:Imut}
\end{equation} 
where $I_p$ and $I_*$ are the inclination  for planetary and binary orbits, $\Omega_p$ and $\Omega_*$ are longitude of ascending node. However, for a transiting planet, we know that the orbital axis lies in the plane of the sky, but not its specific orientation within that plane (for a detailed discussion of the geometry, see \citealt{Zhang2024c}). As a result, while the orbital inclination of an transiting planet can be well constrained, we cannot infer the longitude of the ascending node of a planet from transit observations. This limitation prevents us from determining the true three-dimensional mutual inclination. Instead, we can estimate the mutual inclination as projected along the line of sight direction, using the difference in inclinations: 
\begin{equation}
    \Delta I_{los} = \lvert I_p - I_* \rvert
\end{equation} 
The line-of-sight mutual inclination provides a lower limit on the true mutual inclination. A large value clearly indicates misalignment, while a small value does not guarantee alignment, but is consistent with an aligned configuration. However, if the orbital planes of the planet and stellar companion are independent, the likelihood of both appearing edge-on from our viewpoint is low. If we assume the inclinations of the planet and companion are both drawn independently from an isotropic distribution (i.e., $\cos i$ uniformly distributed between 0 and 1), the chance that both have inclinations between $80^\circ$ and $90^\circ$ is only about $3\%$.  In our specific case—where we select planets with orbital inclinations fixed at $90^\circ$—the probability of the companion also having an inclination between $80^\circ$ and $90^\circ$ under random orientations is a bit higher, about $\sim17\%$. Thus, when both orbits are observed to be edge-on, it strongly suggests they are intrinsically aligned, rather than randomly oriented.



Figure~\ref{fig:dI_pdf} summarizes the posterior distributions for the line-of-sight mutual inclination between transiting planets and stellar companions ($\rm \Delta I_{los}$) for 12 TOI system from our survey. For each system, we report the $97.5\%$ lower credible limit  of the posterior distribution, corresponding to a $2.5\%$ probability that the true value lies below this threshold. In several systems, such as TOI-455 and TOI-4175, we observe well-constrained edge-on orbits for the stellar companions, strongly suggesting alignment with the transiting planets. In systems like TOI-128 and TOI-271, the posteriors are less constrained, but the $97.5\%$ lower credible limits remain below $20^\circ$, which is still consistent with an aligned configuration. If we adopt a $97.5\%$ lower credible limit exceeding $20^\circ$ as the threshold for misalignment, 4 out of the 12 systems in our sample would be classified as misaligned. This corresponds to a misalignment fraction of $33^{+13}_{-12}\%$.


Figure~\ref{fig:dI_ap}a presents the line-of-sight mutual inclination between transiting planets and stellar companions as a function of periastron distance of the stellar companions. We present results from our survey alongside those from the literature, including the $\tau$ Boo system \citep{tauboo2019}, Kepler-444 \citep{Zhangzj2022}, HD 196885 \citep{Chauvin2023}, and 12 systems from \citep{Lester2023}. This combined dataset yields a sample of 26 binary systems hosting S-type transiting planets within 200 AU. On the $x$-axis, we choose to show the periastron distance, as it represents the closest approach of the stellar companion and naturally accounts for the effect of orbital eccentricity. As shown in Figure~\ref{fig:dI_ap}a, while most systems exhibit low values of $\rm{\Delta I_{los}}$, several misaligned systems appear at wider binary periastron distances, with $\rm{\Delta I_{los}}$ becoming increasingly scattered as periastron distance grows. Notably, all five misaligned systems in our sample have periastron distances exceeding 20 AU. To further investigate the distribution of binary$-$planet mutual inclinations and their correlation with binary periastron distances, we perform a series of statistical analyses described in Section~\ref{sec:hba}.

\section{Hierarchical Bayesian Analysis}\label{sec:hba}

\subsection{Single Rayleigh Distribution}\label{sec:srd}

\begin{figure*}
    \centering
    \includegraphics[width=0.9\linewidth]{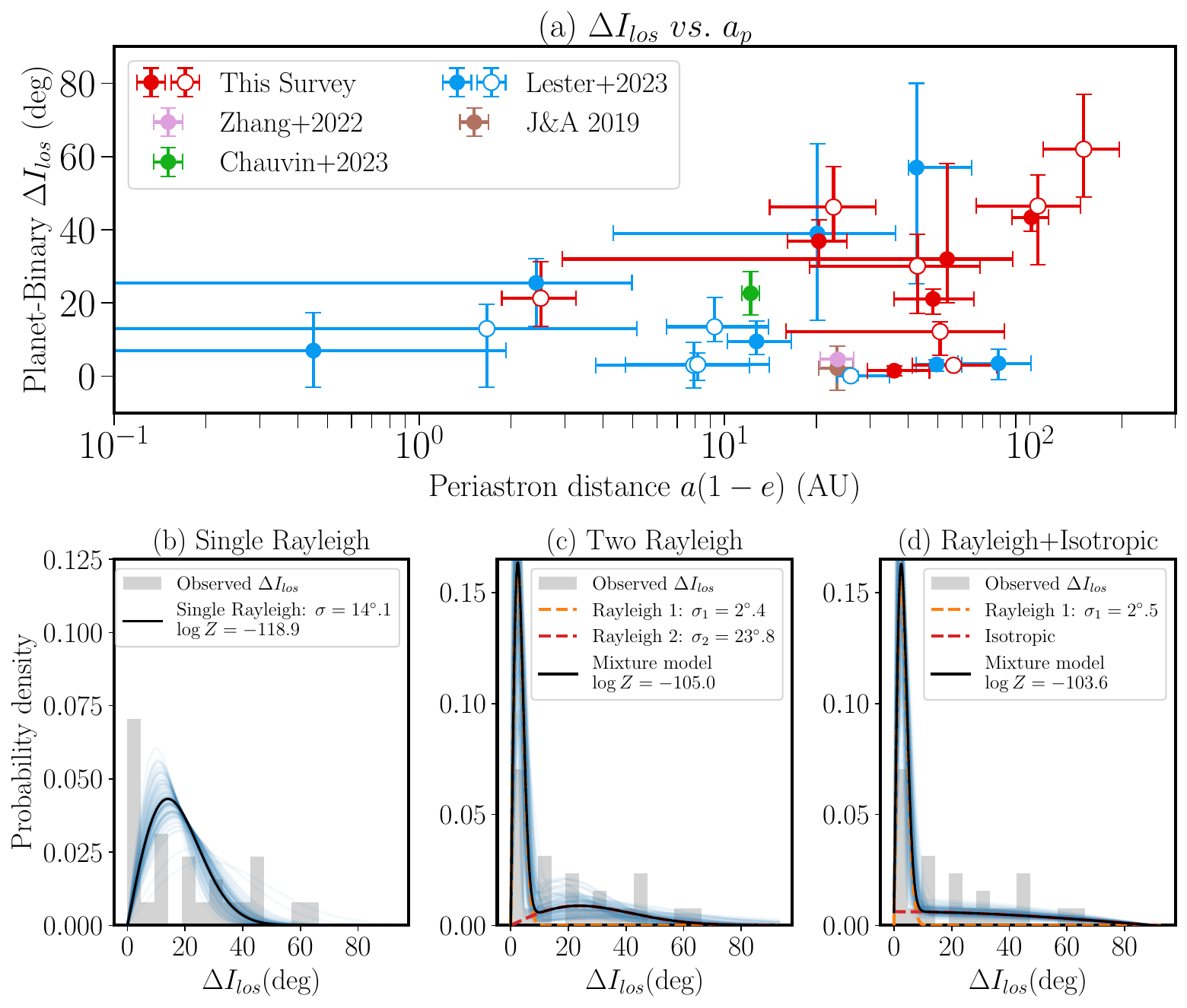}
    \caption{(a): Line-of-sight mutual inclination between transiting planets and stellar companions as the function of periastron distances $a(1-e)$ of the stellar companions.The solid circles are systems with confirmed planets, empty circles are those with planet candidates. Systems with multiple transiting planets are presented only once. (b): the distribution of line-of-sight mutual inclination between transiting planets and stellar companions (gray histogram). Results of BHA modeling for observed $\rm{\Delta I_{los}}$ using a single Rayleigh distribution. The black line show the best-fit model while the solid lies in the background represent random draws from the posterior distribution. (c): same as panel b, but using a mixture of two Rayleigh distributions in the BHA model. (d): same as panel b, but using a mixture model of one Rayleigh distribution and an isotropic distribution. Note: The gray histograms in panel b-d illustrate the distribution and median values of all data points. They do not reflect the uncertainties of individual measurements, which have been incorporated into the model fitting.}\label{fig:dI_ap}
\end{figure*}

Figure~\ref{fig:dI_ap} b shows the distribution of $\rm{\Delta I_{los}}$ for all systems presented in Figure~\ref{fig:dI_ap}a. To study population-level trends in binary-planet mutual inclinations while accounting for the uncertainties in individual measurements, we applied a Hierarchical Bayesian Analysis (HBA) to model the distribution of $\rm{\Delta I_{los}}$ following the approach of \citet{Hogg2010} and \citet{Morgan2024}. In this framework, we assume that each observed value is a noisy measurement of an underlying latent variable drawn from an underlying population distribution. Specifically, we model the underlying distribution as a Rayleigh distribution, which have been widely used to model the mutual inclination between two vectors in the literature \citep{Fabrycky2009,Lissauer2011,Masuda2020,Millholland2021}. We consider the Rayleigh distribution restricted to the physical domain from 0 to $\frac{\pi}{2}$, and normalize it over this interval. This normalization is done by dividing the standard Rayleigh distribution by the integral of its unnormalized form from \( 0 \) to \( \frac{\pi}{2} \). The resulting normalized distribution is given by:   


{\small
\begin{equation}
    \begin{split}
        \mathcal{R}(x \mid \sigma) \;=&\;
        \frac{\dfrac{x}{\sigma^{2}} \exp\!\left(-\tfrac{x^{2}}{2\sigma^{2}}\right)}
             {\displaystyle \int_{0}^{\pi/2} \frac{x'}{\sigma^{2}} 
                \exp\!\left(-\tfrac{x'^2}{2\sigma^{2}}\right)\,dx'}  =\;
        \frac{\dfrac{x}{\sigma^{2}} \exp\!\left(-\tfrac{x^{2}}{2\sigma^{2}}\right)}
             {1 - \dfrac{(\pi/2)^2}{2\sigma^2}}, \\
        & \\
        &  \text{for } 0 \leq x \leq \tfrac{\pi}{2}.
    \end{split}
\end{equation}
}
where $x$ corresponds to the line-of-sight mutual inclination. $\sigma$ is the scale parameter that sets the width of the Rayleigh distribution, representing the typical mutual inclination within the population. The mode of the Rayleigh distribution is $\sigma$, the mean is $\sqrt{2/\pi} \sigma \approx 1.25\sigma$ and the standard deviation is $\sigma\sqrt{2-\pi/2}  \approx 0.655\sigma $. Smaller values of $\sigma$ indicate that the orbits are more closely aligned, while larger $\sigma$ values correspond to systems with greater misalignment and a broader spread in inclination angles.

We then convolve the Rayleigh distribution with a Gaussian distribution centered at the measured value $\Delta I $, with standard deviation $\delta \Delta I $, representing the measurement uncertainty. To ensure that the resulting probability distribution is properly normalized and confined to the physical domain $ [0, \pi/2] $, we normalize the convolution over this interval. The resulting probability distribution as a function of $ \Delta I $ is given by:

{\scriptsize
\begin{equation}
\mathcal{P}(\Delta I,\delta \Delta I) \;=\;
\frac{\displaystyle \int_{0}^{\pi/2} 
        \mathcal{R}(x \mid \sigma)\,
        \mathcal{N}\!\left(\Delta I \mid x, \, \delta \Delta I^{2}\right)\, dx}
     {\displaystyle \int_{0}^{\pi/2} 
        \left[ \int_{0}^{\pi/2} 
            \mathcal{R}(x \mid \sigma)\,
            \mathcal{N}\!\left(\Delta I' \mid x, \, \delta \Delta I^{2}\right)\, dx 
        \right] d\Delta I'}.
\end{equation}
}
where \( \mathcal{N}(\Delta I_{los,i} \mid x, \delta \Delta I_{los,i}^2) \) is the Gaussian distribution centered at \( x \) with standard deviation \( \delta \Delta I_{los,i} \). The dataset includes measurements from 26 systems, where $\Delta I_{\mathrm{los},i}$ represents the line-of-sight mutual inclination for the $i$-th system, and $\delta \Delta I_i$ denotes its associated uncertainty. We model each measurement as a Gaussian distribution. The likelihood for each data point would be

\begin{equation}
\mathcal{L}_i = \mathcal{P}(\Delta I_{los,i}, \  \delta \Delta I_{los,i})
\end{equation}

The total log-likelihood is the sum over all data points:

\begin{equation}
\log \mathcal{L} = \sum_i \log \mathcal{L}_i
\end{equation}

We adopted a log-uniform prior on the scale parameter $\sigma$, spanning the range from 0.1 to 100. We used \texttt{dynesty} \citep{speagle_DYNESTY_2020} to perform the nested sampling with 500 live points. The sampling stops when the estimated contribution of the remaining prior volume to the total evidence is less than 1\%. Figure~\ref{fig:dI_ap}b presents the modeling results for the observed $\Delta I_{\mathrm{los}}$ distribution. We derive a scale parameter of $\sigma = 14^{\circ}.1^{+4^{\circ}.7}_{-3^{\circ}.6}$ (1$\sigma$), representing the typical mutual inclination in the population. The corresponding standard deviation is $0.655\sigma = 9^{\circ}.2$. These results suggest that planet-binary mutual inclinations are generally low, with a typical value around $14^{\circ}$, though not fully coplanar and exhibiting a spread of approximately $9^{\circ}.2$ across the population. As shown in Figure~\ref{fig:dI_ap}b, there is a peak near zero mutual inclination that is not well described by a single Rayleigh distribution. The difficulty with the single-Rayleigh model is that it cannot simultaneously account for the majority of systems near zero while also those in the tails. To account for this, we use a mixture model in the following section.

\subsection{ Mixture Model of Two Rayleigh Distributions}

We then used a mixture of two Rayleigh components to fit the observed $\Delta I_{los}$ distribution by

\begin{equation}
\begin{split}
\mathcal{M}_{1}(x \mid \sigma_1, \sigma_2, w) 
= & \; w \cdot \frac{\dfrac{x}{\sigma_1^{2}} \exp\!\left(-\tfrac{x^{2}}{2\sigma_1^{2}}\right)}
     {\displaystyle \int_{0}^{\pi/2} \frac{x^{\prime}}{\sigma_1^{2}} 
        \exp\!\left(-\tfrac{x^{\prime2}}{2\sigma_1^{2}}\right)\,dx^{\prime}} \\
  & + (1 - w) \cdot \frac{\dfrac{x}{\sigma_2^{2}} \exp\!\left(-\tfrac{x^{2}}{2\sigma_{2}^{2}}\right)}
     {\displaystyle \int_{0}^{\pi/2} \frac{x^{\prime}}{\sigma_2^{2}} 
        \exp\!\left(-\tfrac{x^{\prime2}}{2\sigma_2^{2}}\right)\,dx^{\prime}},\\
\text{for } 0 \leq x \leq \tfrac{\pi}{2}.
\end{split}
\end{equation}

where $x$ corresponds to the line-of-sight mutual inclination.  \( \sigma_1 \) and \( \sigma_2 \) are the scale parameters of two Rayleigh distributions, and \( w \in (0, 1) \) is the mixture weight that determines the fractional contribution of each component. The two Rayleigh distributions are normalized over 0 to $\pi/2$ separately. Same as section~\ref{sec:srd}, we define the likelihood for each observed value \( \Delta I_{los,i} \) with measurement uncertainty \( \delta \Delta I_i \) as the convolution of the underlying distribution with a Gaussian distribution and normalized the convolution from 0 to $\pi/2$:

{\tiny
\begin{equation}
\mathcal{P}(\Delta I,\delta \Delta I) \;=\;
\frac{\displaystyle \int_{0}^{\pi/2} 
        \mathcal{M}_{1}(x \mid \sigma_1,\sigma_2,w)\,
        \mathcal{N}\!\left(\Delta I \mid x, \, \delta \Delta I^{2}\right)\, dx}
     {\displaystyle \int_{0}^{\pi/2} 
        \left[ \int_{0}^{\pi/2} 
            \mathcal{M}_{1}(x \mid \sigma_1,\sigma_2,w)\,
            \mathcal{N}\!\left(\Delta I' \mid x, \, \delta \Delta I^{2}\right)\, dx 
        \right] d\Delta I'}.
\end{equation}
}

\begin{equation}
\log \mathcal{L}  =\sum_i \log \mathcal{P}(\Delta I_{los,i}, \  \delta \Delta I_{los,i})
\end{equation}

where \( \mathcal{N}(\Delta I_{los,i} \mid x, \delta \Delta I_{los,i}^2) \) is the Gaussian distribution centered at \( x \) with standard deviation \( \delta \Delta I_{los,i} \). We adopt log-uniform priors on the scale parameters \( \sigma_1 \) and \( \sigma_2 \), each spanning the range \([0.1, 100]\), and impose the constraint \( \sigma_1 < \sigma_2 \) to ensure the two components remain distinguishable. The mixture weight \( w \) is assigned a uniform prior over the interval \([0, 1]\). We also used \texttt{dynesty} to perform a nested sampling fit.
 
Figure~\ref{fig:dI_ap}c shows the best-fit two-component Rayleigh mixture model. The first component has a scale parameter of \( \sigma_1 = 2^{\circ}.4^{+0.7}_{-0.9} \), indicating a population with small mutual inclinations and a nearly aligned configuration between planets and their stellar companions. This population is also narrow in width, with a standard deviation of $1^{\circ}.6$. The second component has a scale parameter of \( \sigma_2 = 23^{\circ}.6^{+8.8}_{-7.1} \) and a standard deviation of $15^{\circ}.4$, corresponding to a broader, more misaligned population.
We obtained a mixture weight factor of $w = 0.66^{+0.09}_{-0.21}$, which is the fraction of systems that belong to the well-aligned population.

We used the logarithm of the Bayesian evidence, $\rm{log} Z$ from the nested sampling as a metric to compare the single and mixture models. The Bayesian evidence measures how well a model explains the data, taking into account both the fit to the data and the complexity of the model (the number of free parameters). A higher $\rm{log} Z$ value indicates stronger support for one model over another, with differences greater than 5 generally considered strong evidence \citep{Trotta2008}. The two-Rayleigh mixture model is favored over the single-component Rayleigh model with a $\Delta \rm{log} Z$ of 13.9, providing strong statistical evidence for the presence of two distinct populations—one tightly aligned and one significantly more dispersed in mutual inclination.

\subsection{Mixture Model of Rayleigh Distribution and Isotropic Distribution}

We also consider a second mixture model, which assumes that all systems are drawn from either a Rayleigh distribution (with probability $w$) or from an isotropic distribution (with probability $1-w$). For the isotropic population, we assume that stellar companions have randomly oriented orbital planes, corresponding to an isotropic inclination distribution. Under the assumption of isotropy, the inclination $i$ of the stellar companions is uniformly distributed in $\cos i \sim \mathcal{U}(0, 1)$, which implies a probability density function $f(i) = \sin i$ for $i \in [0^\circ, 90^\circ]$\footnote{Details for  isotropic inclination angles see : \url{https://keatonb.github.io/archivers/uniforminclination}}. We also assume the orbital inclination of transiting planets is $90^\circ$.  Therefore, the line-of-sight  mutual inclination between planets and stellar companions is defined as $x = | 90^\circ-i|$, and resulting probability density function of $x$ is:

\[
f(x) = \sin{|90^{\circ}-x|} = \cos x, \quad \text{for } x \in [0^\circ, 90^\circ].
\]

Therefore, this mixture model can be written as 

\begin{equation}
\begin{split}
\mathcal{M}_{2}(x \mid \sigma, w) 
= & \; w \cdot \frac{\dfrac{x}{\sigma^{2}} \exp\!\left(-\tfrac{x^{2}}{2\sigma^{2}}\right)}
     {\displaystyle \int_{0}^{\pi/2} \frac{x^{\prime}}{\sigma^{2}} 
        \exp\!\left(-\tfrac{x^{\prime2}}{2\sigma^{2}}\right)\,dx^{\prime}} \\
  & + (1 - w) \cdot \frac{\cos{x}}{\int_{0}^{\pi/2} \cos{x^{\prime}dx^{\prime}}},\\
\text{for } 0 \leq x \leq \tfrac{\pi}{2}.
\end{split}
\end{equation}

where the model is described by only two parameters: the scale parameter of the Rayleigh component and the weight between two components. The two components have been normalized over 0 to $\pi/2$ separately. As above, the likelihood for each observed value \( \Delta I_{los,i} \) with measurement uncertainty \( \delta \Delta I_i \) is

{\scriptsize
\begin{equation}
\mathcal{P}(\Delta I,\delta \Delta I) \;=\;
\frac{\displaystyle \int_{0}^{\pi/2} 
        \mathcal{M}_{2}(x \mid \sigma,w)\,
        \mathcal{N}\!\left(\Delta I \mid x, \, \delta \Delta I^{2}\right)\, dx}
     {\displaystyle \int_{0}^{\pi/2} 
        \left[ \int_{0}^{\pi/2} 
            \mathcal{M}_{2}(x \mid \sigma,w)\,
            \mathcal{N}\!\left(\Delta I' \mid x, \, \delta \Delta I^{2}\right)\, dx 
        \right] d\Delta I'}.
\end{equation}
}

\begin{equation}
\log \mathcal{L}  =\sum_i \log \mathcal{P}(\Delta I_{los,i}, \  \delta \Delta I_{los,i})
\end{equation}

where \( \mathcal{N}(\Delta I_{los,i} \mid x, \delta \Delta I_{los,i}^2) \) is the Gaussian distribution centered at \( x \) with standard deviation \( \delta \Delta I_{los,i} \). We adopt log-uniform priors on the Rayleigh scale parameters \( \sigma \), spanning the range \([0.1, 100]\), and assign a uniform prior on the mixture weight \( w \) over the interval \([0, 1]\). The model is fit using nested sampling with \texttt{dynesty}. From the fit, we obtain a scale parameter of \( \sigma_1 = 2^{\circ}.5^{+0.6}_{-0.8} \) for the Rayleigh component (see Figure~\ref{fig:dI_ap} d ). This result is consistent with the well-aligned population in the two-component Rayleigh mixture model, with a nearly identical scale parameter. This supports a two-population scenario consisting of a well-aligned component and an isotropic component. The Bayesian evidence for this model is slightly higher than that of the two-component Rayleigh mixture model. However, the difference in log evidence is only 1.4, indicating that the data do not decisively favor one model over the other.

In summary, our analysis supports the presence of two distinct populations of transiting planets in binary systems: one with tightly aligned mutual inclinations between the planetary and binary orbits, and another with more misaligned and broadly scattered inclinations. Our model shows that the misaligned population can be equally well described by either a Rayleigh distribution with $\sigma=23^{\circ}.6$ or an isotropic distribution. Note that  Figure \ref{fig:dI_ap} b-d show histograms of the median values of the data points, which illustrate the overall distribution but do not account for individual measurement uncertainties. Since our model  incorporates these uncertainties, the amplitude of the model curves does not exactly match the histograms.

\begin{figure*}
    \centering
    \includegraphics[width=\linewidth]{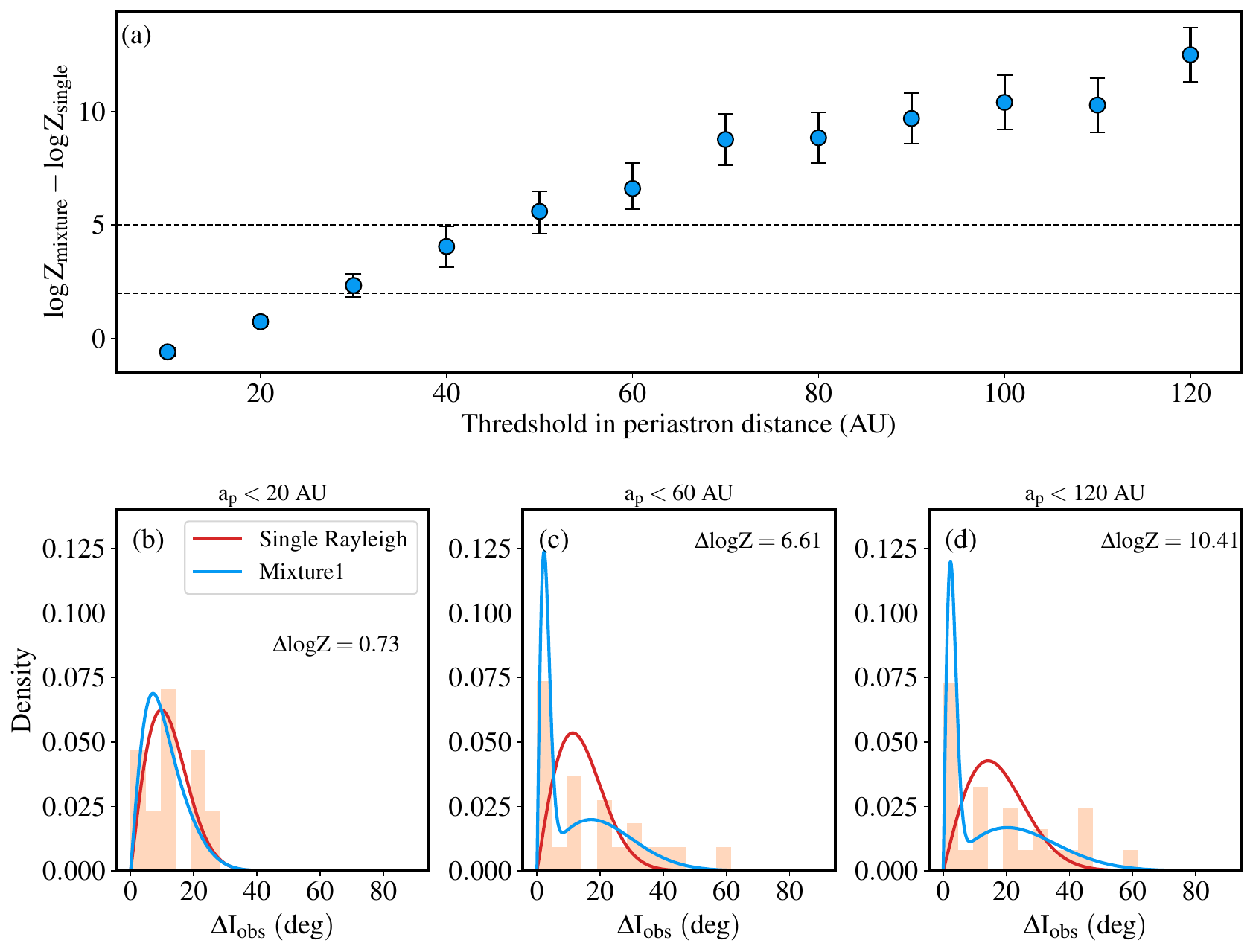}
    \caption{(a) Bayesian evidence ($ \rm \log Z_{mixture}-\log Z_{single}$) difference between a two-component Rayleigh mixture model and a single Rayleigh model, fit to the observed $\rm{\Delta I_{obs}}$ distribution between transiting planets and stellar companions. Each point corresponds to a sub sample restricted to companions with periastron distances below the indicated threshold, which varies from 20 AU to 120 AU. \( \Delta \log Z \)=2 and \( \Delta \log Z \)=5 thresholds are plotted as gray dashed lines. The increasing trend suggests that systems with wider-separation companions increasingly favor the mixture model. (b)-(d)  $\rm{\Delta I_{obs}}$ 
    distributions for systems with periastron distances below 20 AU, 60 AU, and 120 AU. The red and blue curves show the best-fit single Rayleigh and two-Rayleigh mixture models, respectively. }
    \label{fig:movingtest}
\end{figure*}
\subsection{ A Transition in Planet
Alignment for binaries from  40 to 100 AU}


To further investigate the dependence of mutual inclination on  periastron distance of the stellar companions, we perform a Bayesian model comparison by fitting both a single Rayleigh distribution and a two-component Rayleigh mixture model to subsets of systems with periastron distances less than a moving threshold. The threshold varies from 10~AU to 120~AU in fixed steps, with an interval of 10~AU. To account for uncertainties in the periastron distances, we randomly draw the periastron distance of each system from a Gaussian distribution centered on the measured mean value with a standard deviation equal to the reported uncertainty. We then compute the difference in log Bayesian evidence (\( \Delta \log Z \)) between the mixture and single-component models. We perform 50 random draws, and repeat the fits each time to obtain a distribution of \( \Delta \log Z \) values, from which we estimate the mean and uncertainty in \( \Delta \log Z \). 

The results are plotted in Figure~\ref{fig:movingtest}. We find that when considering only systems with stellar companions having periastron distances below 20~AU, the single Rayleigh distribution is favored over the mixture model. The best-fit single Rayleigh model yields a scale parameter of \( \sigma = 9^{\circ}.8^{+4^{\circ}.0}_{-1^{\circ}.4} \), suggesting that planets in these close binaries tend to have low mutual inclinations. However, the Bayesian evidence in favor of the mixture model increases as systems with larger periastron distances are included. Specifically, \( \Delta \log Z \)>3 when we include systems with periastron distances $<40$ AU, and \( \Delta \log Z \)>5 for periastron distances $<60$ AU. \( \Delta \log Z \) between 2 to 5 represents moderate evidence for the mixture model ($2.6-3.6\sigma$; \citealt{Trotta2008}), whereas \( \Delta \log Z \)>5 represents strong evidence ($>3.6\sigma$). This trend suggests a transition in planet-binary alignment with increasing binary separation—from a preference for aligned configurations at periastron distances below 20~AU to the emergence of two distinct populations at wider separations: one nearly aligned and the other exhibiting significantly broader mutual inclination dispersion. Note that most of the data points in our analysis come from our own surveys and from \citet{Lester2023}, with each focusing on different stellar populations—ours on TESS stars and theirs on Kepler stars. The \citet{Lester2023} sample contains tighter binaries, with more systems having periastron distances below 20 AU, while our sample generally consists of wider binaries with larger periastron distances. We share one system in common, TOI-271, for which we obtained consistent results. We observe a similar trend across the two sub-samples. In the regime where periastron distances are below 20 AU, all systems exhibit low mutual inclinations—though our survey includes only one such system. Beyond 20 AU, both our sample and that of \citet{Lester2023} include a mix of aligned and misaligned systems: 4 out of 11 in our sample and 2 out of 5 in \citet{Lester2023} have mutual inclinations above $20^{\circ}$. Expanding the sample will be essential for strengthening the analysis and testing the robustness of this trend.

\begin{figure}
    \centering
    \includegraphics[width=0.85\linewidth]{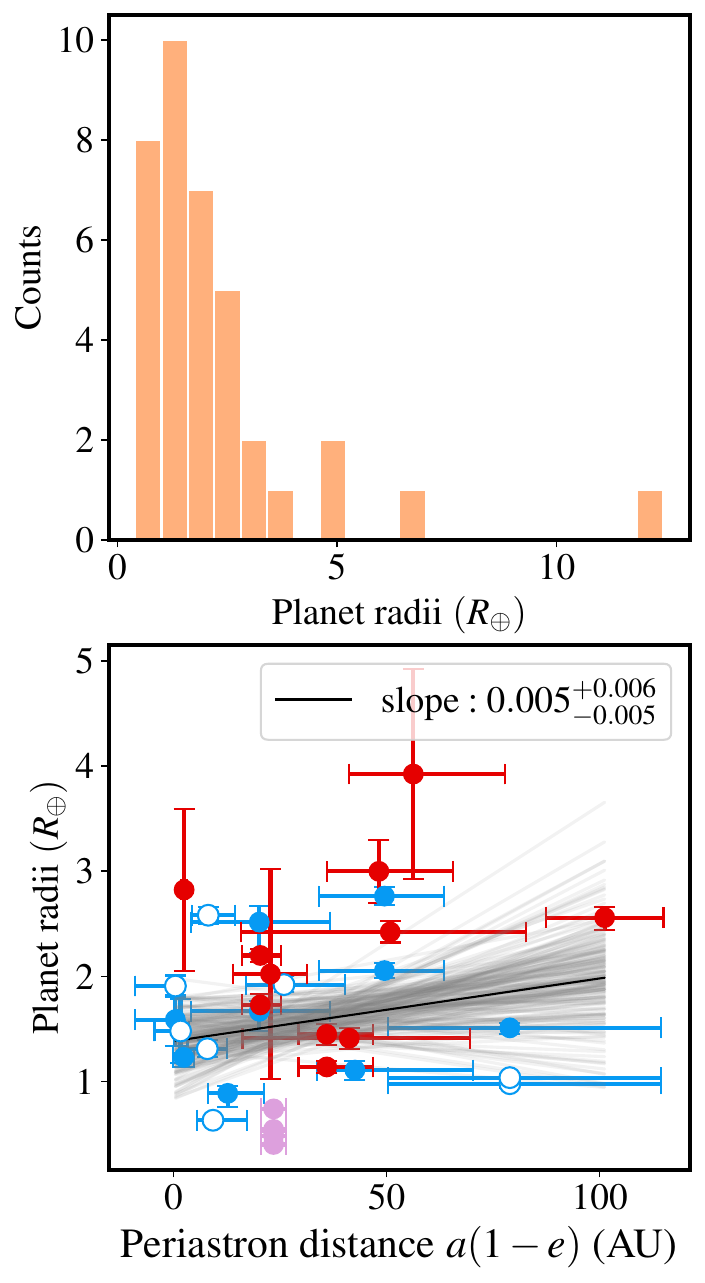}\label{fig:beta1}
    \caption{Top: Planet radius distribution for the systems shown in Figure~\ref{fig:dI_ap} a, all of which host confirmed or candidate planets in binary systems. Most planets have radii smaller than $5M_{\oplus}$. Bottom: Planet radius versus stellar companion periastron distance. The black line shows the best-fit linear trend, with gray lines representing random posterior draws. We find no significant correlation between planet radius and the periastron distance of the stellar companions. }
    \label{fig:rpdist}
\end{figure}

\section{Discussion}
\label{sec:discussion}

\subsection{Distribution of Planet Radii in our sample }

As shown in the previous section, our sample reveals two distinct populations: one with low mutual inclinations between planets and their stellar companions, and another with significantly broader inclination dispersion. A natural question is whether this dichotomy arises for different types of planets—such as small planets versus hot Jupiters—since they may have undergone different formation and evolutionary processes. Figure~\ref{fig:rpdist}a shows the radius distribution of the planets or planet candidates included in Figure~\ref{fig:dI_ap}a. 

To account for flux dilution from the stellar companions, we revise the observed planet radii using the relation:

\begin{equation}
    r_{p,\mathrm{revised}} = r_{p,\mathrm{obs}} \times \sqrt{1 + \frac{F_{\mathrm{comp}}}{F_{\mathrm{host}}}}
\end{equation}

where \( \frac{F_{\mathrm{comp}}}{F_{\mathrm{host}}} \) is the flux ratio between the companion and the host star. We derive this ratio using the observed magnitude difference, assuming the observed band (e.g., K-band or Gaia G-band) approximates the TESS bandpass. In applying the correction, we generally assume that the planet is orbiting the host star, as most of the planets are either confirmed or have been observed directly with ground-based telescopes. The only exception is TOI-5962, for which we provide a more detailed description in Section~\ref{sec:5962}.

The majority of planets in our sample have revised radii below 5~\( R_{\oplus} \), with only one system hosting a hot Jupiter larger than 10~\( R_{\oplus} \). We therefore rule out the possibility that the observed difference in mutual inclination arises from varying planet sizes. Instead, our results highlight this inclination dichotomy specifically within the small-planet population.

Previous studies have suggested that close stellar companions can truncate circumprimary disks, limiting the material available for planet formation. This has been proposed as a possible explanation for the presence of five Mars-sized planets around Kepler-444 A \citep{Dupuy2016}. To test this hypothesis, we examined the correlation between planet radius and the periastron distance of stellar companions using a linear regression. The resulting slope has a $1\sigma$ confidence interval that includes zero (see Figure~\ref{fig:rpdist}b), indicating no significant evidence for a correlation between planet size and companion periastron distance. The lack of correlation is observed in both the Kepler sample from \citet{Lester2021} and the TESS sample from our survey, making it unlikely to be caused by biases from different instruments.


\subsection{Comparison with previous studies}

Numerous surveys have investigated how binary companions affect planet formation, particularly as a function of their projected separation. These studies consistently find a deficit of planets in close binary systems, typically those with separations less than 100 AU \citep{Kraus2016, Ziegler2020, Ziegler2021, Lester2021, Hirsch2021}. This trend supports the theory that close stellar companions can hinder planet formation by truncating circumstellar disks. Observations of disks in binary systems further reinforce this idea: disks in binaries are generally smaller, fainter, and have shorter lifetimes compared to those around single stars. 

The influence of the mutual alignment between planetary orbits and binary orbits on planet formation remains relatively unexplored, as most observational efforts have been limited to individual systems—such as Kepler-444 \citep{Dupuy2016} and $\tau$ Boo \citep{Justesen2019}. Recently, however, a few statistical studies have begun to address this gap. \citet{Dupuy2022} statistically inferred that planet–binary orbits tend to exhibit low mutual inclinations ($<30^{\circ}$), based on orbital arc measurements of 45 binaries hosting \textit{Kepler} planet candidates. \citet{Behmard2022} analyzed 170 TOIs in binary systems with separations between 40 and 10,000 AU. Using Gaia EDR3 parallaxes and proper motions, they inferred 2D relative position and velocity vectors between the host star and its stellar companion. While not a direct measurement of orbital inclinations, this kinematic information provides indirect constraints on the mutual orientation between planetary orbits and binary companions. Their analysis suggests that close-in small planets ($R_p < 4R_{\oplus}$, $a < 1,\mathrm{AU}$) typically have a planet-companion mutual inclination of $35^{+24}_{-24}$ degrees. \citet{Christian2022} also report a preference for alignment between planetary and binary orbits, using a sample of wide resolved binaries identified from \textit{Gaia}, with separations ranging from 100 to $10^{5}$ AU, though their analysis relied only on single-epoch \textit{Gaia} astrometric measurements. They reported an apparent transition between more-aligned systems and isotropic systems at separations of $\sim 700 $ AU. This separation is much wider than in our sample, and their analysis considers only projected separation without accounting for eccentricity, making a direct comparison challenging. However, their findings support the possibility that planet–stellar mutual inclinations tend to lose alignment beyond some characteristic separation. Additionally, \citet{Lester2023} constrained the orbits of 13 binaries hosting transiting planets or candidates (mostly KOIs), mostly with separations below 20 AU, and reported an excess of aligned systems in their sample.

Our statistical analysis of planet–binary mutual inclinations is consistent with previous studies. At the population level, our hierarchical Bayesian modeling yields a typical inclination of $\sigma = 15^{\circ}.1^{+2^{\circ}.6}_{-4^{\circ}.6}$  assuming a single Rayleigh distribution. This result agrees within $1\sigma$ with the results from \citet{Dupuy2022} ($<30^{\circ}$) and \citet{Behmard2022} ($35^{+24}_{-24}$ deg). This implies that planets in binaries tend to have relatively low mutual inclinations compared to an isotropic distribution. However, our analysis further reveals that the inclination distribution is better described by mixture models (two Rayleigh mixture or Rayleigh and isotropic mixture). 

Unlike most previous studies that focused on wide binaries with separations >100 AU \citep{Behmard2022,Christian2022}, our sample includes planet-hosting binaries with separations of $10\sim200$ AU. 
By constraining the full 3D orbits of the stellar companions—including their eccentricities and semi-major axes—we enable the first analysis of how mutual inclination correlates with binary periastron distance in this previously unexplored regime. We find that in very tight binaries (periastron <40 AU), planets tend to be well-aligned with their stellar companions. However, at larger periastron distances, a distinct population with broader inclination dispersion emerges alongside the aligned systems, revealing a possible dichotomy in orbital architecture.

 On the other hand, outer companions—whether giant planets or stellar companions—have been shown to tilt the orbits of inner, small planets, causing them to become misaligned relative to their host stars \citep{Huber2013,Zhang2021}. In extreme cases, such as the KELT-19 system \citep{Sing2009}, planets have even been observed to orbit in a polar orbit, likely due to the influence of stellar companions. Although the planets in our sample are small and thus more challenging targets for Rossiter-McLaughlin (RM) effect measurements, it would be valuable to measure their spin-orbit angles in the future—particularly for systems that appear misaligned with respect to their outer stellar companions. Such measurements would help reveal the full system architecture and provide insight into the dynamical history of these planetary systems.

\subsection{Implication for planet formation in binaries}



Our finding of two distinct $\Delta I_{\mathrm{obs}}$ populations among planet-hosting binaries suggest different physical mechanisms likely govern their dynamical evolution. First, the population of well-aligned systems, characterized by mutual inclinations of $\sigma_1 = 2.4^{+0.7}_{-0.9}$, is consistent with a nearly coplanar configuration. They likely originate from a primordial alignment between the binary orbital plane and the planet-forming protoplanetary disk. One plausible explanation is disk fragmentation during the formation of the binary system \citep{Tokovinin2020}, which naturally leads to such alignment. Another possibility is that these systems have undergone realignment through mechanisms such as disk warping-induced dissipation, which could align the disk and binary orbital planes on timescales shorter than the disk dispersal timescale \citep{Ogilvie2000, Zanazzi2018, Gerbig2024}.

In contrast, a separate population shows larger and more scattered mutual inclinations, which can be described a Rayleigh distribution( $\sigma_2 = 23^{\circ}.6^{+8^{\circ}.8}_{-7^{\circ}.1}$) or isotropic distribution. These systems may represent relatively dynamically hotter configurations, formed with an initial misalignment between the protoplanetary disk and the binary orbital plane. An initially inclined stellar companion can drive dynamical evolution within the planetary system. \citet{zanazzi2018b} find that small close-in planets, influenced by an inclined outer companion, can undergo secular precession and inclination oscillations. This process may lead to misalignment with respect to their host stars. Similarly, \citet{Lubow2016} and \citet{Franchini2020} investigated the dynamical evolution of planet-disk systems with modest initial tilts relative to the binary orbit. Their results show that, even when the planet and disk start out coplanar, gravitational perturbations from a stellar companion can gradually induce significant inclination differences between them over time. In addition, the broader $\Delta I_{\mathrm{obs}}$ distribution in these systems, including misaligned cases, suggests that realignment mechanisms are as efficient as previously thought in at least some binaries.

Furthermore, we found that the dichotomy in planetary alignment arises only in binaries with periastron distances greater than 40 AU. In contrast, tighter binaries are well described by a single Rayleigh distribution with low planet–binary mutual inclinations. One possible explanation is that a misaligned, close-in stellar companion may more strongly disrupt planet formation, allowing only planets in nearly coplanar configurations to survive. Alternatively, closer stellar companions may realign more efficiently with the protoplanetary disk, thereby erasing any initial misalignment between the planets and the binary orbit.

\section{Conclusions}
\label{sec:conclusion}

We have conducted a comprehensive survey for stellar companions to 54 TESS Objects of Interest (TOIs) within 300 pc, selected based on significant astrometric acceleration detected by combining \textit{Hipparcos} and \textit{Gaia} data. 
The astrometric properties of these targets, including their measured acceleration and \textit{Gaia} RUWE values, are summarized in Table~\ref{tab:table1X}. Our main conclusions are as follows:

\begin{enumerate}
    \item We report the detection of 22 stellar companions to 20 TOIs in our sample using high-resolution adaptive optics imaging with Keck/NIRC2, including two hierarchical triple systems. The projected separations of the resolved companions range from $0.1^{\arcsec}$ to $2^{\arcsec}$ (or $10-200$ AU). Additionally, we identified 15 TOIs located in the southern hemisphere—outside the visibility of Keck—that have known stellar companions from previous surveys. In total, our sample contains 35 TOIs with resolved stellar companions, including both binaries and triples.

    \item We identified 7 TOIs that may host unresolved stellar companions at projected separations smaller than $0.1^{\arcsec}$, based on their measured astrometric acceleration.

    \item For remaining 12 TOIs, the observed astrometric acceleration combined with  AO imaging detection limit is consistent with the presence of  substellar companions, including giant planets or brown dwarfs. Among these, six systems are already known to host such substellar companions from previous studies. We report six additional TOIs as new candidate hosts of substellar companions. In particular, for TOI-201 which hosts a eccentric hot Jupiter, we identified an outer brown dwarf companions with semi-major axis of $6.6^{+5.5}_{-2.4}$ AU, and mass of $21.8^{+15.1}_{-7.6}\ \rm{M_{J}}$.

    \item  We presented three-dimensional orbital solutions for 12 stellar companions by combining \textit{Hipparcos}-\textit{Gaia} astrometric accelerations with relative astrometry, RVs (when available), relative RVs (when available), and companion proper motions (when available). Using these orbital solutions, we constrained the line-of-sight mutual inclinations between the transiting planets and their stellar companions $\rm{\Delta I_{los}}$. We found that 4 out of 12 systems exhibit a $97.5\%$ lower credible limit for $\rm{\Delta I_{los}}$ above $20^\circ$, corresponding to a misalignment fraction of $33^{+13}_{-12}\%$.

    \item We combined our results with previous surveys to assemble 26 systems with measured line-of-sight mutual inclinations $\Delta I_{\mathrm{los}}$ between transiting planets and their stellar companions. Using Bayesian Hierarchical Analysis (HBA), we fit the observed $\Delta I_{\mathrm{los}}$ distribution with three models: a single Rayleigh distribution, a two-component Rayleigh mixture, and a Rayleigh-isotropic mixture (Figure~\ref{fig:dI_ap}). Bayesian evidence strongly favors the mixture models over the single Rayleigh ($\log Z \gtrsim 13.9$, or $\approx4\sigma$), indicating the presence of two distinct populations (see Figure~\ref{fig:dI_ap}).  One population exhibits well-aligned orbits with outer stellar companions, characterized by a Rayleigh scale parameter of \( \sigma_1 = 2^{\circ}.4^{+0.7}_{-0.9} \); the other shows broader, more scattered mutual inclinations which can be equally described by a Rayleigh distribution with
    $ \sigma_2 = 23^{\circ}.6^{+8.8}_{-7.1}$ or isotropic distribution. This dichotomy likely indicates that distinct dynamical mechanisms are responsible for shaping the observed systems. 
    

    \item To study whether planet-binary alignment changes as a function of the stellar companion's periastron distance, we repeat the above Bayesian Hierarchical Analysis on subsets of systems with a moving threshold in periastron distance from 10 to 120 AU. We find that the emergence of a misaligned population only becomes significant when the periastron distance is above 40 AU. In other words, systems with close-in or eccentric stellar companions with small periastron distances ($<40$ AU) tend to favor planet-binary alignment. Overall, the tendency for planet-binary misalignment increases as a function of binary periastron distance over the entire range (40 to 120 AU) of our systems (see Figure~\ref{fig:movingtest}). 
    

\end{enumerate}

The upcoming Gaia DR4 time-series astrometry will enable mutual inclination measurements between transiting planets and binary companions for a significantly larger sample. This will provide an opportunity to further test this observational trend. 

\acknowledgments
The authors wish to recognize and acknowledge the very significant cultural role and reverence that the summit of Maunakea has always had within the Native Hawaiian community. We are most fortunate to have the opportunity to conduct observations from this mountain.

J.Z. would like to thank for Simon Albrecht and Trent Dupuy
helpful discussions. J.Z. and D.H. acknowledge support from the National Aeronautics and Space Administration through the Astrophysics Data Analysis Program (80NSSC19K0597) and the  Exoplanets Research Program (80NSSC22K0781).  J.Z. and L.M.W. acknowledge support from the NASA-Keck PI Data Award (80NSSC19K1475).  L.M.W. acknowledges support from the NASA Exoplanet Research Program (80NSSC23K0269).
K.H. is supported by the National Science Foundation Graduate Research Fellowship Program under Grant No. 2139433. J.X. is supported by the NASA Future Investigators in NASA Earth and Space Science and Technology (FINESST) award \#80NSSC23K1434.  This research was supported in part by grant NSF PHY-2309135 to the Kavli Institute for Theoretical Physics (KITP).

Funding for KPIC has been provided by the California Institute of Technology, the Jet Propulsion Laboratory, the Heising-Simons Foundation (grants \#2015-129, \#2017-318, \#2019-1312, \#2023-4598), the Simons Foundation, and the NSF under grant AST-1611623. 

The W. M. Keck Observatory is operated as a scientific partnership among the California Institute of Technology, the University of California, and NASA. The Keck Observatory was made possible by the generous financial support of the W. M. Keck Foundation. 

Funding for the TESS mission is provided by NASA's Science Mission Directorate. This research has made use of the Exoplanet Follow-up Observation Program website, which is operated by the California Institute of Technology, under contract with the National Aeronautics and Space Administration under the Exoplanet Exploration Program.

\vspace{5mm}
\facilities{ \TESS, \Gaia, \HIP, Keck:10m}


\software{\texttt{orvara} \citep{Brandt2021b}, \textit{dynesty}\citep{speagle_DYNESTY_2020}
          }



\newpage

\appendix
\counterwithin{figure}{section}
\counterwithin{table}{section}

\section{Planet or planet candidate properties of TOIs with significant proper motion anomalies ($>3\sigma$) }\label{sec:toilow}


\begin{center}
\begin{longtable*}{lccccc}
\caption{Planet properties in our sample}\label{tab:table1Y}\\
\hline
TOI & HIP  & $r_{pl}$ &$P_{pl}$ & Disposition$^{*}$ &Reference\\
 &  Number &   ($R_{\oplus}$) &(days) & &   \\ 
\hline
\hline
\endhead
\hline
\endfoot
\multicolumn{6}{p{0.7\textwidth}}{
{\footnotesize
\textsuperscript{1} CP: Confirmed Planet, PC: Planet Candidate, FP: False Positive, FA: False Alarm. \newline
\textsuperscript{2} The EXFOP disposition for   522.01 is listed as a PC, but four RV measurements obtained with HIRES indicate that it is actually an eclipsing binary (EB).\newline
\textsuperscript{3} The EXFOP disposition for TOI-1684.01 is listed as a planet candidate (PC), but our HIRES RV measurements indicate the transits originate from the stellar companions. Additionally, the two RVs of the companions, taken one day apart, show variations exceeding $10\,\mathrm{km\,s^{-1}}$, consistent with an eclipsing binary (EB).
}}
\endlastfoot
\multicolumn{6}{c}{\textit{Confirmed and Candidate Planet Host Stars}} \\
\hline
  128.01 & 24718  & 2.32 & 4.94 & PC &  \cite{TOI_ref}\\
  141.01 & 111553  & 1.72 & 1.01 & CP &  \cite{Espinoza2020}\\
  144.01 & 26394  & 2.0 & 6.27 & CP &  \cite{Gandolfi2018}\\
  179.01 & 13754  & 2.68 & 4.14 & CP &  \cite{Desidera2023} \\
  201.01 & 27515  & 11.45 & 52.98 & CP & \cite{Hobson2021}  \\
  201.02 & 27515  & 1.19 & 5.85 & PC & \cite{TOI_ref} \\
  271.01 & 21952  & 2.81 & 2.48 & PC & \cite{TOI_ref} \\
  369.01 & 115594  & 10.9 & 5.46 & PC & \cite{TOI_ref} \\
  402.01 & 11433  & 2.2 & 17.18 & CP & \cite{Gandolfi2019} \\
  402.02 & 11433  & 1.73 & 4.76 & CP & \cite{Gandolfi2019} \\
  455.01 & 14101  & 1.45 & 5.36 & CP & \cite{Winters2019}  \\
  455.02 & 14101  & 1.14 & 3.12 & CP & \cite{Winters2022}  \\
  906.01 & 78301  & 5.05 & 1.66 & PC & \cite{TOI_ref} \\
  930.01 & 16881  & 6.57 & 4.9 & PC &  \cite{TOI_ref}\\
  1099.01 & 108162  & 2.56 & 6.44 & CP & \cite{Barros2023} \\
  1131.01 & 81087  & 4.91 & 0.59 & PC &\cite{TOI_ref}  \\
  1144.01 & 97657  & 5.28 & 4.89 &  CP & \cite{Bakos2010}  \\
  1151.01 & 96618  & 21.29 & 3.47 &  CP & \cite{Talens2018} \\
  1204.01 & 55069  & 2.02 & 1.38 & PC & \cite{TOI_ref}  \\
  1271.01 & 66192  & 13.14 & 6.13 &  CP & \cite{daSilva2006} \\
  1339.01 & 99175  & 2.56 & 28.58 & CP & \cite{Badenas2020} \\
  1339.02 & 99175  & 3.2 & 8.88 & CP & \cite{Badenas2020}  \\
  1339.03 & 99175  & 2.62 & 38.35 & CP & \cite{Badenas2020}  \\
  1730.01 & 34730  & 1.48 & 2.16 & PC & \cite{TOI_ref}  \\
  1730.02 & 34730  & 2.44 & 12.57 & PC & \cite{TOI_ref}  \\
  1730.03 & 34730  & 2.76 & 6.23 & PC & \cite{TOI_ref}  \\
  1799.01 & 54491  & 1.63 & 7.09 & CP & \cite{Polanski2024}  \\
  1831.01 & 65205  & 4.4 & 0.56 & PC & \cite{TOI_ref} \\
  4175.01 & 64573  & -- & 2.16 & PC &\cite{TOI_ref}  \\
  4399.01 & 94235  & 3.0 & 7.71 & CP & \cite{zhou2022} \\
  4568.01 & 77921  & 2.26 & 14.01 & PC & \cite{TOI_ref} \\
  4603.01 & 26250  & 14.24 & 7.25 & CP & \cite{Khandelwal2023} \\
  5140.01 & 42403  & 2.9 & 15.61 & CP & \cite{Sulis2024}  \\
  5383.01 & 69275  & 1.4 & 2.81 & PC & \cite{TOI_ref} \\
  5521.01 & 57386  & 6.56 & 18.51 & PC & \cite{TOI_ref} \\
  5811.01 & 102295  & 10.4 & 6.26 & PC & \cite{TOI_ref} \\
  5962.01 & 112486  & 4.86 & 1.93 & PC & \cite{TOI_ref} \\
\hline
\multicolumn{6}{c}{\textit{Transit False Positive Stars}} \\
\hline
  230.01 & 114003  & 3.93 & 13.34 & FA & \cite{TOI_ref} \\
  394.01 & 15053  & 5.96 & 0.39 & FP &  \cite{TOI_ref}\\
  510.01 & 33681  & 2.83 & 1.35 & FP & \cite{TOI_ref}  \\
  522.01 & 40694  & 2.11 & 0.4 & FP$^{2}$ &  \cite{TOI_ref}  \\
  575.01 & 41849  & 17.46 & 19.37 & FP & \cite{TOI_ref} \\
  621.01 & 44094  & 18.92 & 3.11 & FP & \cite{TOI_ref} \\
  635.01 & 47990  & 1.5 & 0.49 & FA & \cite{TOI_ref} \\
  680.01 & 58234  & 4.88 & 0.43 & FP & \cite{TOI_ref} \\
  896.01 & 28122  & 3.21 & 3.53 & FA & \cite{TOI_ref} \\
  909.01 & 82032  & 1.81 & 3.9 & FP & \cite{TOI_ref}  \\
  951.01 & 21000  & 20.19 & 2.97 & FP & \cite{TOI_ref} \\
  1124.01 & 98516  & 14.01 & 3.52 & FP & \cite{TOI_ref} \\
  1418.01 & 83168  & 1.7 & 0.68 & FA & \cite{TOI_ref} \\
  1665.01 & 27844  & 12.38 & 1.76 & FP & \cite{TOI_ref} \\
  1684.01 & 20334  & 9.37 & 1.15 & FP$^{3}$ & \cite{TOI_ref} \\
  1719.01 & 44289  & 9.66 & 2.76 & PC & \cite{TOI_ref}  \\
  1837.01 & 67650  & 7.82 & 5.82 & FP & \cite{TOI_ref} \\
  1946.01 & 70833  & 21.68 & 10.85 & FP & \cite{TOI_ref} \\
  2017.01 & 74685  & 16.89 & 5.5 & FP & \cite{TOI_ref} \\
  2118.01 & 79105  & 4.26 & 2.34 & FP & \cite{TOI_ref} \\
  2299.01 & 93711  & 3.69 & 165.02 & FA & \cite{TOI_ref} \\
  2666.01 & 45621  & 16.66 &-- & FP & \cite{TOI_ref} \\
  4314.01 & 22084  & 3.4 & 73.58 & FA & \cite{TOI_ref} \\
  5079.01 & 24007  & 2.23 & 1.49 & FP & \cite{TOI_ref}   \\
\hline
\end{longtable*}
\end{center}

\clearpage

\section{Unresolved stellar companions inferred from astrometric acceleration}
\begin{figure}[ht]
    \centering
    \includegraphics[width=\linewidth]{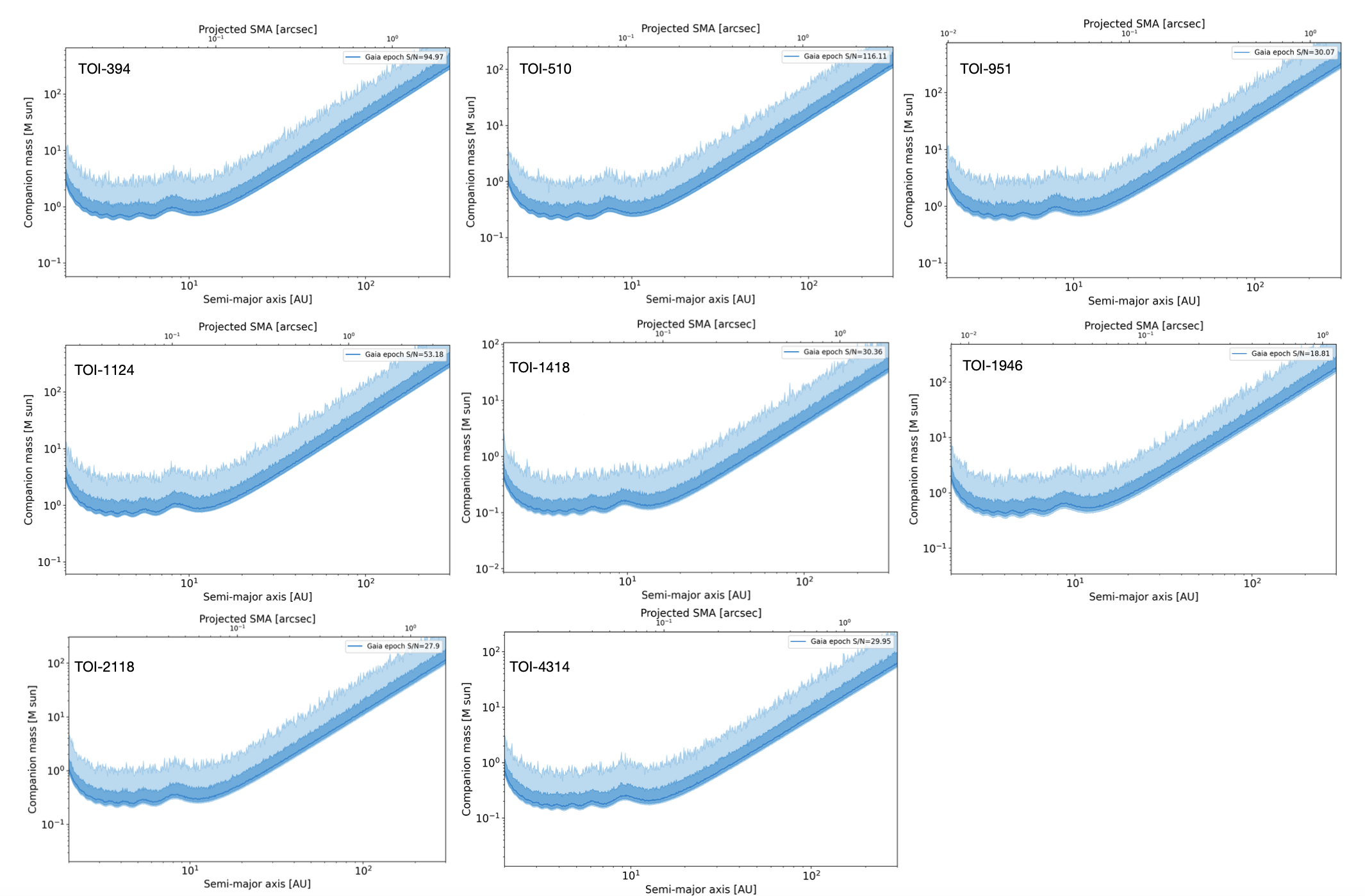}
    \caption{Astrometric acceleration from \textit{Hipparcos}-\textit{Gaia} reveals the presence of unresolved stellar companions within $0.1^{\arcsec}$ of 8 TOIs. This plot shows the inferred constraints on companion mass and semi-major axis.  }
    \label{fig:unres}
\end{figure}

\section{Relative RVs and proper motions of TOIs}\label{sec:revel_pm}
\begin{center}
\begin{longtable*}{llcccccccc}
\caption{Relative RVs used in our analysis}\label{tab:table_relRV}\\
\hline
Name & Time & Primary RV &  Primary RV error & Comp. RV  & Comp. RV error & Inst.   \\
 &  UT Date &  $\rm km\ s^{-1}$ & $\rm km\ s^{-1}$& $\rm km\ s^{-1}$& $\rm km\ s^{-1}$ &  \\
\hline
\hline
\endfirsthead
\hline
Name & Time & Primary RV &  Primary RV error & Comp. RV  & Comp. RV error & Inst.   \\
 &   &  $\rm km\ s^{-1}$ & $\rm km\ s^{-1}$& $\rm km\ s^{-1}$& $\rm km\ s^{-1}$ &  \\
\hline
\hline
\endhead
\hline
\endfoot
TOI-402 & Sep. 26 2024  &  -3.31& 0.10 &-5.58 & 0.10 & KPIC \\
TOI-1684 & Jul. 22 2022  & 33.87 & 0.10 & 26.57 & 0.30& KPIC \\
TOI-1684 & Jul. 23 2022  & 33.65 & 0.10& 53.05& 0.30& KPIC \\
TOI-5962 & Nov. 22 2024  & -278.77 & 3.10 & -270.66 & 0.33& KPIC \\
TOI-128 & Jan. 01 2016  & 11.04  & 0.14& 8.76 & 0.65 & Gaia DR3 \\
TOI-1099 & Jan. 01 2016  & -11.51  & 0.12 & -9.91 & 1.75 & Gaia DR3 \\
TOI-4175 & Jan. 01 2016  & 42.21  & 0.25 & 44.19 & 0.54& Gaia DR3 \\
\hline
\end{longtable*}
\end{center}

\begin{center}
\begin{longtable*}{llcccccc}
\caption{Relative proper motions from Gaia DR3 used in our analysis}\label{tab:table_relpm}\\
\hline
Name &  Primary $\mu_{\alpha}$ &  Primary $\mu_{\delta*}$ & Comp. $\mu_{\alpha}$  & Comp. $\mu_{\delta*}$   \\
 &   $\rm mas\ yr^{-1}$ & $\rm mas\ yr^{-1}$& $\rm mas\ yr^{-1}$& $\rm mas\ yr^{-1}$   \\
\hline
\hline
\endfirsthead
\hline
Name &  Primary $\mu_{\alpha}$ &  Primary $\mu_{\delta*}$ & Comp. $\mu_{\alpha}$  & Comp. $\mu_{\delta*}$   \\
 &    $\rm mas\ yr^{-1}$ & $\rm mas\ yr^{-1}$& $\rm mas\ yr^{-1}$& $\rm mas\ yr^{-1}$   \\
\hline
\hline
\endhead
\hline
\endfoot
TOI-128   & $-48.97\pm{0.02}$ & $45.38\pm{0.02}$& $-49.43\pm{0.03}$ & $42.89\pm{0.03}$  \\
TOI-906 & $-28.76\pm{0.02}$ & $-37.95\pm{0.02}$ & $-27.19\pm{0.18}$& $-41.10\pm{0.21}$ \\
TOI-4175 & $140.23\pm{0.04}$ &  $-140.05\pm{0.03}$& $130.36\pm{0.04}$ &  $-143.00\pm{0.04}$\\ 
TOI-5962 & $456.93\pm{0.02}$ &  $-57.85\pm{0.01}$& $456.20\pm{0.03}$& $-57.05\pm{0.06}$\\
\hline
\end{longtable*}
\end{center}

\subsection{Measuring RVs from KPIC data}
We fit the KPIC spectra of both the primary and secondary  stars using BT-Settl models \citep{Allard2012} to measure their RVs. Our grid of BT-Settl models assumes solar metallicity and covers between $3000-6000~$K in \Teff and $3.5-5.5$ in \logg, with 100~$K$ spacing in \Teff and 0.5 dex spacing in \logg. We use a forward model framework to fit the data and measure the RVs, which is described in \citet{Xuan2024}. The fitted parameters are RV, \vsini, \Teff, \logg, the flux scale, and an error multiple term to account for underestimation of the data uncertainties. The forward model is
\begin{equation}
    FM_A = \alpha_A T M_A
     \label{eqn:star}
\end{equation}
where $\alpha_A$ is the flux scale of the star in units of data numbers, $M_A$ is the stellar template from BT-Settl, and $T$ is the telluric and instrumental response, which is obtained from the telluric star observations. Following \citet{Xuan2022}, we perform continuum subtraction using a 100-pixel box median to estimate the continuum, before computing the residuals and log likelihood. The posterior is sampled with the nested sampling package \texttt{DYNESTY} \citep{speagle_DYNESTY_2020}, and we use 500 live points. 


\section{Binary properties from isochrone fitting}\label{sec:binary_mass}
\begin{center}
\begin{longtable*}{llcccccccc}
\caption{Binary properties from isochrone fitting}\label{tab:table_maas}\\
\hline
Name & HIP   & $ M_{\mathrm{A}}$ &  $M_{\mathrm{B}}$ & $ R_{\mathrm{A}}$ & $R_{\mathrm{B}}$ & $T_{\rm{eff, }A}$ &  $ T_{\rm{eff, }B}$ &  $\log g_{A}$ & $\log g_{B}$  \\
 &  Number &   $\rm M_{\odot}$ & $\rm M_{\odot}$ &$\rm R_{\odot}$  &$\rm R_{\odot}$  & $\rm K$ & $\rm K$  &  $\rm cgs$ & $\rm cgs$   \\
\hline
\hline
\endfirsthead
\hline
Name & HIP   &  $T_{\rm{eff, }A}$ &  $T_{\rm{eff, }B}$ & $M_{A}$ &  $M_{B}$ & $logg_{A}$ & $logg_{B}$ &  \\
 &  Number &    $\rm K$ & $\rm K$ & $\rm M_{\odot}$ &$\rm M_{\odot}$ & & & \\
\hline
\hline
\endhead
\hline
\endfoot
\multicolumn{10}{p{0.9\textwidth}}{
{\footnotesize
\textsuperscript{\dag} Note: The stellar properties are derived using isochrone fitting, which is model-dependent. We inflate the uncertainties ($5\%$ in mass, $4\%$ in radius, $2\%$ in $\rm{T_{eff}}$ ) listed in the table following the recommendations from \citet{Tayar2022}.
}}
\endlastfoot
TOI-128 & 24718& $1.08^{+0.05}_{-0.05}$ &$0.71^{+0.05}_{-0.05}$ &$1.08^{+0.04}_{-0.04}$ &$0.66^{+0.04}_{-0.04}$ &$6031^{+120}_{-120}$ &$4459^{+90}_{-90}$ &$4.4^{+0.05}_{-0.05}$ &$4.64^{+0.05}_{-0.05}$ \\
TOI-271 & 21952 & $1.09^{+0.07}_{-0.07}$ &$0.42^{+0.05}_{-0.05}$ &$1.36^{+0.05}_{-0.05}$ &$0.43^{+0.05}_{-0.05}$ &$5952^{+110}_{-110}$ &$3633^{+70}_{-70}$ &$4.2^{+0.05}_{-0.05}$ &$4.79^{+0.05}_{-0.05}$ \\
TOI-369 & 115594 & 
$1.38^{+0.07}_{-0.07}$ &$1.12^{+0.06}_{-0.06}$ &$1.93^{+0.07}_{-0.07}$ &$1.18^{+0.04}_{-0.04}$ &$6325^{+120}_{-120}$ &$6149^{+93}_{-96}$ &$4.00^{+0.05}_{-0.05}$ &$4.35^{+0.05}_{-0.05}$  \\
TOI-402 & 11433 & $0.82^{+0.04}_{-0.04}$ &$0.22^{+0.02}_{-0.02}$ &$0.86^{+0.03}_{-0.03}$ &$0.25^{+0.02}_{-0.02}$ &$5091^{+100}_{-100}$ &$3252^{+60}_{-60}$ &$4.48^{+0.05}_{-0.05}$ &$4.97^{+0.05}_{-0.05}$  \\
TOI-906 & 78301 & $1.14^{+0.06}_{-0.06}$ &$0.73^{+0.04}_{-0.04}$ &$1.87^{+0.08}_{-0.07}$ &$0.7^{+0.03}_{-0.03}$ &$5962^{+120}_{-120}$ &$4754^{+90}_{-90}$ &$3.94^{+0.05}_{-0.05}$ &$4.61^{+0.05}_{-0.05}$  \\
TOI-930 & 16881 & $1.87^{+0.22}_{-0.08}$ &$1.11^{+0.05}_{-0.04}$ &$3.98^{+0.16}_{-0.12}$ &$1.06^{+0.04}_{-0.04}$ &$6417^{+120}_{-120}$ &$5993^{+159}_{-125}$ &$3.5^{+0.07}_{-0.04}$ &$4.43^{+0.05}_{-0.05}$    \\
TOI-1099 & 108162 & $0.80^{+0.04}_{-0.04}$ &$0.55^{+0.03}_{-0.03}$ &$0.90^{+0.03}_{-0.03}$ &$0.56^{+0.02}_{-0.02}$ &$5092^{+100}_{-100}$ &$3858^{+77}_{-77}$ &$4.43^{+0.05}_{-0.05}$ &$4.69^{+0.05}_{-0.05}$   \\
TOI-1204 & 55069 & $1.19^{+0.10}_{-0.03}$ &$0.44^{+0.02}_{-0.02}$ &$1.78^{+0.06}_{-0.05}$ &$0.46^{+0.02}_{-0.02}$ &$5931^{+100}_{-100}$ &$3641^{+70}_{-70}$ &$4.01^{+0.05}_{-0.04}$ &$4.77^{+0.05}_{-0.05}$ \\
TOI-4175 & 64573 & $0.86^{+0.05}_{-0.05}$ & $0.82^{+0.05}_{-0.05}$ &$0.90^{+0.03}_{-0.03}$ &$0.88^{+0.03}_{-0.03}$ & $5298^{+120}_{-120}$ & $5118^{+120}_{-120}$ &$4.46^{+0.05}_{-0.05}$& $4.49^{+0.05}_{-0.05}$  \\
TOI-4399 & 94235 &$1.04^{+0.05}_{-0.05}$ &$0.31^{+0.02}_{-0.02}$ &$1.08^{+0.04}_{-0.04}$ &$0.33^{+0.02}_{-0.02}$ &$5998^{+120}_{-120}$ &$3534^{+70}_{-70}$ &$4.38^{+0.05}_{-0.05}$ &$4.88^{+0.05}_{-0.05}$   \\
TOI-5962 & 112486 & $0.65^{+0.03}_{-0.03}$ &$0.26^{+0.02}_{-0.02}$ &$0.63^{+0.03}_{-0.03}$ &$0.26^{+0.02}_{-0.02}$ &$5378^{+120}_{-120}$ &$3826^{+70}_{-70}$ &$4.65^{+0.05}_{-0.05}$ &$5.02^{+0.05}_{-0.05}$ \\
\hline
\end{longtable*}
\end{center}

\newpage
\section{Orbital fitting results}\label{sec:orbital_fitting}
\begin{figure*}[ht]
    \centering
\includegraphics[width=0.9\linewidth]{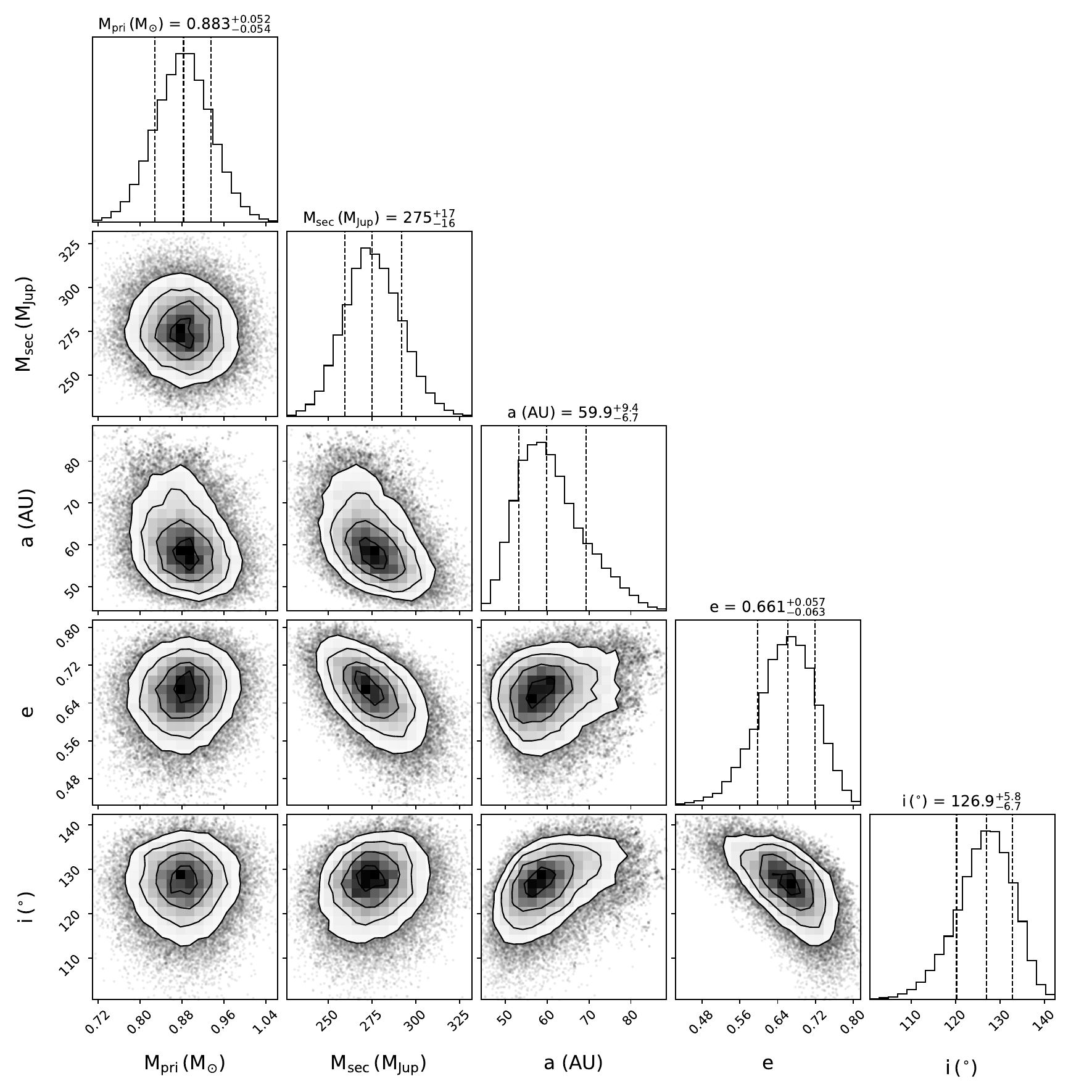}
    \caption{Joint posterior distributions for selected orbital parameters of TOI-402 B. These are the host star's mass ($\rm M_{pri}$), the companion mass ($\rm M_{sec}$), the semi-major axis $a$, the orbital eccentricity $e$, and the orbital inclination $I$. The values and histogram distributions of the posteriors of selected parameters are shown, along with 1 $\sigma$ uncertainties. The complete set of joint posterior distributions (12 figures) is available as a figure set in the online Journal}
    \label{fig:figure402_corner}
\end{figure*}

\newpage
\begin{figure*}[ht]
    \centering
\includegraphics[width=0.9\linewidth]{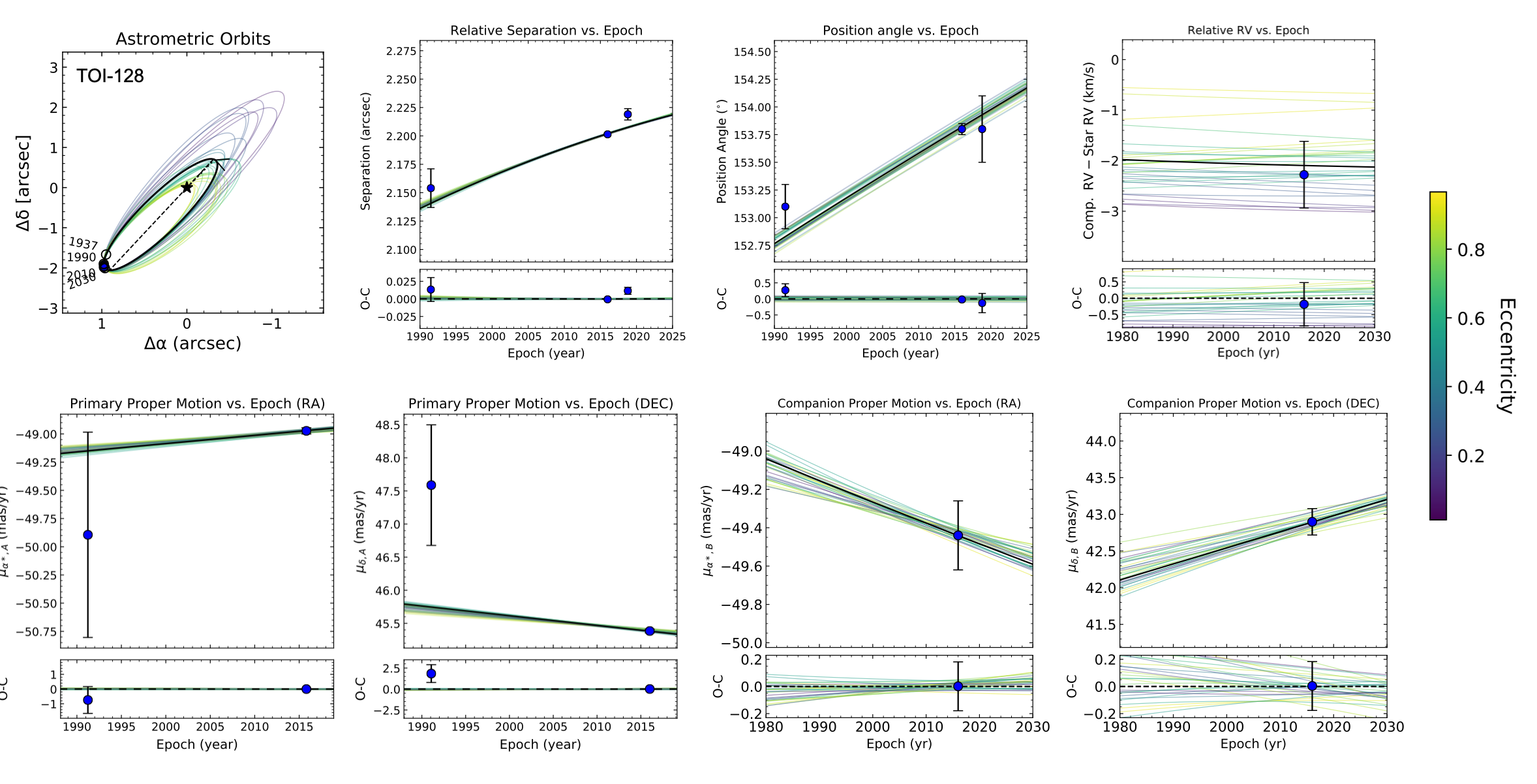}
    \caption{Orbital characterization of TOI-128 B  around A using relative astrometry and absolute astrometry from Hipparcos and Gaia. Similar to figure~\ref{fig:figure402}}
    \label{fig:figure128}
\end{figure*}
\begin{figure*}[!b]
    \centering
\includegraphics[width=0.62\linewidth]{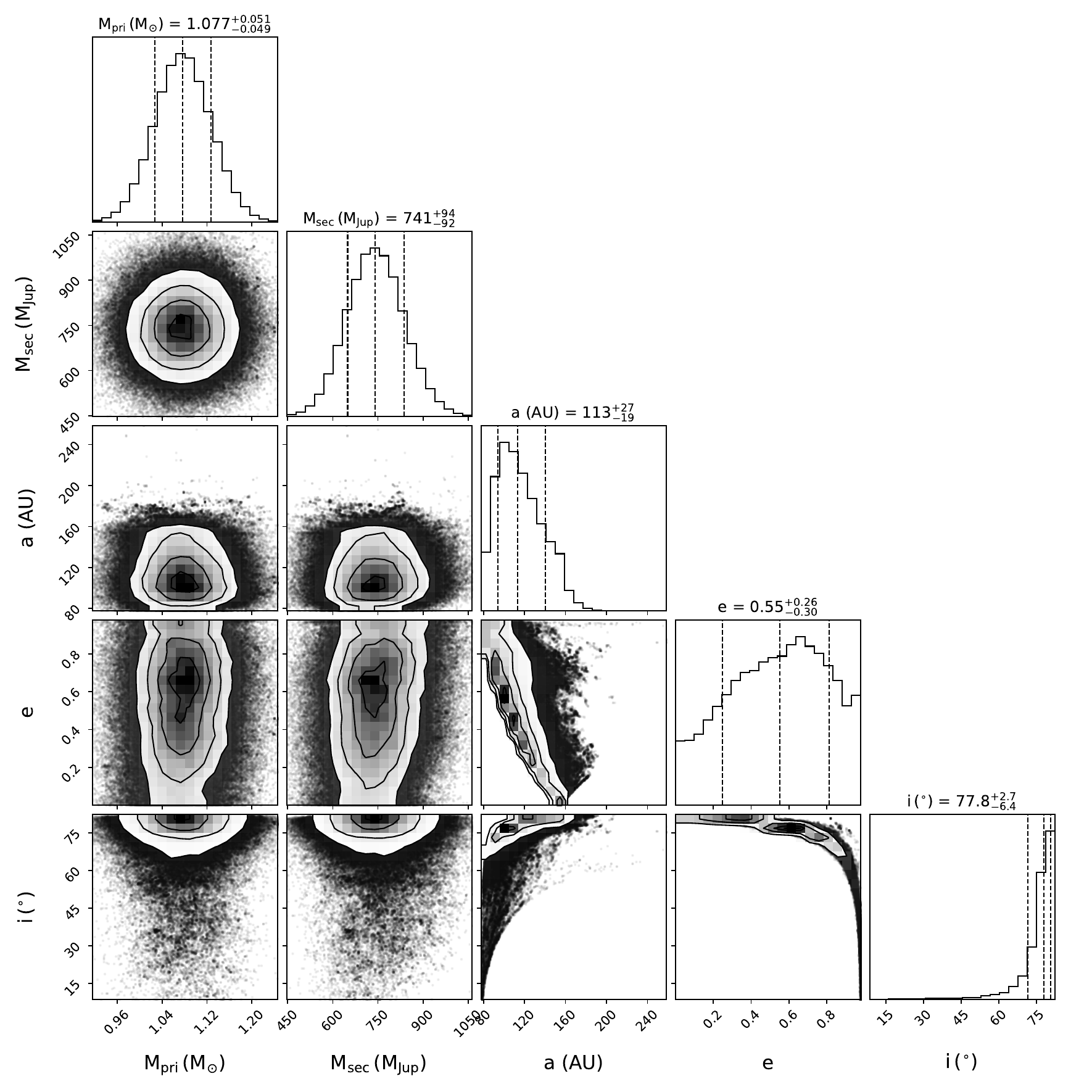}
    \caption{Joint posterior distributions for selected orbital parameters of TOI-128 B. Same as Figure~\ref{fig:figure402_corner}}
    \label{fig:figure128_corner}
\end{figure*}

\newpage
\begin{figure*}[ht]
    \centering
\includegraphics[width=0.75\linewidth]{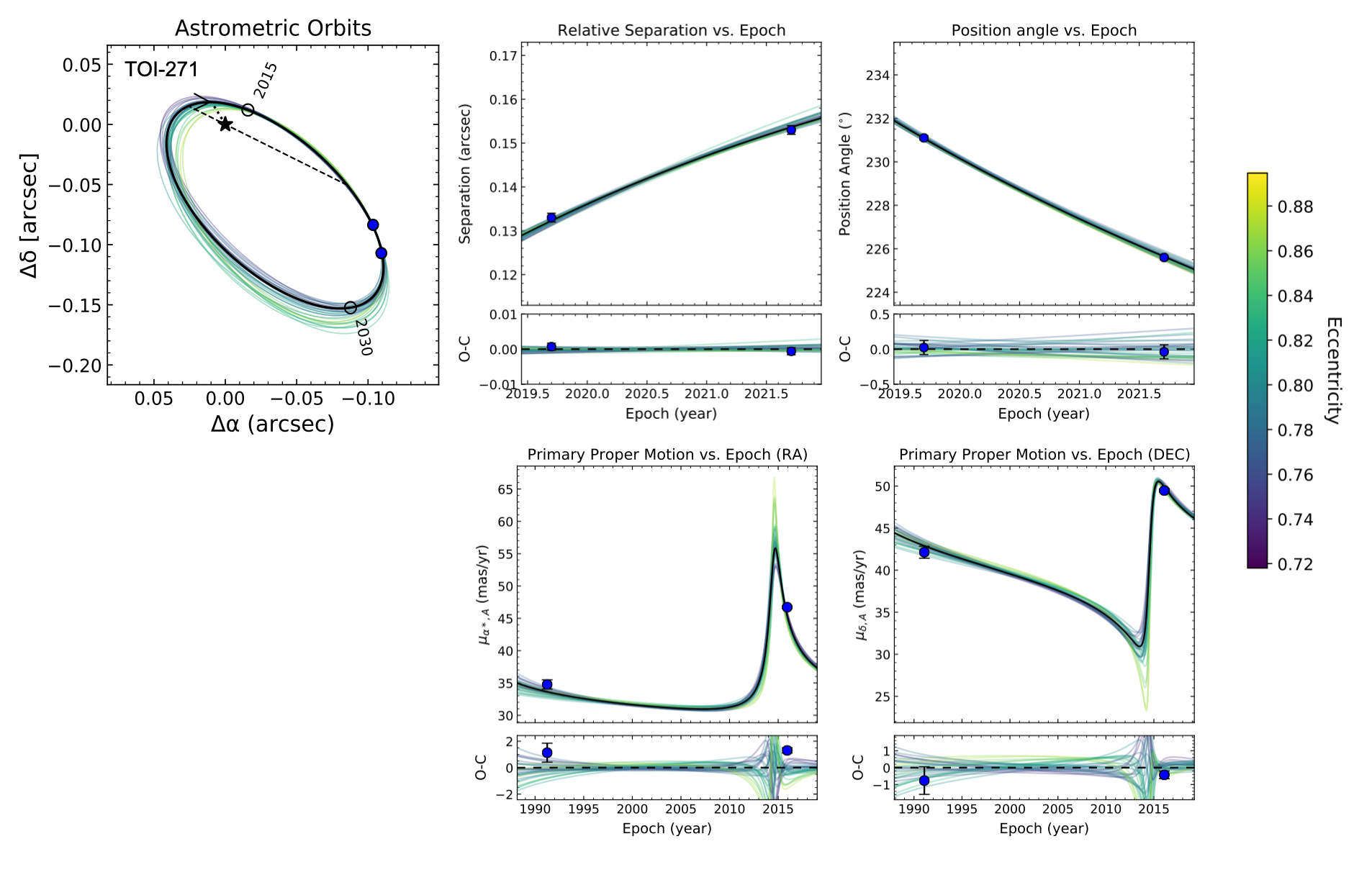}
    \caption{Orbital characterization of TOI-271 B  around A using relative astrometry and absolute astrometry from Hipparcos and Gaia. Similar to figure~\ref{fig:figure402}}
    \label{fig:figure271}
\end{figure*}
\begin{figure*}[!b]
    \centering
\includegraphics[width=0.62\linewidth]{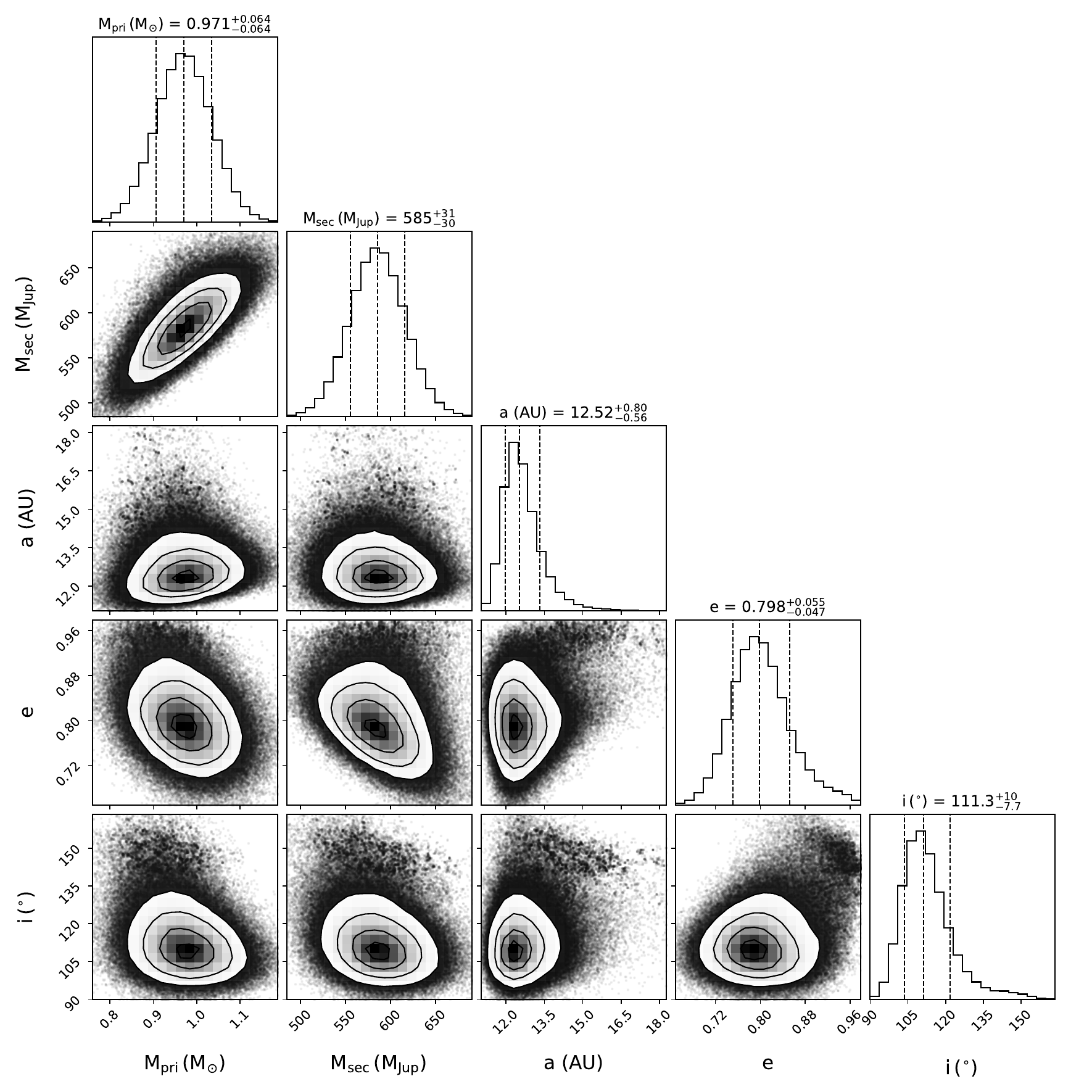}
    \caption{Joint posterior distributions for selected orbital parameters of TOI-271 B. Same as Figure~\ref{fig:figure402_corner}}
    \label{fig:figure271_corner}
\end{figure*}

\newpage
\begin{figure*}[ht]
    \centering
\includegraphics[width=0.83\linewidth]{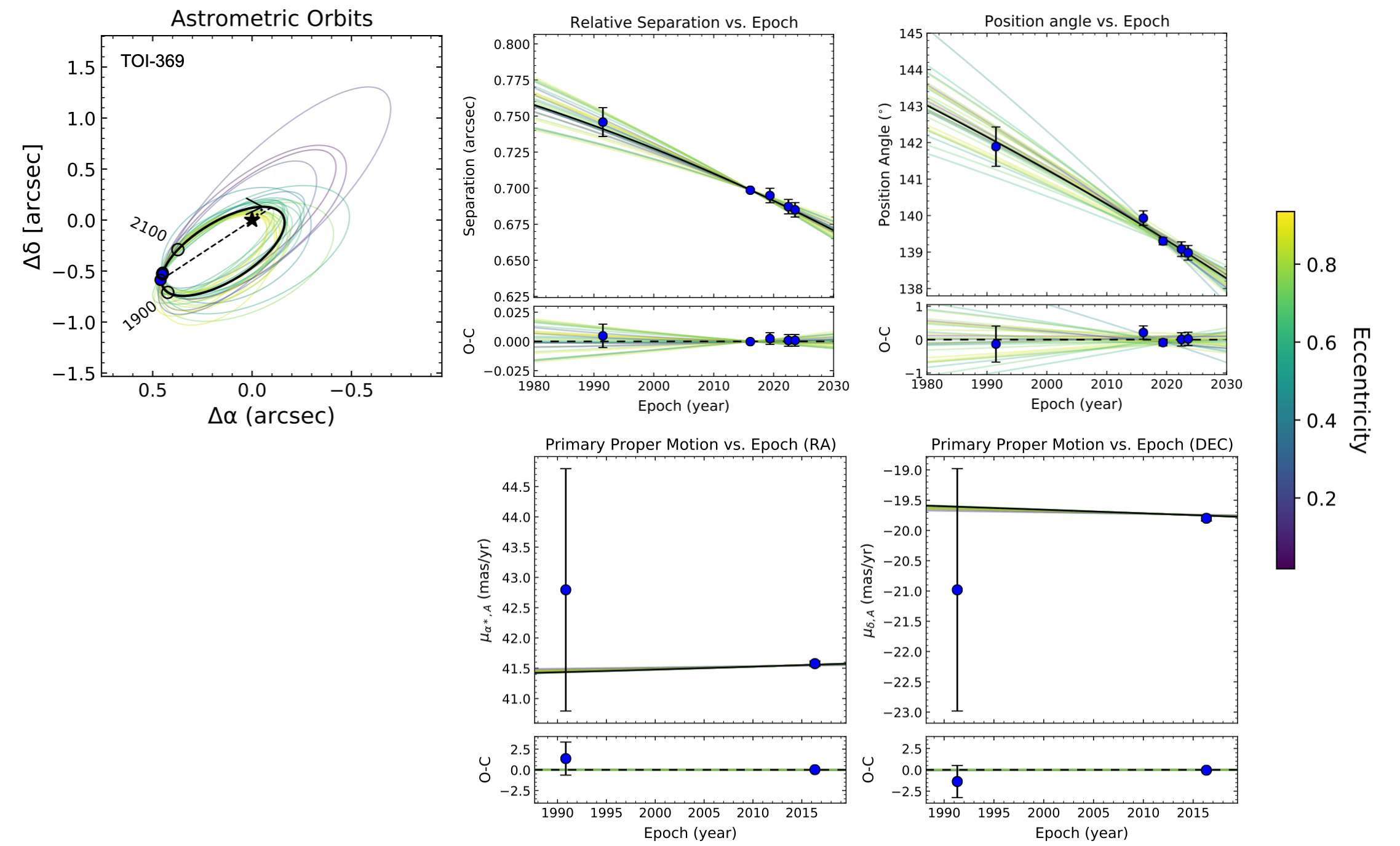}
    \caption{Orbital characterization of TOI-369 B  around A using relative astrometry and absolute astrometry from Hipparcos and Gaia. Similar to figure~\ref{fig:figure402}}
    \label{fig:figure369}
\end{figure*}
\begin{figure*}[!b]
    \centering
\includegraphics[width=0.6\linewidth]{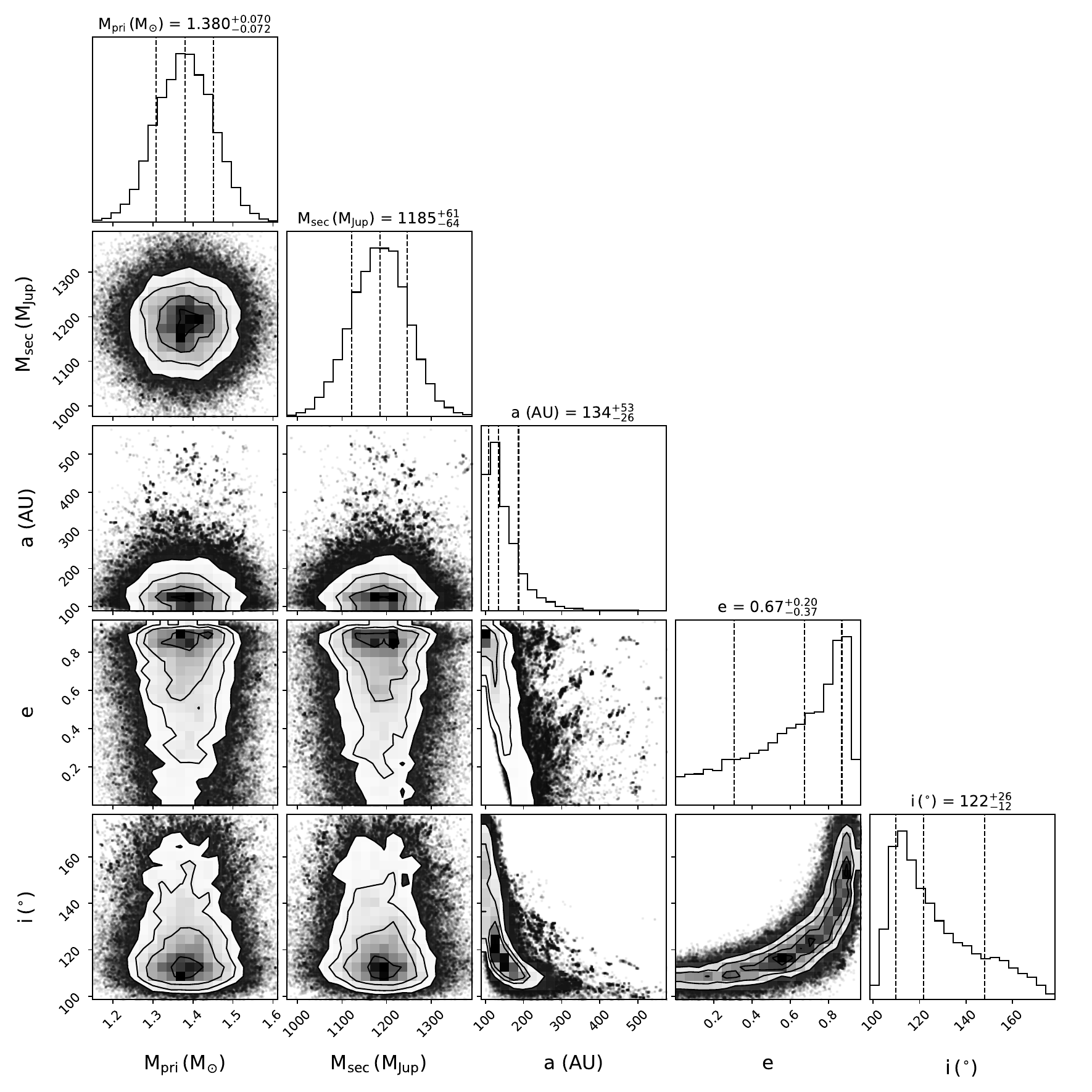}
    \caption{Joint posterior distributions for selected orbital parameters of TOI-369 B. Same as Figure~\ref{fig:figure402_corner}}
    \label{fig:figure369_corner}
\end{figure*}

\newpage
\begin{figure*}[ht]
    \centering
\includegraphics[width=0.9\linewidth]{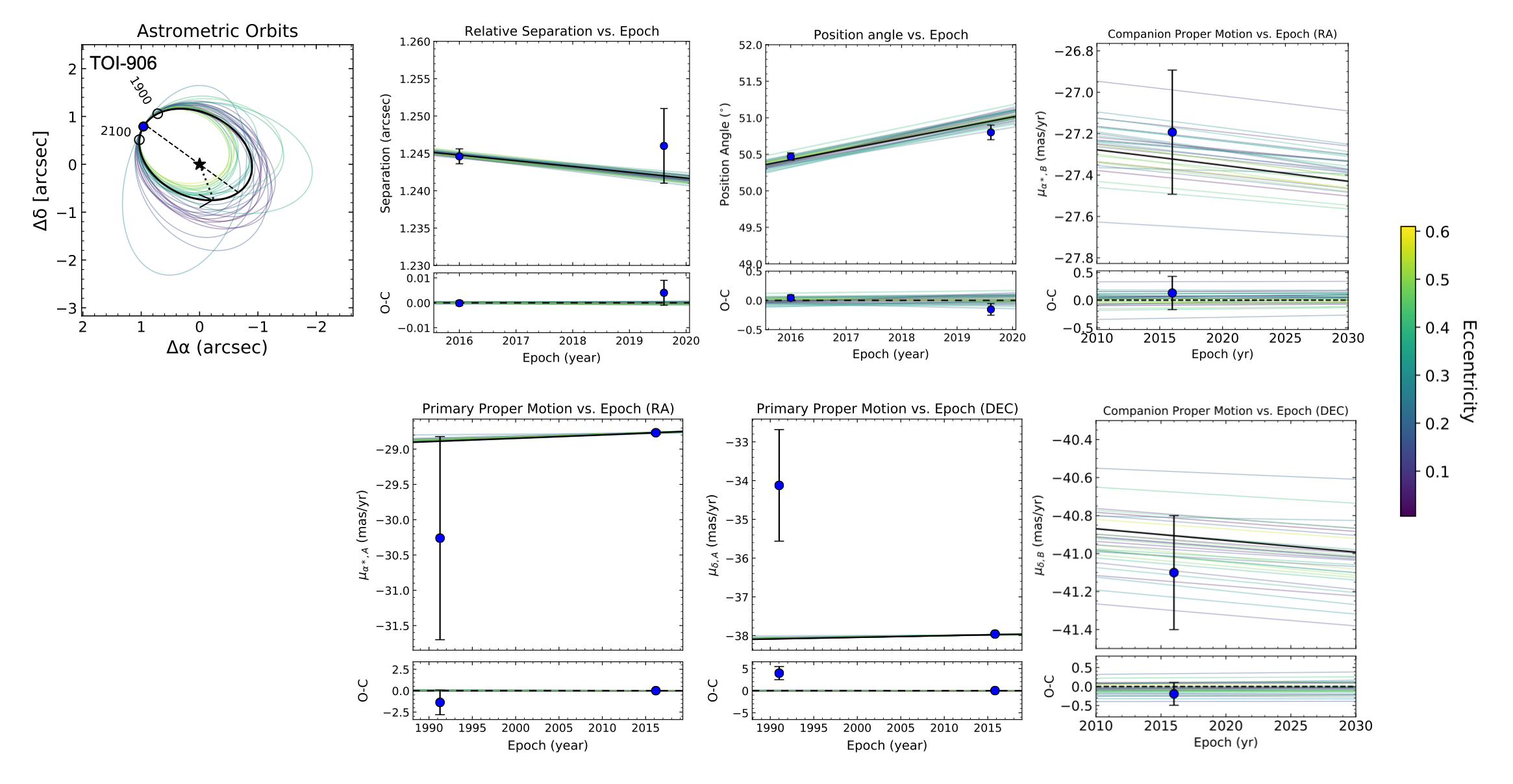}
    \caption{Orbital characterization of TOI-906 B  around A using relative astrometry and absolute astrometry from Hipparcos and Gaia. Similar to figure~\ref{fig:figure402}}
    \label{fig:figure906}
\end{figure*}
\begin{figure*}[!b]
    \centering
\includegraphics[width=0.62\linewidth]{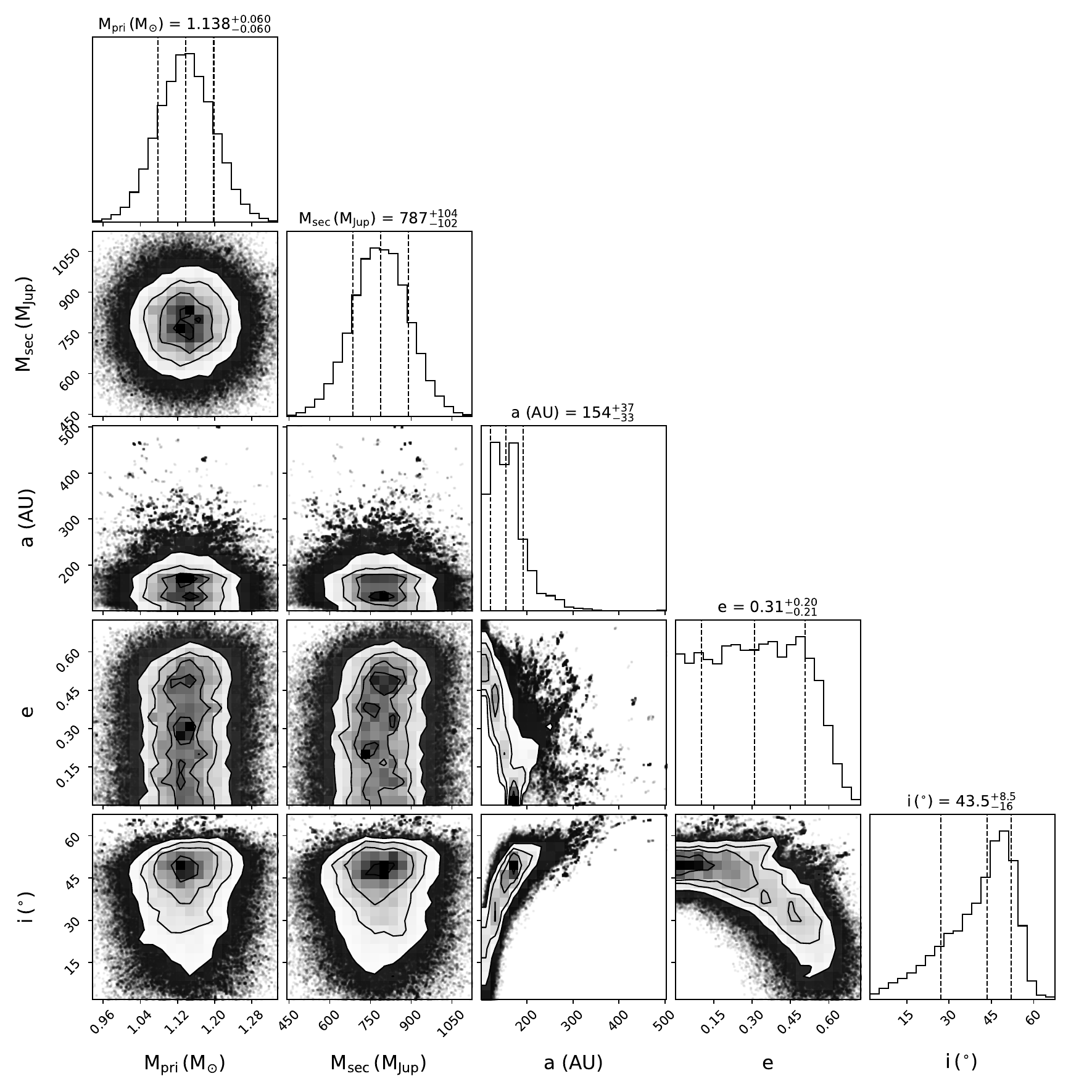}
    \caption{Joint posterior distributions for selected orbital parameters of TOI-906 B. Same as Figure~\ref{fig:figure402_corner}}
    \label{fig:figure906_corner}
\end{figure*}

\newpage
\begin{figure*}[ht]
    \centering
\includegraphics[width=0.75\linewidth]{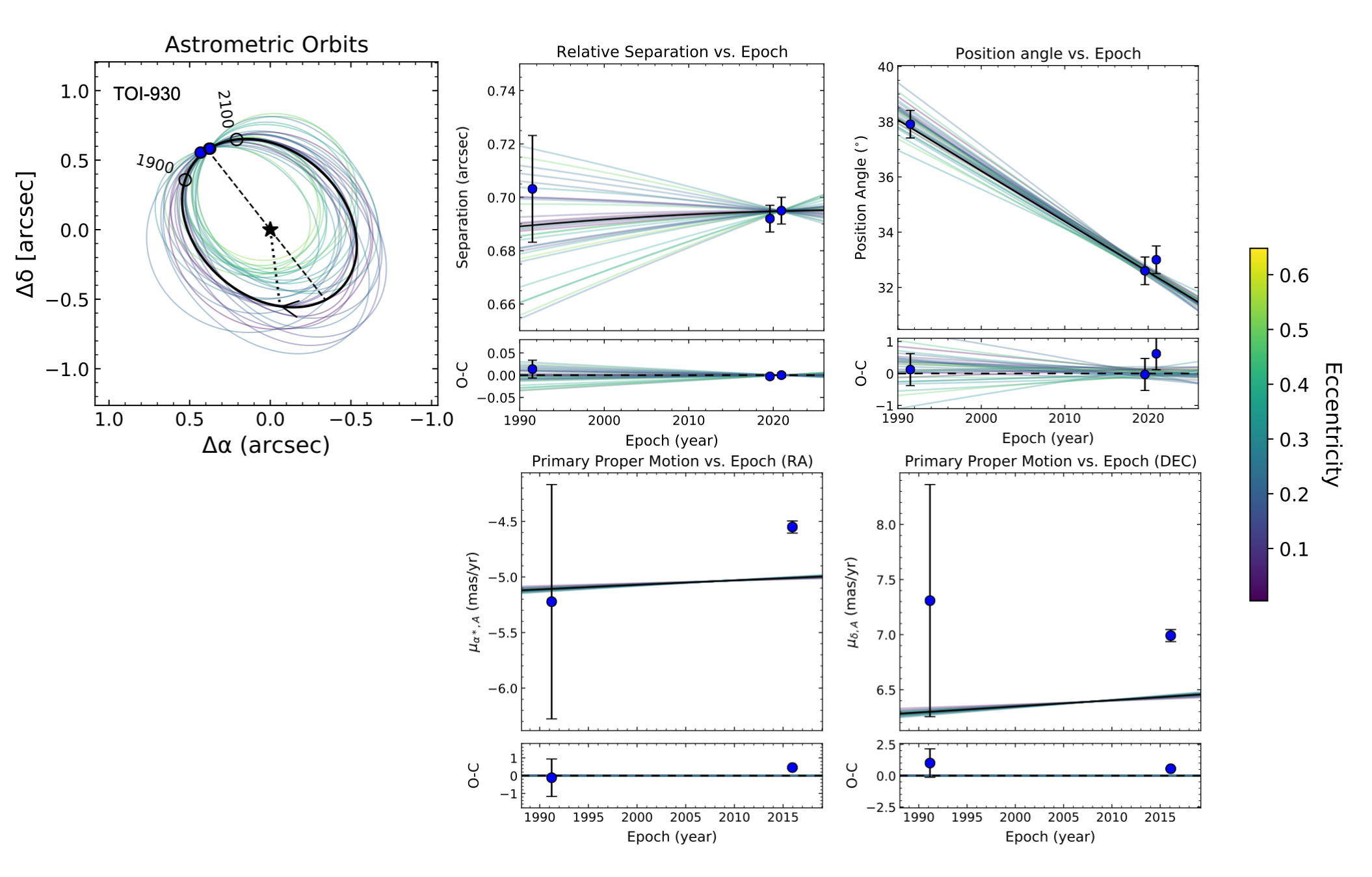}
    \caption{Orbital characterization of TOI-930 B  around A using relative astrometry and absolute astrometry from Hipparcos and Gaia. Similar to figure~\ref{fig:figure402}}
    \label{fig:figure930}
\end{figure*}
\begin{figure*}[!b]
    \centering
\includegraphics[width=0.62\linewidth]{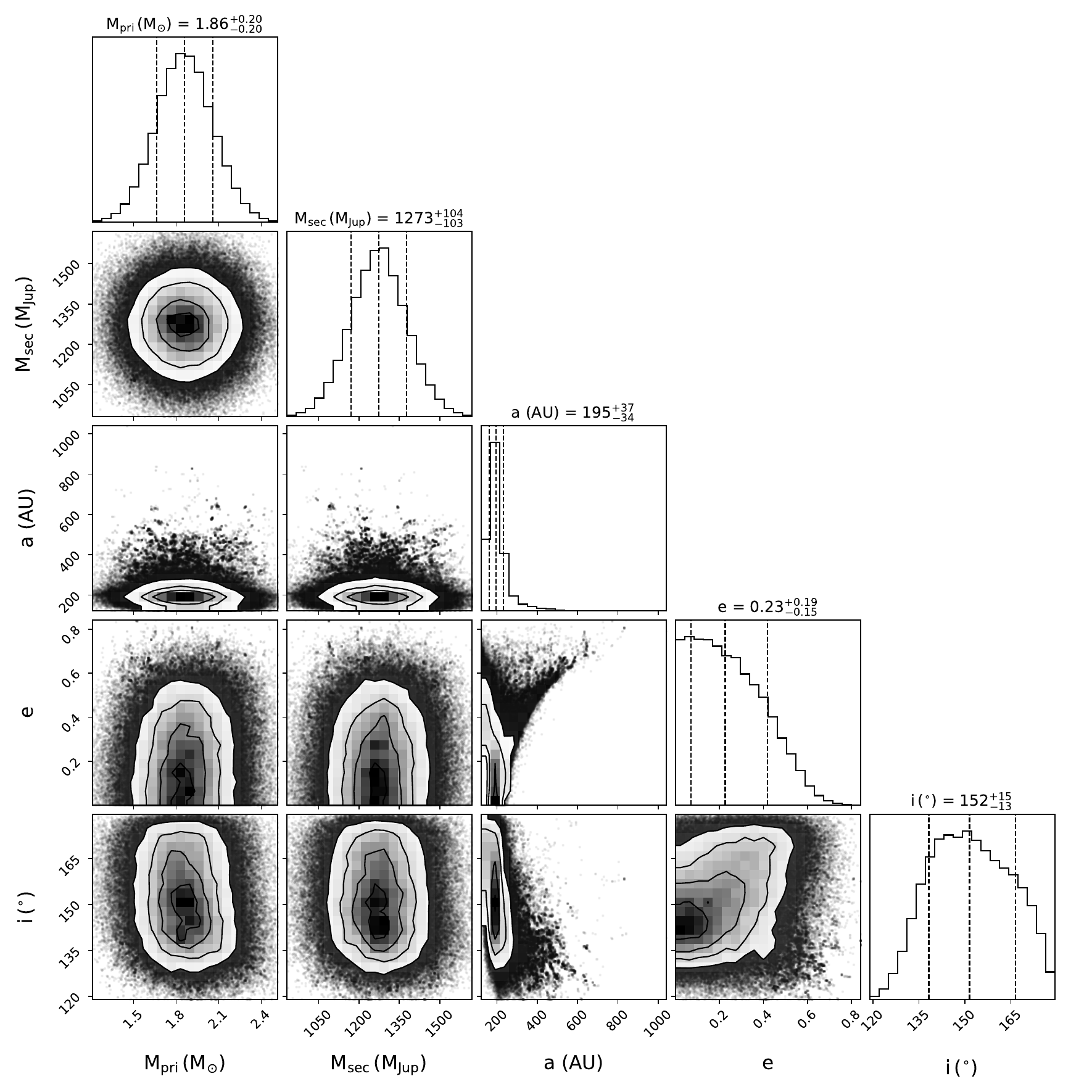}
    \caption{Joint posterior distributions for selected orbital parameters of TOI-930 B. Same as Figure~\ref{fig:figure402_corner}}
    \label{fig:figure930_corner}
\end{figure*}

\newpage
\begin{figure*}[ht]
    \centering
\includegraphics[width=0.75\linewidth]{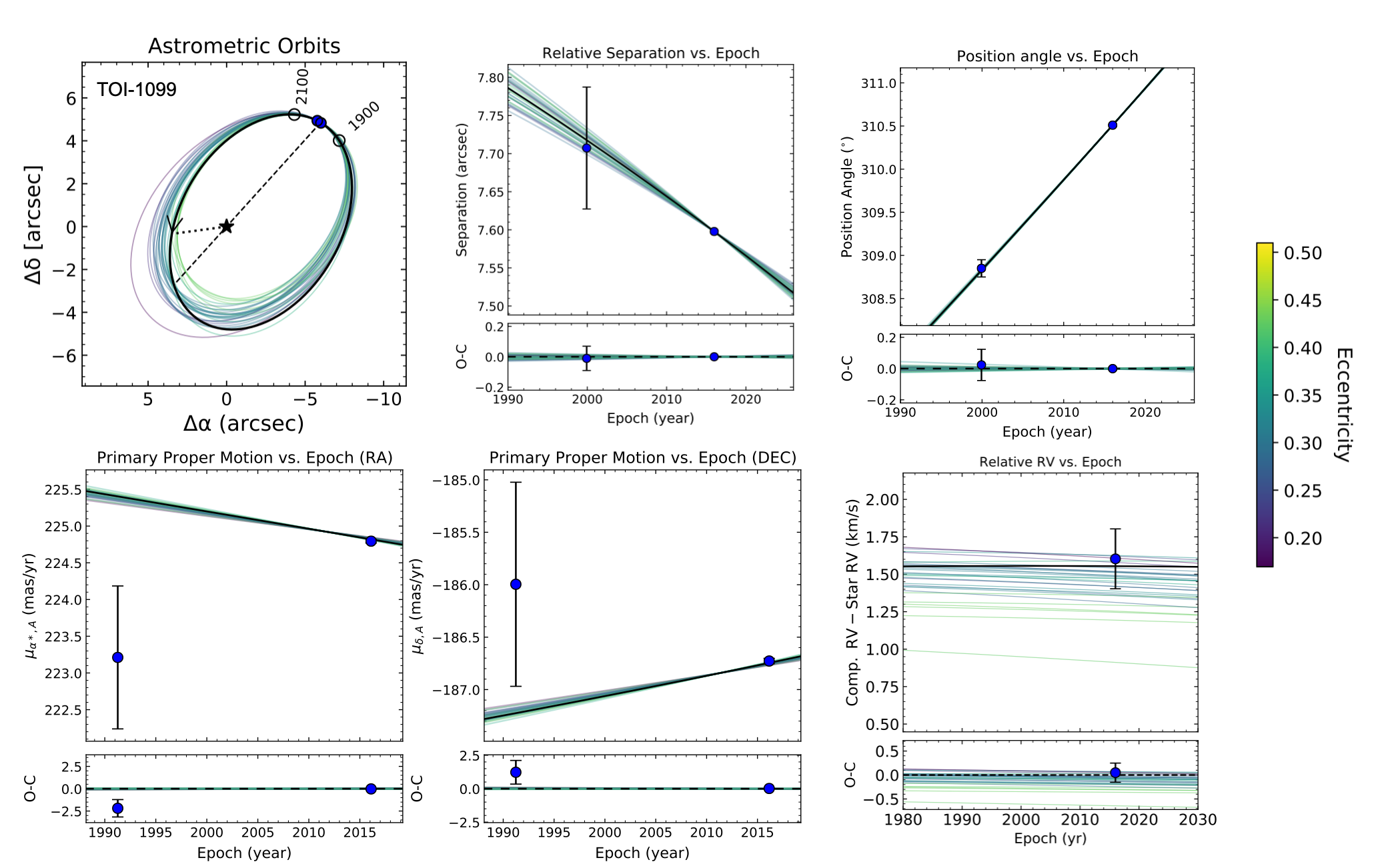}
    \caption{Orbital characterization of TOI-1099 B  around A using relative astrometry and absolute astrometry from Hipparcos and Gaia. Similar to figure~\ref{fig:figure402}}
    \label{fig:figure1099}
\end{figure*}
\begin{figure*}[!b]
    \centering
\includegraphics[width=0.62\linewidth]{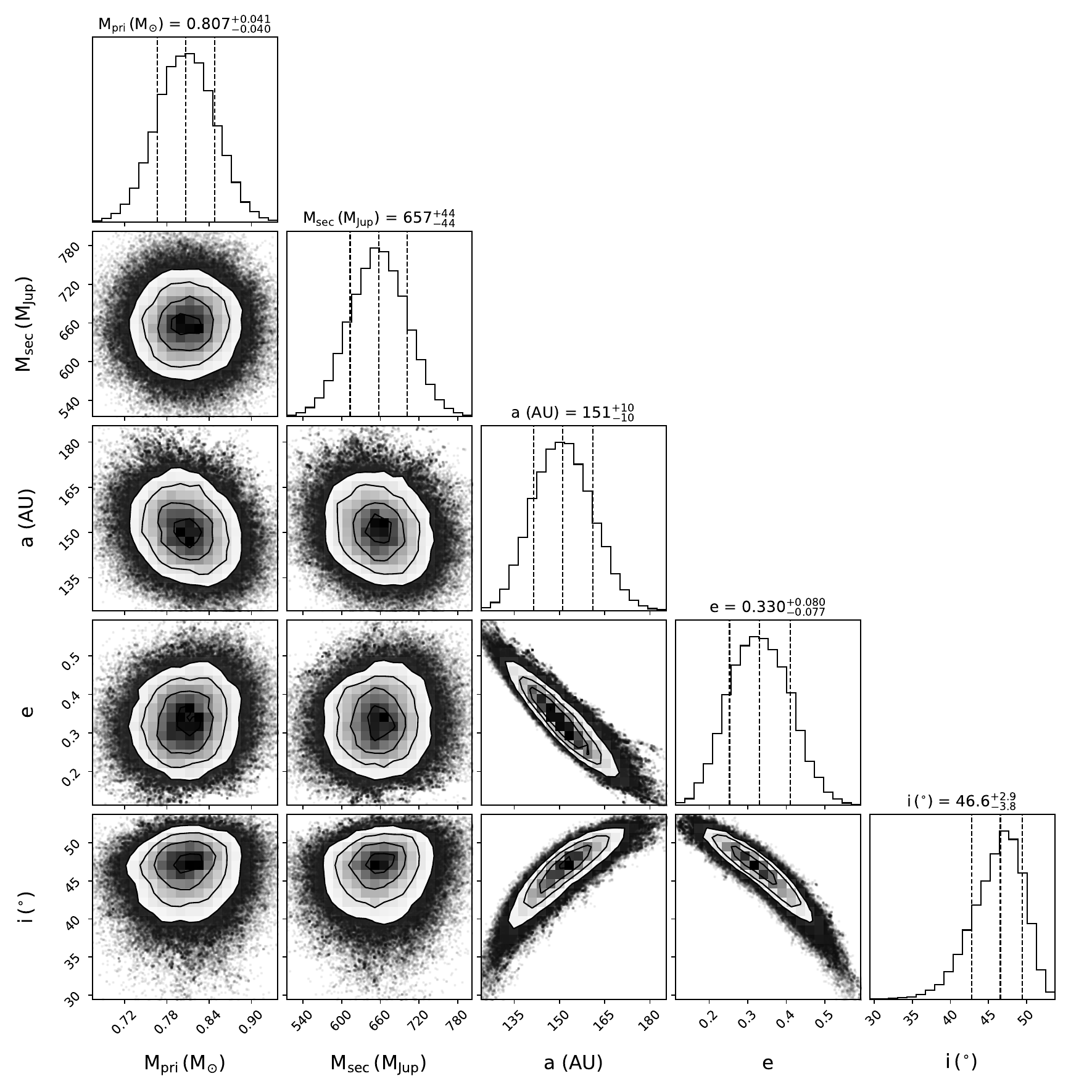}
    \caption{Joint posterior distributions for selected orbital parameters of TOI-1099 B. Same as Figure~\ref{fig:figure402_corner}}
    \label{fig:figure1099_corner}
\end{figure*}

\newpage
\begin{figure*}[ht]
    \centering
\includegraphics[width=0.75\linewidth]{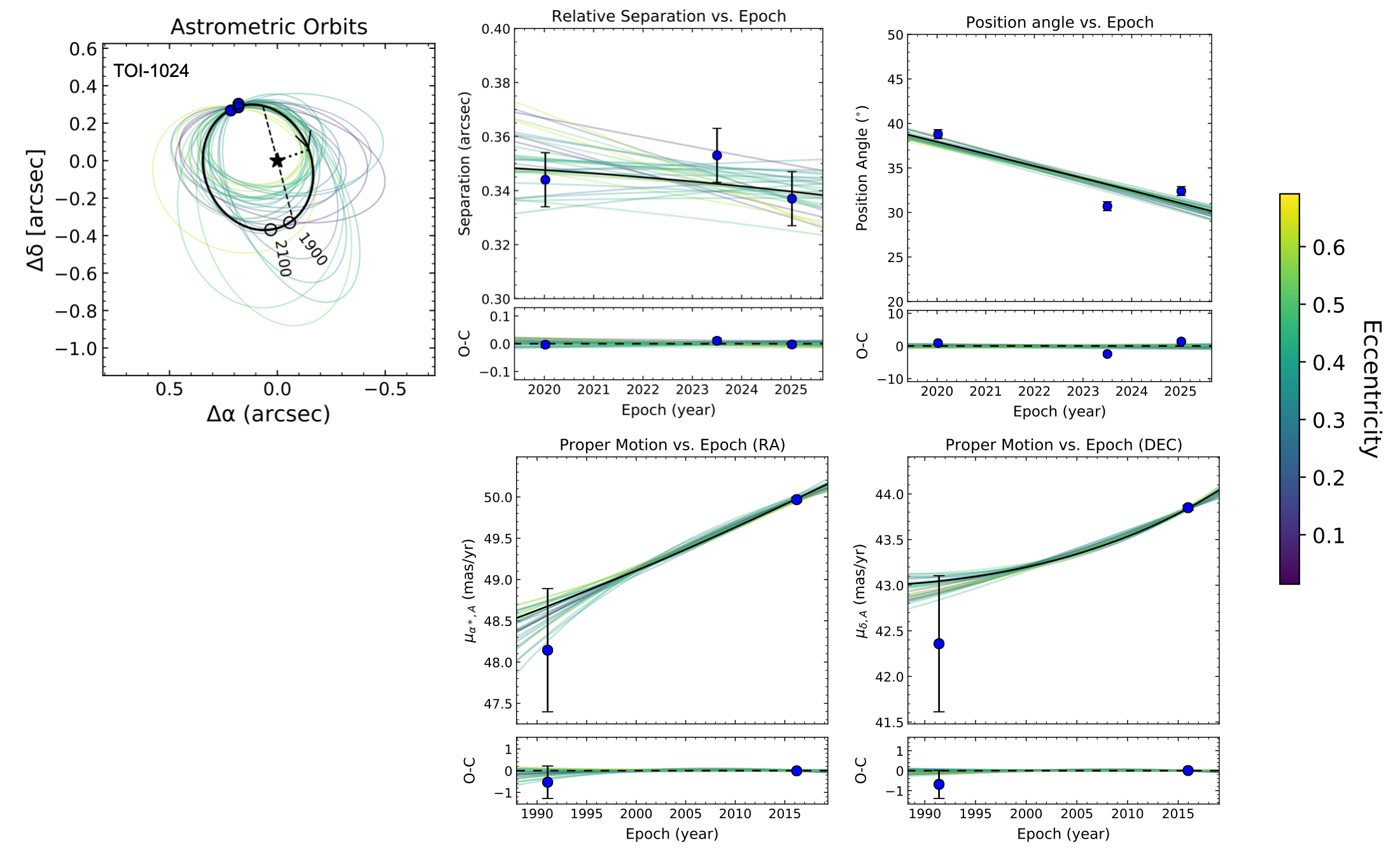}
    \caption{Orbital characterization of TOI-1204 B  around A using relative astrometry and absolute astrometry from Hipparcos and Gaia. Similar to figure~\ref{fig:figure402}}
    \label{fig:figure1204}
\end{figure*}
\begin{figure*}[!b]
    \centering
\includegraphics[width=0.62\linewidth]{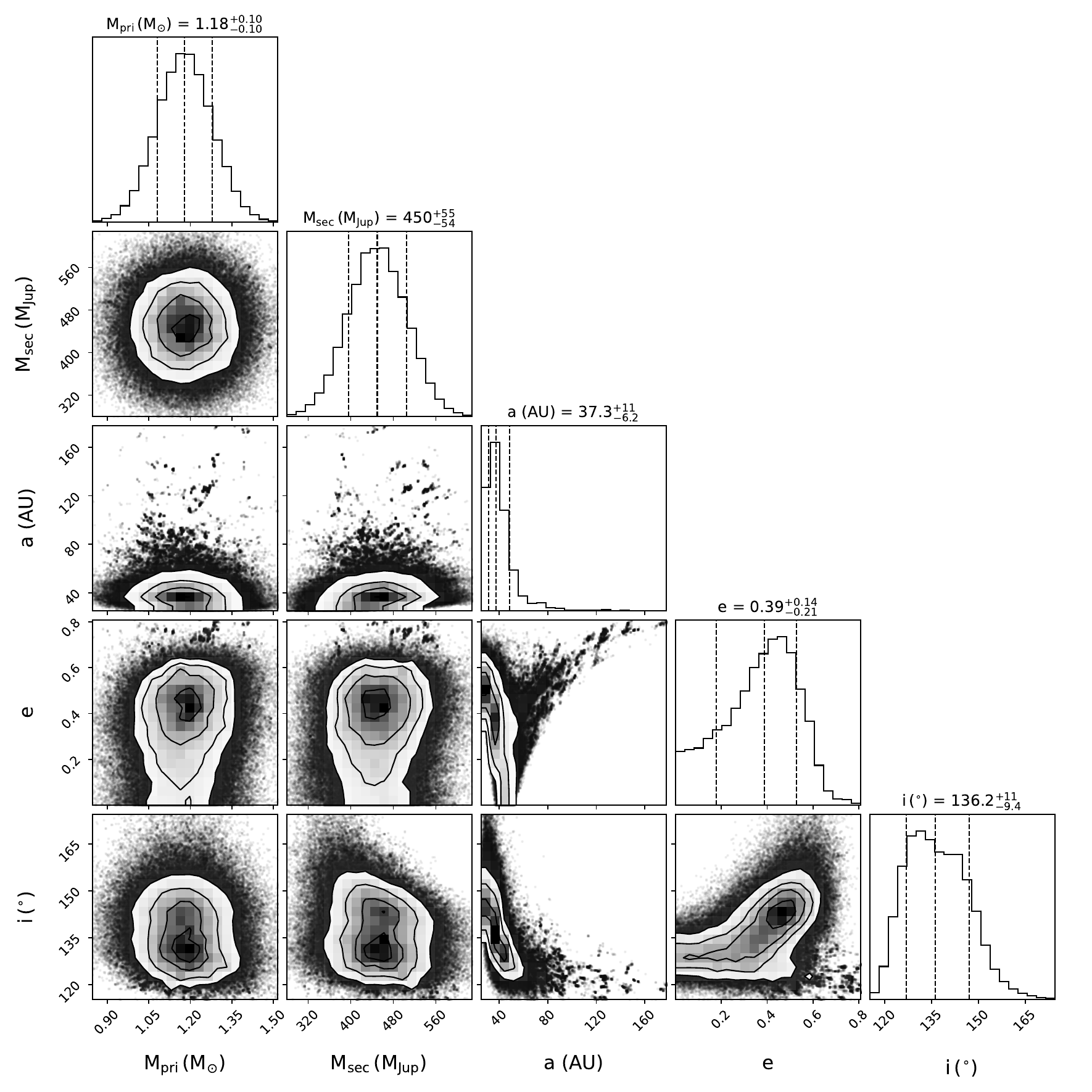}
    \caption{Joint posterior distributions for selected orbital parameters of TOI-1024 B. Same as Figure~\ref{fig:figure402_corner}}
    \label{fig:figure1024_corner}
\end{figure*}

\newpage
\begin{figure*}[ht]
    \centering
\includegraphics[width=0.9\linewidth]{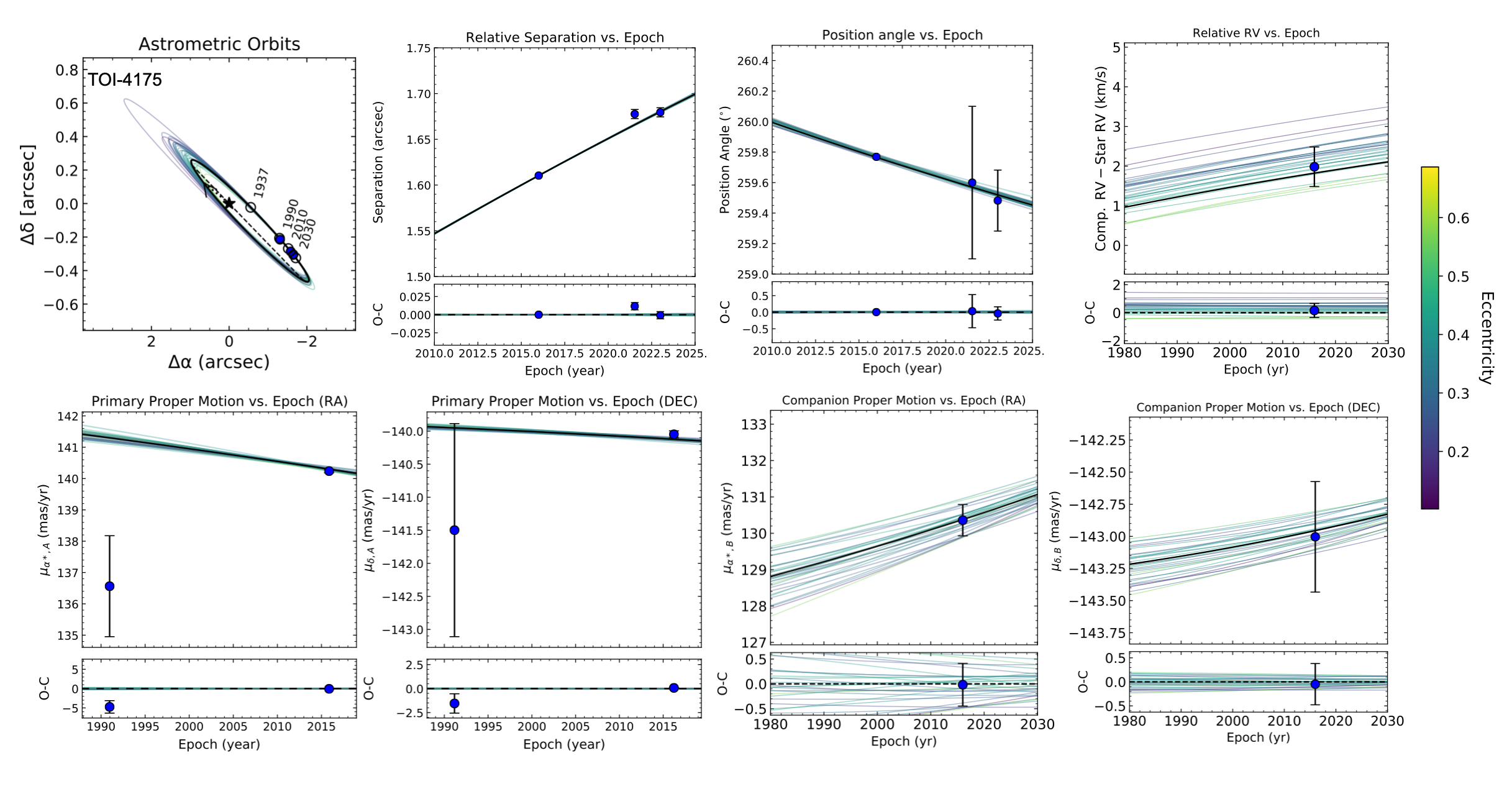}
    \caption{Orbital characterization of TOI-4175 B  around A using relative astrometry and absolute astrometry from Hipparcos and Gaia. Similar to figure~\ref{fig:figure402}}
    \label{fig:figure4175}
\end{figure*}
\begin{figure*}[!b]
    \centering
    \includegraphics[width=0.62\linewidth]{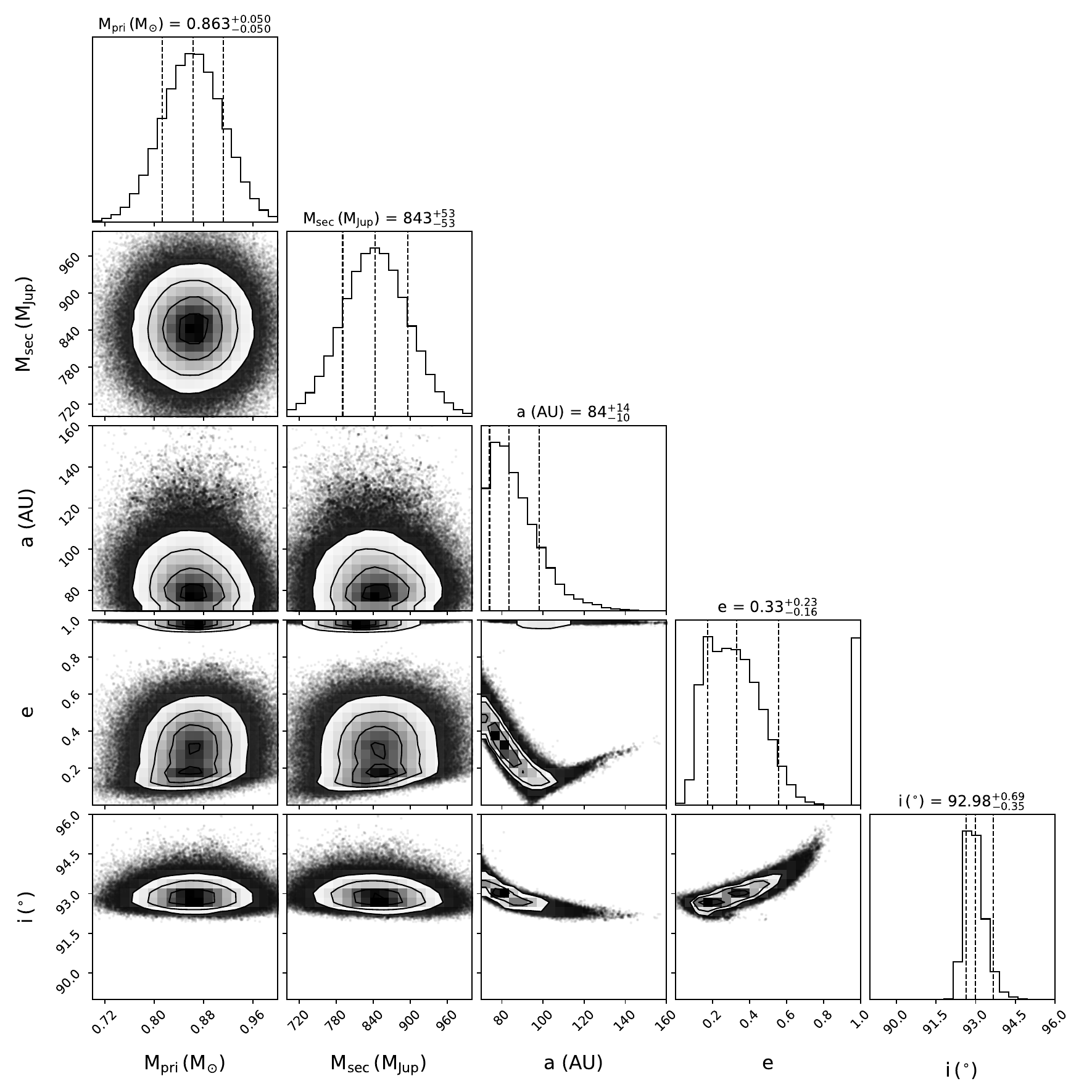}
    \caption{Joint posterior distributions for selected orbital parameters of TOI-4175 B. Same as Figure~\ref{fig:figure402_corner}}
    \label{fig:figure4175_corner}
\end{figure*}

\newpage
\begin{figure*}[ht]
    \centering
\includegraphics[width=0.75\linewidth]{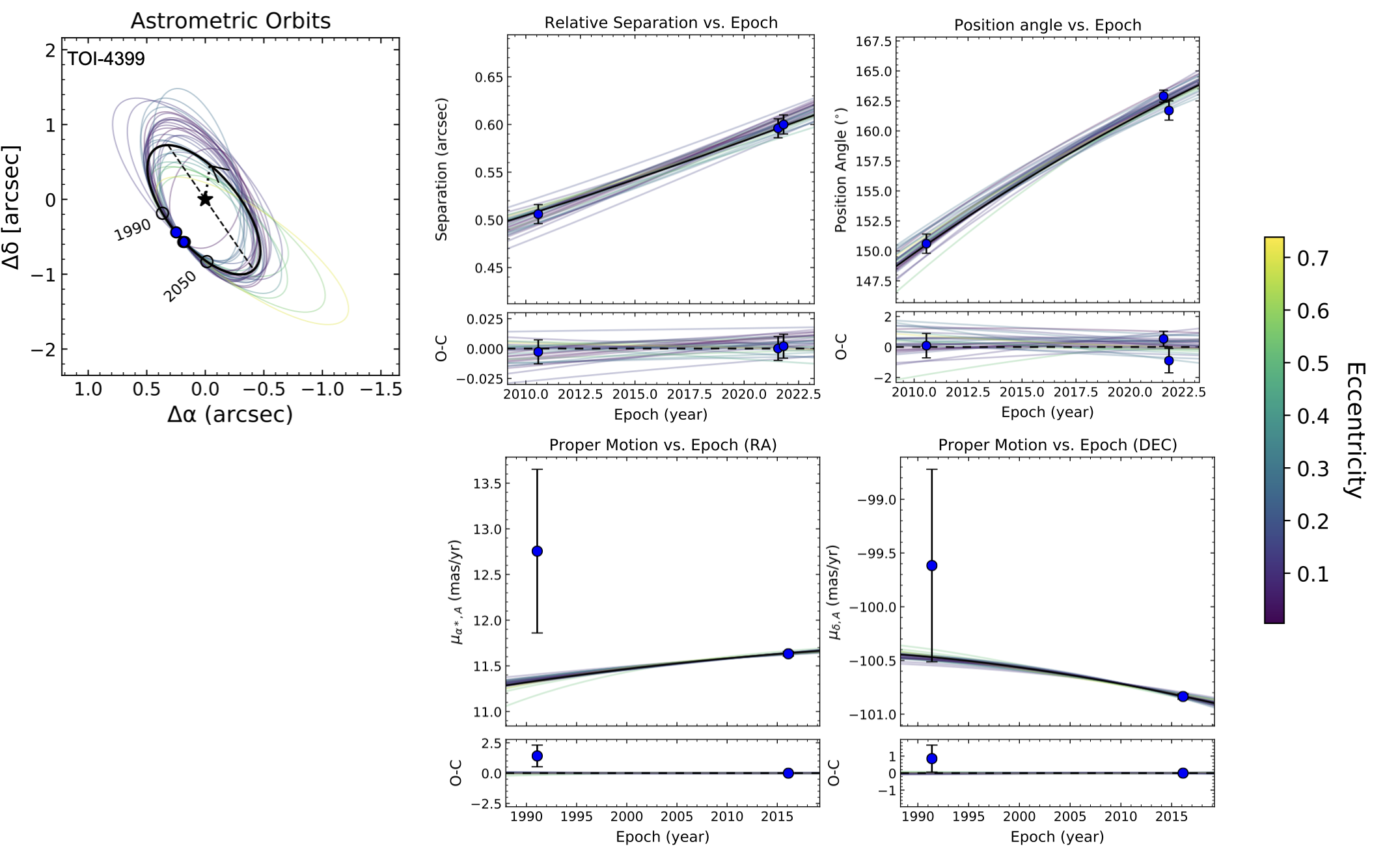}
    \caption{Orbital characterization of TOI-4399 B  around A using relative astrometry and absolute astrometry from Hipparcos and Gaia. Similar to figure~\ref{fig:figure402}}
    \label{fig:figure4399}
\end{figure*}
\begin{figure*}[!b]
    \centering
    \includegraphics[width=0.62\linewidth]{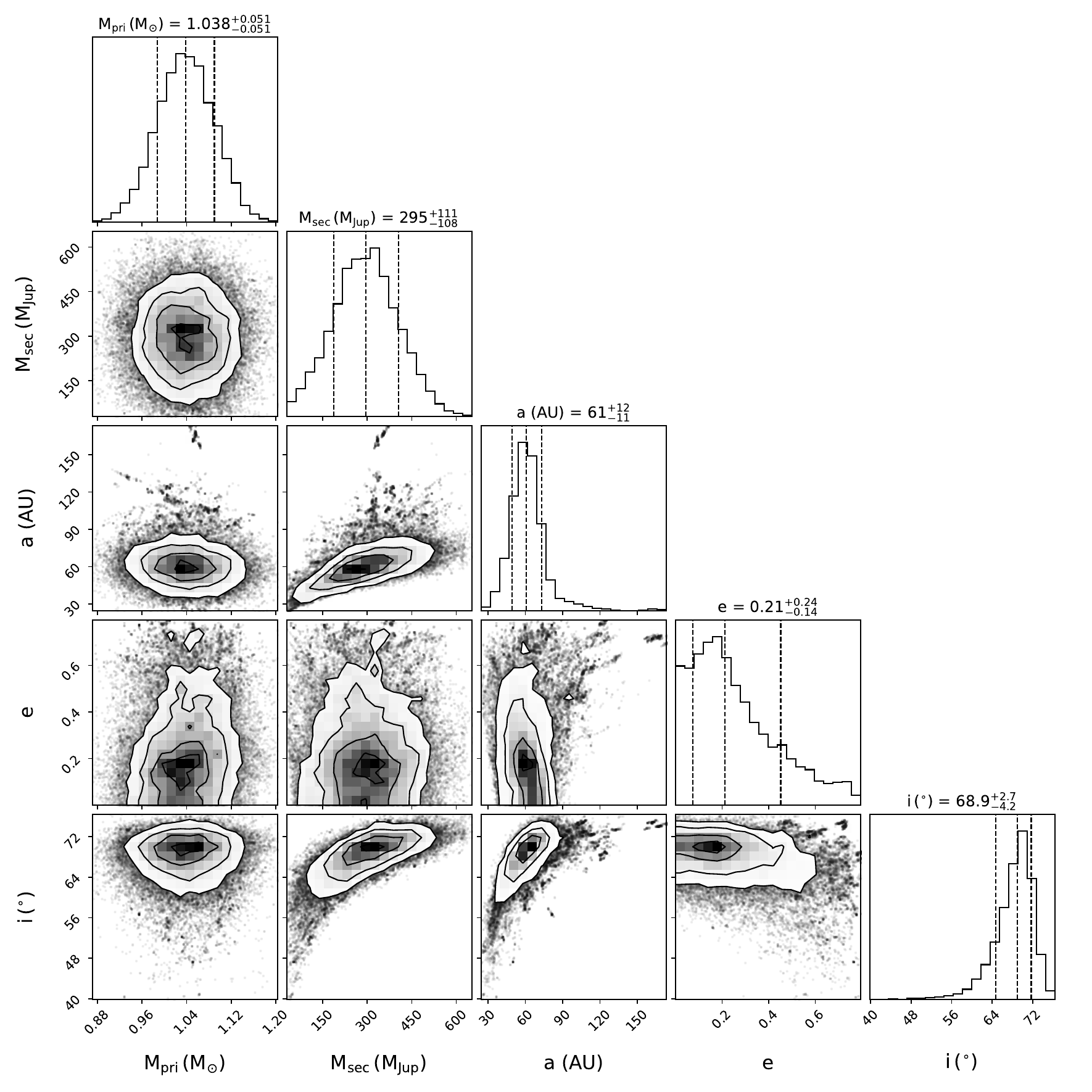}
    \caption{Joint posterior distributions for selected orbital parameters of TOI-4399 B. Same as Figure~\ref{fig:figure402_corner}}
    \label{fig:figure4399_corner}
\end{figure*}


\newpage
\begin{figure*}[ht]
    \centering
\includegraphics[width=0.9\linewidth]{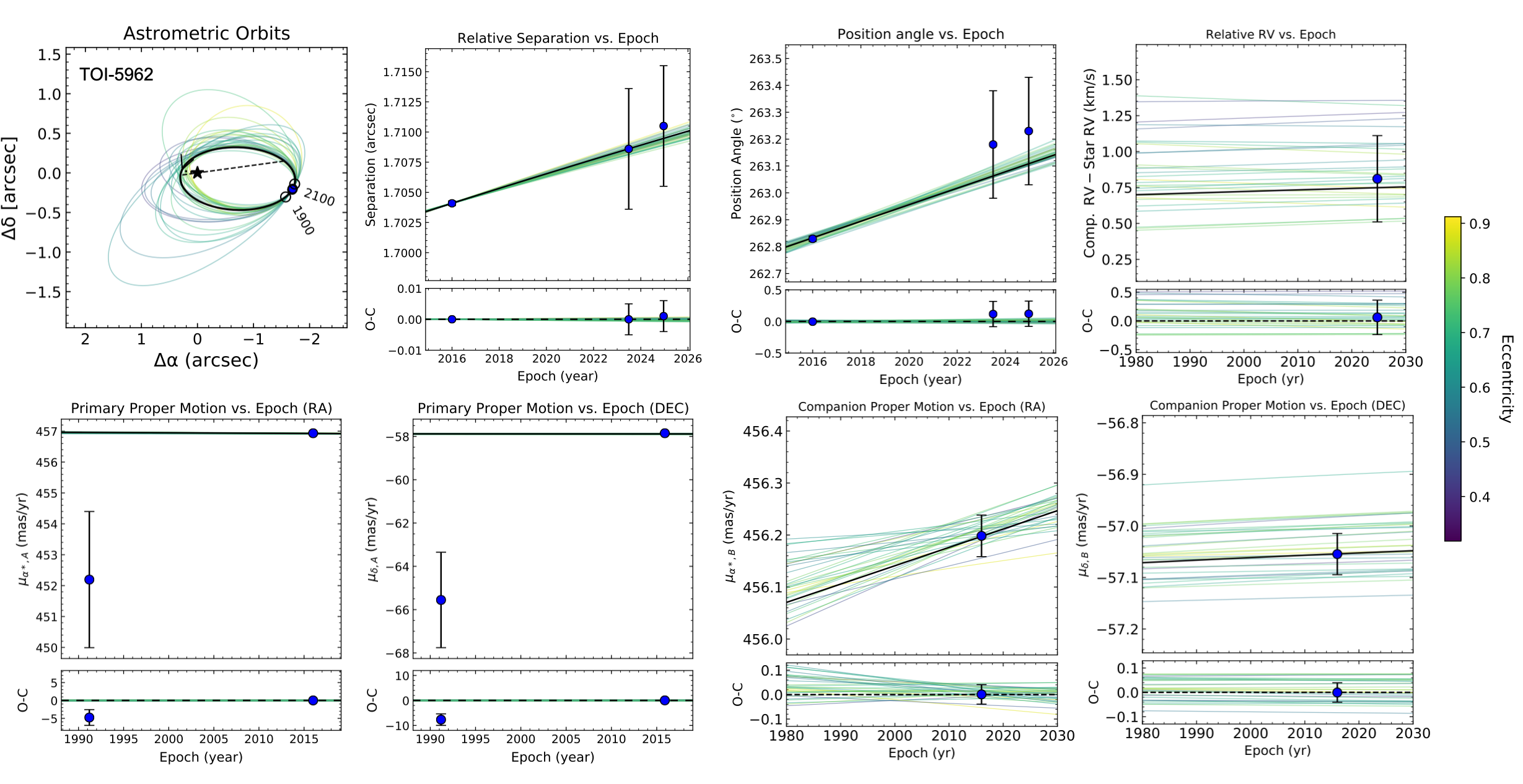}
    \caption{Orbital characterization of TOI-5962 B  around A using relative astrometry and absolute astrometry from Hipparcos and Gaia. Similar to figure~\ref{fig:figure402}}
    \label{fig:figure5962}
\end{figure*}
\begin{figure*}[!b]
    \centering
\includegraphics[width=0.62\linewidth]{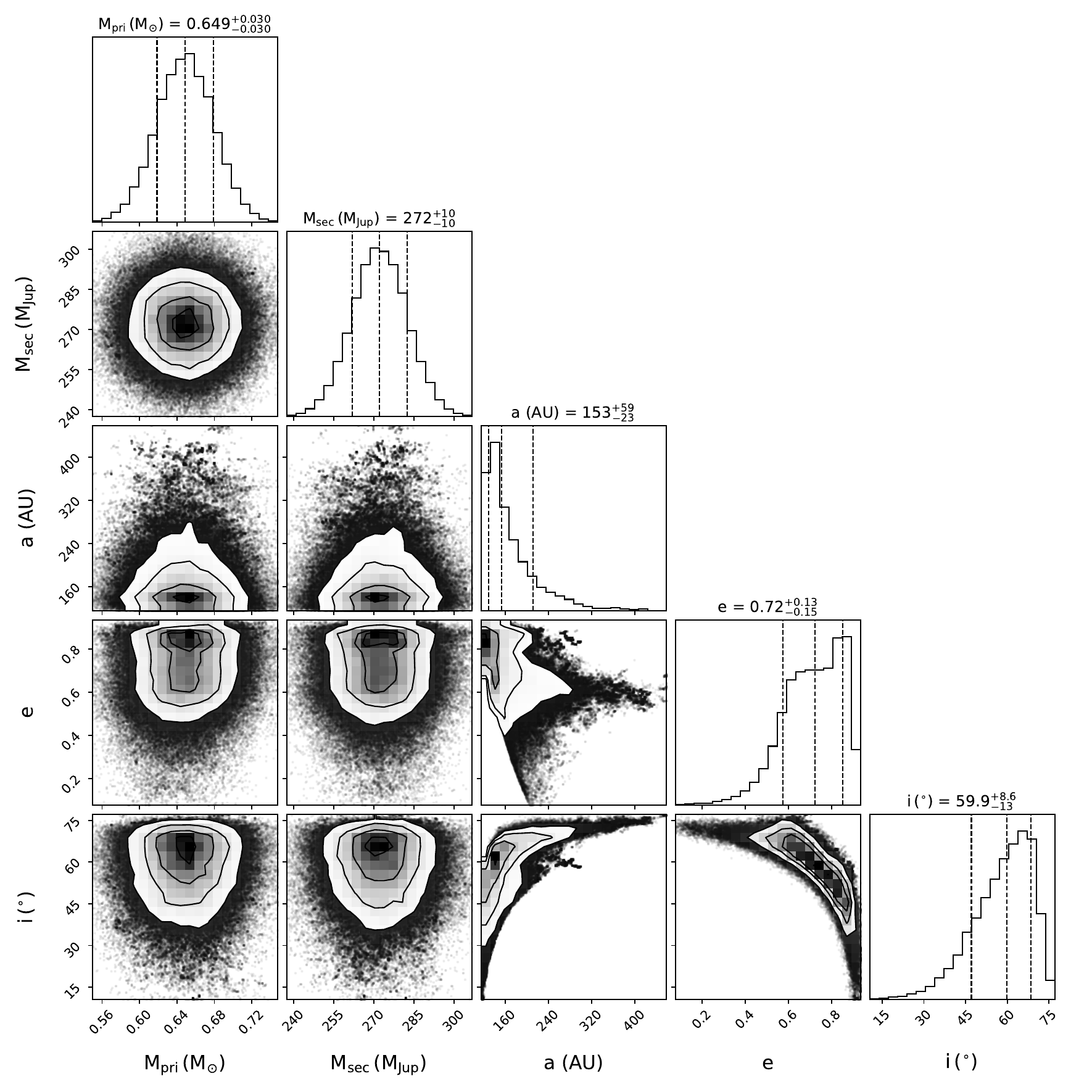}
    \caption{Joint posterior distributions for selected orbital parameters of TOI-5962 B. Same as Figure~\ref{fig:figure402_corner}}
    \label{fig:figure5962_corner}
\end{figure*}


\bibliography{sample63}{}

\begin{thebibliography}{}
\expandafter\ifx\csname natexlab\endcsname\relax\def\natexlab#1{#1}\fi
\providecommand{\url}[1]{\href{#1}{#1}}
\providecommand{\dodoi}[1]{doi:~\href{http://doi.org/#1}{\nolinkurl{#1}}}
\providecommand{\doeprint}[1]{\href{http://ascl.net/#1}{\nolinkurl{http://ascl.net/#1}}}
\providecommand{\doarXiv}[1]{\href{https://arxiv.org/abs/#1}{\nolinkurl{https://arxiv.org/abs/#1}}}

\bibitem[{ESA(1997)}]{ESA}
 1997, ESA Special Publication, Vol. 1200, {The HIPPARCOS and TYCHO catalogues. Astrometric and photometric star catalogues derived from the ESA HIPPARCOS Space Astrometry Mission}

\bibitem[{{Akeson} {et~al.}(2019){Akeson}, {Jensen}, {Carpenter}, {Ricci}, {Laos}, {Nogueira}, \& {Suen-Lewis}}]{Akeson2019}
{Akeson}, R.~L., {Jensen}, E. L.~N., {Carpenter}, J., {et~al.} 2019, \apj, 872, 158, \dodoi{10.3847/1538-4357/aaff6a}

\bibitem[{{Allard} {et~al.}(2012){Allard}, {Homeier}, \& {Freytag}}]{Allard2012}
{Allard}, F., {Homeier}, D., \& {Freytag}, B. 2012, Philosophical Transactions of the Royal Society of London Series A, 370, 2765, \dodoi{10.1098/rsta.2011.0269}

\bibitem[{{Artymowicz} \& {Lubow}(1994)}]{AandL1994}
{Artymowicz}, P., \& {Lubow}, S.~H. 1994, \apj, 421, 651, \dodoi{10.1086/173679}

\bibitem[{{Badenas-Agusti} {et~al.}(2020){Badenas-Agusti}, {G{\"u}nther}, {Daylan}, {Mikal-Evans}, {Vanderburg}, {Huang}, {Matthews}, {Rackham}, {Bieryla}, {Stassun}, {Kane}, {Shporer}, {Fulton}, {Hill}, {Nowak}, {Ribas}, {Pall{\'e}}, {Jenkins}, {Latham}, {Seager}, {Ricker}, {Vanderspek}, {Winn}, {Abril-Pla}, {Collins}, {Serra}, {Niraula}, {Rustamkulov}, {Barclay}, {Crossfield}, {Howell}, {Ciardi}, {Gonzales}, {Schlieder}, {Caldwell}, {Fausnaugh}, {McDermott}, {Paegert}, {Pepper}, {Rose}, \& {Twicken}}]{Badenas2020}
{Badenas-Agusti}, M., {G{\"u}nther}, M.~N., {Daylan}, T., {et~al.} 2020, \aj, 160, 113, \dodoi{10.3847/1538-3881/aba0b5}

\bibitem[{{Bakos} {et~al.}(2010){Bakos}, {Torres}, {P{\'a}l}, {Hartman}, {Kov{\'a}cs}, {Noyes}, {Latham}, {Sasselov}, {Sip{\H{o}}cz}, {Esquerdo}, {Fischer}, {Johnson}, {Marcy}, {Butler}, {Isaacson}, {Howard}, {Vogt}, {Kov{\'a}cs}, {Fernandez}, {Mo{\'o}r}, {Stefanik}, {L{\'a}z{\'a}r}, {Papp}, \& {S{\'a}ri}}]{Bakos2010}
{Bakos}, G.~{\'A}., {Torres}, G., {P{\'a}l}, A., {et~al.} 2010, \apj, 710, 1724, \dodoi{10.1088/0004-637X/710/2/1724}

\bibitem[{{Barros} {et~al.}(2023){Barros}, {Demangeon}, {Armstrong}, {Delgado Mena}, {Acu{\~n}a}, {Fern{\'a}ndez Fern{\'a}ndez}, {Deleuil}, {Collins}, {Howell}, {Ziegler}, {Adibekyan}, {Sousa}, {Stassun}, {Grieves}, {Lillo-Box}, {Hellier}, {Wheatley}, {Brice{\~n}o}, {Collins}, {Hawthorn}, {Hoyer}, {Jenkins}, {Law}, {Mann}, {Matson}, {Mousis}, {Nielsen}, {Osborn}, {Osborn}, {Paegert}, {Papini}, {Ricker}, {Rudat}, {Santos}, {Seager}, {Stockdale}, {Str{\o}m}, {Twicken}, {Udry}, {Wang}, {Vanderspek}, \& {Winn}}]{Barros2023}
{Barros}, S.~C.~C., {Demangeon}, O.~D.~S., {Armstrong}, D.~J., {et~al.} 2023, \aap, 673, A4, \dodoi{10.1051/0004-6361/202245741}

\bibitem[{{Behmard} {et~al.}(2022){Behmard}, {Dai}, \& {Howard}}]{Behmard2022}
{Behmard}, A., {Dai}, F., \& {Howard}, A.~W. 2022, \aj, 163, 160, \dodoi{10.3847/1538-3881/ac53a7}

\bibitem[{{Berger} {et~al.}(2020){Berger}, {Huber}, {van Saders}, {Gaidos}, {Tayar}, \& {Kraus}}]{berger2020}
{Berger}, T.~A., {Huber}, D., {van Saders}, J.~L., {et~al.} 2020, \aj, 159, 280, \dodoi{10.3847/1538-3881/159/6/280}

\bibitem[{{Brandt}(2021)}]{Brandt2021}
{Brandt}, T.~D. 2021, \apjs, 254, 42, \dodoi{10.3847/1538-4365/abf93c}

\bibitem[{{Brandt} {et~al.}(2019){Brandt}, {Dupuy}, \& {Bowler}}]{Brandt2019a}
{Brandt}, T.~D., {Dupuy}, T.~J., \& {Bowler}, B.~P. 2019, \aj, 158, 140, \dodoi{10.3847/1538-3881/ab04a8}

\bibitem[{{Brandt} {et~al.}(2021){Brandt}, {Dupuy}, {Li}, {Brandt}, {Zeng}, {Michalik}, {Bardalez Gagliuffi}, \& {Raposo-Pulido}}]{Brandt2021b}
{Brandt}, T.~D., {Dupuy}, T.~J., {Li}, Y., {et~al.} 2021, {orvara: Orbits from Radial Velocity, Absolute, and/or Relative Astrometry}.
\newblock \doeprint{2105.012}

\bibitem[{{Chauvin} {et~al.}(2023){Chauvin}, {Videla}, {Beust}, {Mendez}, {Correia}, {Lacour}, {Tokovinin}, {Hagelberg}, {Bouchy}, {Boisse}, {Villegas}, {Bonavita}, {Desidera}, {Faramaz}, {Forveille}, {Gallenne}, {Haubois}, {Jenkins}, {Kervella}, {Lagrange}, {Melo}, {Thebault}, {Udry}, \& {Segransan}}]{Chauvin2023}
{Chauvin}, G., {Videla}, M., {Beust}, H., {et~al.} 2023, \aap, 675, A114, \dodoi{10.1051/0004-6361/202244502}

\bibitem[{{Christiaens} {et~al.}(2023){Christiaens}, {Gonzalez}, {Farkas}, {Dahlqvist}, {Nasedkin}, {Milli}, {Absil}, {Ngo}, {Cantero}, {Rainot}, {Hammond}, {Bonse}, {Cantalloube}, {Vigan}, {Kompella}, \& {Hancock}}]{Christiaens2023}
{Christiaens}, V., {Gonzalez}, C., {Farkas}, R., {et~al.} 2023, The Journal of Open Source Software, 8, 4774, \dodoi{10.21105/joss.04774}

\bibitem[{{Christian} {et~al.}(2022){Christian}, {Vanderburg}, {Becker}, {Yahalomi}, {Pearce}, {Zhou}, {Collins}, {Kraus}, {Stassun}, {de Beurs}, {Ricker}, {Vanderspek}, {Latham}, {Winn}, {Seager}, {Jenkins}, {Abe}, {Agabi}, {Amado}, {Baker}, {Barkaoui}, {Benkhaldoun}, {Benni}, {Berberian}, {Berlind}, {Bieryla}, {Esparza-Borges}, {Bowen}, {Brown}, {Buchhave}, {Burke}, {Buttu}, {Cadieux}, {Caldwell}, {Charbonneau}, {Chazov}, {Chimaladinne}, {Collins}, {Combs}, {Conti}, {Crouzet}, {de Leon}, {Deljookorani}, {Diamond}, {Doyon}, {Dragomir}, {Dransfield}, {Essack}, {Evans}, {Fukui}, {Gan}, {Esquerdo}, {Gillon}, {Girardin}, {Guerra}, {Guillot}, {K. Habich}, {Henriksen}, {Hoch}, {Isogai}, {Jehin}, {Jensen}, {Johnson}, {Livingston}, {Kielkopf}, {Kim}, {Kawauchi}, {Krushinsky}, {Kunzle}, {Laloum}, {Leger}, {Lewin}, {Mallia}, {Massey}, {Mori}, {McLeod}, {M{\'e}karnia}, {Mireles}, {Mishevskiy}, {Tamura}, {Murgas}, {Narita}, {Naves}, {Nelson}, {Osborn}, {Palle}, {Parviainen}, {Plavchan}, {Pozuelos}, {Rabus}, {Relles},
  {Rodr{\'\i}guez L{\'o}pez}, {Quinn}, {Schmider}, {Schlieder}, {Schwarz}, {Shporer}, {Sibbald}, {Srdoc}, {Stibbards}, {Stickler}, {Suarez}, {Stockdale}, {Tan}, {Terada}, {Triaud}, {Tronsgaard}, {Waalkes}, {Wang}, {Watanabe}, {Wenceslas}, {Wingham}, {Wittrock}, \& {Ziegler}}]{Christian2022}
{Christian}, S., {Vanderburg}, A., {Becker}, J., {et~al.} 2022, \aj, 163, 207, \dodoi{10.3847/1538-3881/ac517f}

\bibitem[{{Correia} {et~al.}(2008){Correia}, {Udry}, {Mayor}, {Eggenberger}, {Naef}, {Beuzit}, {Perrier}, {Queloz}, {Sivan}, {Pepe}, {Santos}, \& {S{\'e}gransan}}]{Correia2008}
{Correia}, A.~C.~M., {Udry}, S., {Mayor}, M., {et~al.} 2008, \aap, 479, 271, \dodoi{10.1051/0004-6361:20078908}

\bibitem[{{Cox} {et~al.}(2017){Cox}, {Harris}, {Looney}, {Chiang}, {Chandler}, {Kratter}, {Li}, {Perez}, \& {Tobin}}]{Cox2017}
{Cox}, E.~G., {Harris}, R.~J., {Looney}, L.~W., {et~al.} 2017, \apj, 851, 83, \dodoi{10.3847/1538-4357/aa97e2}

\bibitem[{{Cutri} {et~al.}(2003){Cutri}, {Skrutskie}, {van Dyk}, {Beichman}, {Carpenter}, {Chester}, {Cambresy}, {Evans}, {Fowler}, {Gizis}, {Howard}, {Huchra}, {Jarrett}, {Kopan}, {Kirkpatrick}, {Light}, {Marsh}, {McCallon}, {Schneider}, {Stiening}, {Sykes}, {Weinberg}, {Wheaton}, {Wheelock}, \& {Zacarias}}]{Cutri2003}
{Cutri}, R.~M., {Skrutskie}, M.~F., {van Dyk}, S., {et~al.} 2003, VizieR Online Data Catalog, II/246

\bibitem[{{da Silva} {et~al.}(2006){da Silva}, {Udry}, {Bouchy}, {Mayor}, {Moutou}, {Pont}, {Queloz}, {Santos}, {S{\'e}gransan}, \& {Zucker}}]{daSilva2006}
{da Silva}, R., {Udry}, S., {Bouchy}, F., {et~al.} 2006, \aap, 446, 717, \dodoi{10.1051/0004-6361:20054116}

\bibitem[{{Damasso} {et~al.}(2020){Damasso}, {Sozzetti}, {Lovis}, {Barros}, {Sousa}, {Demangeon}, {Faria}, {Lillo-Box}, {Cristiani}, {Pepe}, {Rebolo}, {Santos}, {Zapatero Osorio}, {Gonz{\'a}lez Hern{\'a}ndez}, {Amate}, {Pasquini}, {Zerbi}, {Adibekyan}, {Abreu}, {Affolter}, {Alibert}, {Aliverti}, {Allart}, {Allende Prieto}, {{\'A}lvarez}, {Alves}, {Avila}, {Baldini}, {Bandy}, {Benz}, {Bianco}, {Borsa}, {Bossini}, {Bourrier}, {Bouchy}, {Broeg}, {Cabral}, {Calderone}, {Cirami}, {Coelho}, {Conconi}, {Coretti}, {Cumani}, {Cupani}, {D'Odorico}, {Deiries}, {Dekker}, {Delabre}, {Di Marcantonio}, {Dumusque}, {Ehrenreich}, {Figueira}, {Fragoso}, {Genolet}, {Genoni}, {G{\'e}nova Santos}, {Hughes}, {Iwert}, {Kerber}, {Knudstrup}, {Landoni}, {Lavie}, {Lizon}, {Lo Curto}, {Maire}, {Martins}, {M{\'e}gevand}, {Mehner}, {Micela}, {Modigliani}, {Molaro}, {Monteiro}, {Monteiro}, {Moschetti}, {Mueller}, {Murphy}, {Nunes}, {Oggioni}, {Oliveira}, {Oshagh}, {Pall{\'e}}, {Pariani}, {Poretti}, {Rasilla}, {Rebord{\~a}o}, {Redaelli},
  {Riva}, {Santana Tschudi}, {Santin}, {Santos}, {S{\'e}gransan}, {Schmidt}, {Segovia}, {Sosnowska}, {Span{\`o}}, {Su{\'a}rez Mascare{\~n}o}, {Tabernero}, {Tenegi}, {Udry}, \& {Zanutta}}]{Damasso2020}
{Damasso}, M., {Sozzetti}, A., {Lovis}, C., {et~al.} 2020, \aap, 642, A31, \dodoi{10.1051/0004-6361/202038416}

\bibitem[{{De Rosa} {et~al.}(2020){De Rosa}, {Dawson}, \& {Nielsen}}]{DR2020}
{De Rosa}, R.~J., {Dawson}, R., \& {Nielsen}, E.~L. 2020, \aap, 640, A73, \dodoi{10.1051/0004-6361/202038496}

\bibitem[{Delorme {et~al.}(2021)Delorme, Jovanovic, Echeverri, Mawet, Kent~Wallace, Bartos, Cetre, Wizinowich, Ragland, Lilley, Wetherell, Doppmann, Wang, Morris, Ruffio, Martin, Fitzgerald, Ruane, Schofield, Suominen, Calvin, Wang, Magnone, Johnson, Sohn, L{\'o}pez, Bond, Pezzato, Sayson, Chun, \& Skemer}]{delorme_Keck_2021a}
Delorme, J.-R., Jovanovic, N., Echeverri, D., {et~al.} 2021, JATIS, 7, 035006, \dodoi{10.1117/1.JATIS.7.3.035006}

\bibitem[{{Desidera} {et~al.}(2023){Desidera}, {Damasso}, {Gratton}, {Benatti}, {Nardiello}, {D'Orazi}, {Lanza}, {Locci}, {Marzari}, {Mesa}, {Messina}, {Pillitteri}, {Sozzetti}, {Girard}, {Maggio}, {Micela}, {Malavolta}, {Nascimbeni}, {Pinamonti}, {Squicciarini}, {Alcal{\'a}}, {Biazzo}, {Bohn}, {Bonavita}, {Brooks}, {Chauvin}, {Covino}, {Delorme}, {Hagelberg}, {Janson}, {Lagrange}, \& {Lazzoni}}]{Desidera2023}
{Desidera}, S., {Damasso}, M., {Gratton}, R., {et~al.} 2023, \aap, 675, A158, \dodoi{10.1051/0004-6361/202244611}

\bibitem[{{Dieterich} {et~al.}(2012){Dieterich}, {Henry}, {Golimowski}, {Krist}, \& {Tanner}}]{Dieterich2012}
{Dieterich}, S.~B., {Henry}, T.~J., {Golimowski}, D.~A., {Krist}, J.~E., \& {Tanner}, A.~M. 2012, \aj, 144, 64, \dodoi{10.1088/0004-6256/144/2/64}

\bibitem[{{Do {\'O}} {et~al.}(2024){Do {\'O}}, {Sappey}, {Konopacky}, {Ruffio}, {O'Neil}, {Do}, {Martinez}, {Barman}, {Nguyen}, {Xuan}, {Theissen}, {Blunt}, {Thompson}, {Hsu}, {Baker}, {Bartos}, {Blake}, {Calvin}, {Cetre}, {Delorme}, {Doppmann}, {Echeverri}, {Finnerty}, {Fitzgerald}, {Inglis}, {Jovanovic}, {L{\'o}pez}, {Mawet}, {Morris}, {Pezzato}, {Schofield}, {Skemer}, {Wallace}, {Wang}, {Wang}, \& {Liberman}}]{DoO2024}
{Do {\'O}}, C.~R., {Sappey}, B., {Konopacky}, Q.~M., {et~al.} 2024, \aj, 167, 278, \dodoi{10.3847/1538-3881/ad3df3}

\bibitem[{{Dumusque} {et~al.}(2019){Dumusque}, {Turner}, {Dorn}, {Eastman}, {Allart}, {Adibekyan}, {Sousa}, {Santos}, {Mordasini}, {Bourrier}, {Bouchy}, {Coffinet}, {Davies}, {D{\'\i}az}, {Fausnaugh}, {Glidden}, {Guerrero}, {Henze}, {Jenkins}, {Latham}, {Lovis}, {Mayor}, {Pepe}, {Quintana}, {Ricker}, {Rowden}, {Segransan}, {Su{\'a}rez Mascare{\~n}o}, {Seager}, {Twicken}, {Udry}, {Vanderspek}, \& {Winn}}]{Dumusque2019}
{Dumusque}, X., {Turner}, O., {Dorn}, C., {et~al.} 2019, \aap, 627, A43, \dodoi{10.1051/0004-6361/201935457}

\bibitem[{{Dupuy} {et~al.}(2016){Dupuy}, {Kratter}, {Kraus}, {Isaacson}, {Mann}, {Ireland}, {Howard}, \& {Huber}}]{Dupuy2016}
{Dupuy}, T.~J., {Kratter}, K.~M., {Kraus}, A.~L., {et~al.} 2016, \apj, 817, 80, \dodoi{10.3847/0004-637X/817/1/80}

\bibitem[{{Dupuy} {et~al.}(2022){Dupuy}, {Kraus}, {Kratter}, {Rizzuto}, {Mann}, {Huber}, \& {Ireland}}]{Dupuy2022}
{Dupuy}, T.~J., {Kraus}, A.~L., {Kratter}, K.~M., {et~al.} 2022, \mnras, 512, 648, \dodoi{10.1093/mnras/stac306}

\bibitem[{{Echeverri} {et~al.}(2022){Echeverri}, {Jovanovic}, {Delorme}, {Xin}, {Schofield}, {Finnerty}, {Wang}, {Xuan}, {Mawet}, {Baker}, {Bartos}, {Bond}, {Bryan}, {Calvin}, {Cetre}, {Doppmann}, {Fitzgerald}, {Fucik}, {Horstman}, {Lopez}, {Martin}, {Martin}, {Mennesson}, {Morris}, {Nash}, {Pezzato}, {Porter}, {Ragland}, {Roberts}, {Ruane}, {Ruffio}, {Sappey}, {Serabyn}, {Skemer}, {Venenciano}, {Wallace}, {Wang}, \& {Wizinowich}}]{Echeverri2022}
{Echeverri}, D., {Jovanovic}, N., {Delorme}, J.-R., {et~al.} 2022, in Society of Photo-Optical Instrumentation Engineers (SPIE) Conference Series, Vol. 12184, Ground-based and Airborne Instrumentation for Astronomy IX, ed. C.~J. {Evans}, J.~J. {Bryant}, \& K.~{Motohara}, 121841W, \dodoi{10.1117/12.2630518}

\bibitem[{{ESA}(1997)}]{ESA1997}
{ESA}, ed. 1997, ESA Special Publication, Vol. 1200, {The HIPPARCOS and TYCHO catalogues. Astrometric and photometric star catalogues derived from the ESA HIPPARCOS Space Astrometry Mission}

\bibitem[{{Espinoza} {et~al.}(2020){Espinoza}, {Brahm}, {Henning}, {Jord{\'a}n}, {Dorn}, {Rojas}, {Sarkis}, {Kossakowski}, {Schlecker}, {D{\'\i}az}, {Jenkins}, {Aguilera-Gomez}, {Jenkins}, {Twicken}, {Collins}, {Lissauer}, {Armstrong}, {Adibekyan}, {Barrado}, {Barros}, {Battley}, {Bayliss}, {Bouchy}, {Bryant}, {Cooke}, {Demangeon}, {Dumusque}, {Figueira}, {Giles}, {Lillo-Box}, {Lovis}, {Nielsen}, {Pepe}, {Pollacco}, {Santos}, {Sousa}, {Udry}, {Wheatley}, {Turner}, {Marmier}, {S{\'e}gransan}, {Ricker}, {Latham}, {Seager}, {Winn}, {Kielkopf}, {Hart}, {Wingham}, {Jensen}, {He{\l}miniak}, {Tokovinin}, {Brice{\~n}o}, {Ziegler}, {Law}, {Mann}, {Daylan}, {Doty}, {Guerrero}, {Boyd}, {Crossfield}, {Morris}, {Henze}, \& {Chacon}}]{Espinoza2020}
{Espinoza}, N., {Brahm}, R., {Henning}, T., {et~al.} 2020, \mnras, 491, 2982, \dodoi{10.1093/mnras/stz3150}

\bibitem[{{Fabrycky} \& {Winn}(2009)}]{Fabrycky2009}
{Fabrycky}, D.~C., \& {Winn}, J.~N. 2009, \apj, 696, 1230, \dodoi{10.1088/0004-637X/696/2/1230}

\bibitem[{{Ferrer-Ch{\'a}vez} {et~al.}(2021){Ferrer-Ch{\'a}vez}, {Blunt}, \& {Wang}}]{Rodrigo2021}
{Ferrer-Ch{\'a}vez}, R., {Blunt}, S., \& {Wang}, J.~J. 2021, Research Notes of the American Astronomical Society, 5, 162, \dodoi{10.3847/2515-5172/ac151d}

\bibitem[{{Franchini} {et~al.}(2020){Franchini}, {Martin}, \& {Lubow}}]{Franchini2020}
{Franchini}, A., {Martin}, R.~G., \& {Lubow}, S.~H. 2020, \mnras, 491, 5351, \dodoi{10.1093/mnras/stz3175}

\bibitem[{{Gaia Collaboration} {et~al.}(2022){Gaia Collaboration}, {Klioner}, {Lindegren}, {Mignard}, {Hern{\'a}ndez}, {Ramos-Lerate}, {Bastian}, {Biermann}, {Bombrun}, {de Torres}, {Gerlach}, {Geyer}, {Hilger}, {Hobbs}, {Lammers}, {McMillan}, {Steidelm{\"u}ller}, {Teyssier}, {Raiteri}, {Bartolom{\'e}}, {Bernet}, {Casta{\~n}eda}, {Clotet}, {Davidson}, {Fabricius}, {Garralda Torres}, {Gonz{\'a}lez-Vidal}, {Portell}, {Rowell}, {Torra}, {Torra}, {Brown}, {Vallenari}, {Prusti}, {de Bruijne}, {Arenou}, {Babusiaux}, {Creevey}, {Ducourant}, {Evans}, {Eyer}, {Guerra}, {Hutton}, {Jordi}, {Luri}, {Panem}, {Pourbaix}, {Randich}, {Sartoretti}, {Soubiran}, {Tanga}, {Walton}, {Bailer-Jones}, {Drimmel}, {Jansen}, {Katz}, {Lattanzi}, {van Leeuwen}, {Bakker}, {Cacciari}, {De Angeli}, {Fouesneau}, {Fr{\'e}mat}, {Galluccio}, {Guerrier}, {Heiter}, {Masana}, {Messineo}, {Mowlavi}, {Nicolas}, {Nienartowicz}, {Pailler}, {Panuzzo}, {Riclet}, {Roux}, {Seabroke}, {Sordo}, {Th{\'e}venin}, {Gracia-Abril}, {Altmann}, {Andrae}, {Audard},
  {Bellas-Velidis}, {Benson}, {Berthier}, {Blomme}, {Burgess}, {Busonero}, {Busso}, {C{\'a}novas}, {Carry}, {Cellino}, {Cheek}, {Clementini}, {Damerdji}, {de Teodoro}, {Nu{\~n}ez Campos}, {Delchambre}, {Dell'Oro}, {Esquej}, {Fern{\'a}ndez-Hern{\'a}ndez}, {Fraile}, {Garabato}, {Garc{\'\i}a-Lario}, {Gosset}, {Haigron}, {Halbwachs}, {Hambly}, {Harrison}, {Hestroffer}, {Hodgkin}, {Holl}, {Jan{\ss}en}, {Jevardat de Fombelle}, {Jordan}, {Krone-Martins}, {Lanzafame}, {L{\"o}ffler}, {Marchal}, {Marrese}, {Moitinho}, {Muinonen}, {Osborne}, {Pancino}, {Pauwels}, {Recio-Blanco}, {Reyl{\'e}}, {Riello}, {Rimoldini}, {Roegiers}, {Rybizki}, {Sarro}, {Siopis}, {Smith}, {Sozzetti}, {Utrilla}, {van Leeuwen}, {Abbas}, {{\'A}brah{\'a}m}, {Abreu Aramburu}, {Aerts}, {Aguado}, {Ajaj}, {Aldea-Montero}, {Altavilla}, {{\'A}lvarez}, {Alves}, {Anderson}, {Anglada Varela}, {Antoja}, {Baines}, {Baker}, {Balaguer-N{\'u}{\~n}ez}, {Balbinot}, {Balog}, {Barache}, {Barbato}, {Barros}, {Barstow}, {Bassilana}, {Bauchet}, {Becciani},
  {Bellazzini}, {Berihuete}, {Bertone}, {Bianchi}, {Binnenfeld}, {Blanco-Cuaresma}, {Boch}, {Bossini}, {Bouquillon}, {Bragaglia}, {Bramante}, {Breedt}, {Bressan}, {Brouillet}, {Brugaletta}, {Bucciarelli}, {Burlacu}, {Butkevich}, {Buzzi}, {Caffau}, {Cancelliere}, {Cantat-Gaudin}, {Carballo}, {Carlucci}, {Carnerero}, {Carrasco}, {Casamiquela}, {Castellani}, {Castro-Ginard}, {Chaoul}, {Charlot}, {Chemin}, {Chiaramida}, {Chiavassa}, {Chornay}, {Comoretto}, {Contursi}, {Cooper}, {Cornez}, {Cowell}, {Crifo}, {Cropper}, {Crosta}, {Crowley}, {Dafonte}, {Dapergolas}, {David}, {de Laverny}, {De Luise}, {De March}, {De Ridder}, {de Souza}, {del Peloso}, {del Pozo}, {Delbo}, {Delgado}, {Delisle}, {Demouchy}, {Dharmawardena}, {Diakite}, {Diener}, {Distefano}, {Dolding}, {Enke}, {Fabre}, {Fabrizio}, {Faigler}, {Fedorets}, {Fernique}, {Fienga}, {Figueras}, {Fournier}, {Fouron}, {Fragkoudi}, {Gai}, {Garcia-Gutierrez}, {Garcia-Reinaldos}, {Garc{\'\i}a-Torres}, {Garofalo}, {Gavel}, {Gavras}, {Giacobbe}, {Gilmore}, {Girona},
  {Giuffrida}, {Gomel}, {Gomez}, {Gonz{\'a}lez-N{\'u}{\~n}ez}, {Gonz{\'a}lez-Santamar{\'\i}a}, {Granvik}, {Guillout}, {Guiraud}, {Guti{\'e}rrez-S{\'a}nchez}, {Guy}, {Hatzidimitriou}, {Hauser}, {Haywood}, {Helmer}, {Helmi}, {Sarmiento}, {Hidalgo}, {H{\l}adczuk}, {Holland}, {Huckle}, {Jardine}, {Jasniewicz}, {Jean-Antoine Piccolo}, {Jim{\'e}nez-Arranz}, {Juaristi Campillo}, {Julbe}, {Karbevska}, {Kervella}, {Khanna}, {Kordopatis}, {Korn}, {K{\'o}sp{\'a}l}, {Kostrzewa-Rutkowska}, {Kruszy{\'n}ska}, {Kun}, {Laizeau}, {Lambert}, {Lanza}, {Lasne}, {Le Campion}, {Lebreton}, {Lebzelter}, {Leccia}, {Leclerc}, {Lecoeur-Taibi}, {Liao}, {Licata}, {Lindstr{\o}m}, {Lister}, {Livanou}, {Lobel}, {Lorca}, {Loup}, {Madrero Pardo}, {Magdaleno Romeo}, {Managau}, {Mann}, {Manteiga}, {Marchant}, {Marconi}, {Marcos}, {Marcos Santos}, {Mar{\'\i}n Pina}, {Marinoni}, {Marocco}, {Marshall}, {Polo}, {Mart{\'\i}n-Fleitas}, {Marton}, {Mary}, {Masip}, {Massari}, {Mastrobuono-Battisti}, {Mazeh}, {Messina}, {Michalik}, {Millar}, {Mints},
  {Molina}, {Molinaro}, {Moln{\'a}r}, {Monari}, {Mongui{\'o}}, {Montegriffo}, {Montero}, {Mor}, {Mora}, {Morbidelli}, {Morel}, {Morris}, {Muraveva}, {Murphy}, {Musella}, {Nagy}, {Noval}, {Oca{\~n}a}, {Ogden}, {Ordenovic}, {Osinde}, {Pagani}, {Pagano}, {Palaversa}, {Palicio}, {Pallas-Quintela}, {Panahi}, {Payne-Wardenaar}, {Pe{\~n}alosa Esteller}, {Penttil{\"a}}, {Pichon}, {Piersimoni}, {Pineau}, {Plachy}, {Plum}, {Poggio}, {Pr{\v{s}}a}, {Pulone}, {Racero}, {Ragaini}, {Rainer}, {Rambaux}, {Ramos}, {Re Fiorentin}, {Regibo}, {Richards}, {Rios Diaz}, {Ripepi}, {Riva}, {Rix}, {Rixon}, {Robichon}, {Robin}, {Robin}, {Roelens}, {Rogues}, {Rohrbasser}, {Romero-G{\'o}mez}, {Royer}, {Ruz Mieres}, {Rybicki}, {Sadowski}, {S{\'a}ez N{\'u}{\~n}ez}, {Sagrist{\`a} Sell{\'e}s}, {Sahlmann}, {Salguero}, {Samaras}, {Sanchez Gimenez}, {Sanna}, {Santove{\~n}a}, {Sarasso}, {Schultheis}, {Sciacca}, {Segol}, {Segovia}, {S{\'e}gransan}, {Semeux}, {Shahaf}, {Siddiqui}, {Siebert}, {Siltala}, {Silvelo}, {Slezak}, {Slezak}, {Smart},
  {Snaith}, {Solano}, {Solitro}, {Souami}, {Souchay}, {Spagna}, {Spina}, {Spoto}, {Steele}, {Stephenson}, {S{\"u}veges}, {Surdej}, {Szabados}, {Szegedi-Elek}, {Taris}, {Taylor}, {Teixeira}, {Tolomei}, {Tonello}, {Torralba Elipe}, {Trabucchi}, {Tsounis}, {Turon}, {Ulla}, {Unger}, {Vaillant}, {van Dillen}, {van Reeven}, {Vanel}, {Vecchiato}, {Viala}, {Vicente}, {Voutsinas}, {Weiler}, {Wevers}, {Wyrzykowski}, {Yoldas}, {Yvard}, {Zhao}, {Zorec}, {Zucker}, \& {Zwitter}}]{Gaia}
{Gaia Collaboration}, {Klioner}, S.~A., {Lindegren}, L., {et~al.} 2022, arXiv e-prints, arXiv:2204.12574.
\newblock \doarXiv{2204.12574}

\bibitem[{{Gaia Collaboration} {et~al.}(2023){Gaia Collaboration}, {Arenou}, {Babusiaux}, {Barstow}, {Faigler}, {Jorissen}, {Kervella}, {Mazeh}, {Mowlavi}, {Panuzzo}, {Sahlmann}, {Shahaf}, {Sozzetti}, {Bauchet}, {Damerdji}, {Gavras}, {Giacobbe}, {Gosset}, {Halbwachs}, {Holl}, {Lattanzi}, {Leclerc}, {Morel}, {Pourbaix}, {Re Fiorentin}, {Sadowski}, {S{\'e}gransan}, {Siopis}, {Teyssier}, {Zwitter}, {Planquart}, {Brown}, {Vallenari}, {Prusti}, {de Bruijne}, {Biermann}, {Creevey}, {Ducourant}, {Evans}, {Eyer}, {Guerra}, {Hutton}, {Jordi}, {Klioner}, {Lammers}, {Lindegren}, {Luri}, {Mignard}, {Panem}, {Randich}, {Sartoretti}, {Soubiran}, {Tanga}, {Walton}, {Bailer-Jones}, {Bastian}, {Drimmel}, {Jansen}, {Katz}, {van Leeuwen}, {Bakker}, {Cacciari}, {Casta{\~n}eda}, {De Angeli}, {Fabricius}, {Fouesneau}, {Fr{\'e}mat}, {Galluccio}, {Guerrier}, {Heiter}, {Masana}, {Messineo}, {Nicolas}, {Nienartowicz}, {Pailler}, {Riclet}, {Roux}, {Seabroke}, {Sordo}, {Th{\'e}venin}, {Gracia-Abril}, {Portell}, {Altmann}, {Andrae},
  {Audard}, {Bellas-Velidis}, {Benson}, {Berthier}, {Blomme}, {Burgess}, {Busonero}, {Busso}, {C{\'a}novas}, {Carry}, {Cellino}, {Cheek}, {Clementini}, {Davidson}, {de Teodoro}, {Nu{\~n}ez Campos}, {Delchambre}, {Dell'Oro}, {Esquej}, {Fern{\'a}ndez-Hern{\'a}ndez}, {Fraile}, {Garabato}, {Garc{\'\i}a-Lario}, {Haigron}, {Hambly}, {Harrison}, {Hern{\'a}ndez}, {Hestroffer}, {Hodgkin}, {Jan{\ss}en}, {Jevardat de Fombelle}, {Jordan}, {Krone-Martins}, {Lanzafame}, {L{\"o}ffler}, {Marchal}, {Marrese}, {Moitinho}, {Muinonen}, {Osborne}, {Pancino}, {Pauwels}, {Recio-Blanco}, {Reyl{\'e}}, {Riello}, {Rimoldini}, {Roegiers}, {Rybizki}, {Sarro}, {Smith}, {Utrilla}, {van Leeuwen}, {Abbas}, {{\'A}brah{\'a}m}, {Abreu Aramburu}, {Aerts}, {Aguado}, {Ajaj}, {Aldea-Montero}, {Altavilla}, {{\'A}lvarez}, {Alves}, {Anders}, {Anderson}, {Anglada Varela}, {Antoja}, {Baines}, {Baker}, {Balaguer-N{\'u}{\~n}ez}, {Balbinot}, {Balog}, {Barache}, {Barbato}, {Barros}, {Bartolom{\'e}}, {Bassilana}, {Becciani}, {Bellazzini}, {Berihuete},
  {Bernet}, {Bertone}, {Bianchi}, {Binnenfeld}, {Blanco-Cuaresma}, {Blazere}, {Boch}, {Bombrun}, {Bossini}, {Bouquillon}, {Bragaglia}, {Bramante}, {Breedt}, {Bressan}, {Brouillet}, {Brugaletta}, {Bucciarelli}, {Burlacu}, {Butkevich}, {Buzzi}, {Caffau}, {Cancelliere}, {Cantat-Gaudin}, {Carballo}, {Carlucci}, {Carnerero}, {Carrasco}, {Casamiquela}, {Castellani}, {Castro-Ginard}, {Chaoul}, {Charlot}, {Chemin}, {Chiaramida}, {Chiavassa}, {Chornay}, \& {Comoretto}}]{NSS2023}
{Gaia Collaboration}, {Arenou}, F., {Babusiaux}, C., {et~al.} 2023, \aap, 674, A34, \dodoi{10.1051/0004-6361/202243782}

\bibitem[{{Gandolfi} {et~al.}(2018){Gandolfi}, {Barrag{\'a}n}, {Livingston}, {Fridlund}, {Justesen}, {Redfield}, {Fossati}, {Mathur}, {Grziwa}, {Cabrera}, {Garc{\'\i}a}, {Persson}, {Van Eylen}, {Hatzes}, {Hidalgo}, {Albrecht}, {Bugnet}, {Cochran}, {Csizmadia}, {Deeg}, {Eigm{\"u}ller}, {Endl}, {Erikson}, {Esposito}, {Guenther}, {Korth}, {Luque}, {Monta{\~n}es Rodr{\'\i}guez}, {Nespral}, {Nowak}, {P{\"a}tzold}, \& {Prieto-Arranz}}]{Gandolfi2018}
{Gandolfi}, D., {Barrag{\'a}n}, O., {Livingston}, J.~H., {et~al.} 2018, \aap, 619, L10, \dodoi{10.1051/0004-6361/201834289}

\bibitem[{{Gandolfi} {et~al.}(2019){Gandolfi}, {Fossati}, {Livingston}, {Stassun}, {Grziwa}, {Barrag{\'a}n}, {Fridlund}, {Kubyshkina}, {Persson}, {Dai}, {Lam}, {Albrecht}, {Batalha}, {Beck}, {Justesen}, {Cabrera}, {Cartwright}, {Cochran}, {Csizmadia}, {Davies}, {Deeg}, {Eigm{\"u}ller}, {Endl}, {Erikson}, {Esposito}, {Garc{\'\i}a}, {Goeke}, {Gonz{\'a}lez-Cuesta}, {Guenther}, {Hatzes}, {Hidalgo}, {Hirano}, {Hjorth}, {Kabath}, {Knudstrup}, {Korth}, {Li}, {Luque}, {Mathur}, {Monta{\~n}es Rodr{\'\i}guez}, {Narita}, {Nespral}, {Niraula}, {Nowak}, {Palle}, {P{\"a}tzold}, {Prieto-Arranz}, {Rauer}, {Redfield}, {Ribas}, {Skarka}, {Smith}, {Rowden}, {Torres}, {Van Eylen}, \& {Vezie}}]{Gandolfi2019}
{Gandolfi}, D., {Fossati}, L., {Livingston}, J.~H., {et~al.} 2019, \apjl, 876, L24, \dodoi{10.3847/2041-8213/ab17d9}

\bibitem[{{Gerbig} {et~al.}(2024){Gerbig}, {Rice}, {Zanazzi}, {Christian}, \& {Vanderburg}}]{Gerbig2024}
{Gerbig}, K., {Rice}, M., {Zanazzi}, J.~J., {Christian}, S., \& {Vanderburg}, A. 2024, \apj, 972, 161, \dodoi{10.3847/1538-4357/ad5f2b}

\bibitem[{{Gomez Gonzalez} {et~al.}(2017){Gomez Gonzalez}, {Wertz}, {Absil}, {Christiaens}, {Defr{\`e}re}, {Mawet}, {Milli}, {Absil}, {Van Droogenbroeck}, {Cantalloube}, {Hinz}, {Skemer}, {Karlsson}, \& {Surdej}}]{Gomez2017}
{Gomez Gonzalez}, C.~A., {Wertz}, O., {Absil}, O., {et~al.} 2017, \aj, 154, 7, \dodoi{10.3847/1538-3881/aa73d7}

\bibitem[{{Guerrero} {et~al.}(2021){Guerrero}, {Seager}, {Huang}, {Vanderburg}, {Garcia Soto}, {Mireles}, {Hesse}, {Fong}, {Glidden}, {Shporer}, {Latham}, {Collins}, {Quinn}, {Burt}, {Dragomir}, {Crossfield}, {Vanderspek}, {Fausnaugh}, {Burke}, {Ricker}, {Daylan}, {Essack}, {G{\"u}nther}, {Osborn}, {Pepper}, {Rowden}, {Sha}, {Villanueva}, {Yahalomi}, {Yu}, {Ballard}, {Batalha}, {Berardo}, {Chontos}, {Dittmann}, {Esquerdo}, {Mikal-Evans}, {Jayaraman}, {Krishnamurthy}, {Louie}, {Mehrle}, {Niraula}, {Rackham}, {Rodriguez}, {Rowden}, {Sousa-Silva}, {Watanabe}, {Wong}, {Zhan}, {Zivanovic}, {Christiansen}, {Ciardi}, {Swain}, {Lund}, {Mullally}, {Fleming}, {Rodriguez}, {Boyd}, {Quintana}, {Barclay}, {Col{\'o}n}, {Rinehart}, {Schlieder}, {Clampin}, {Jenkins}, {Twicken}, {Caldwell}, {Coughlin}, {Henze}, {Lissauer}, {Morris}, {Rose}, {Smith}, {Tenenbaum}, {Ting}, {Wohler}, {Bakos}, {Bean}, {Berta-Thompson}, {Bieryla}, {Bouma}, {Buchhave}, {Butler}, {Charbonneau}, {Doty}, {Ge}, {Holman}, {Howard}, {Kaltenegger}, {Kane},
  {Kjeldsen}, {Kreidberg}, {Lin}, {Minsky}, {Narita}, {Paegert}, {P{\'a}l}, {Palle}, {Sasselov}, {Spencer}, {Sozzetti}, {Stassun}, {Torres}, {Udry}, \& {Winn}}]{TOI_ref}
{Guerrero}, N.~M., {Seager}, S., {Huang}, C.~X., {et~al.} 2021, \apjs, 254, 39, \dodoi{10.3847/1538-4365/abefe1}

\bibitem[{{Halbwachs} {et~al.}(2023){Halbwachs}, {Pourbaix}, {Arenou}, {Galluccio}, {Guillout}, {Bauchet}, {Marchal}, {Sadowski}, \& {Teyssier}}]{Halbwachs2023}
{Halbwachs}, J.-L., {Pourbaix}, D., {Arenou}, F., {et~al.} 2023, Astronomy and Astrophysics, 674, A9, \dodoi{10.1051/0004-6361/202243969}

\bibitem[{{Hatzes} {et~al.}(2003){Hatzes}, {Cochran}, {Endl}, {McArthur}, {Paulson}, {Walker}, {Campbell}, \& {Yang}}]{Hatzes2003}
{Hatzes}, A.~P., {Cochran}, W.~D., {Endl}, M., {et~al.} 2003, \apj, 599, 1383, \dodoi{10.1086/379281}

\bibitem[{{Hirsch} {et~al.}(2021){Hirsch}, {Rosenthal}, {Fulton}, {Howard}, {Ciardi}, {Marcy}, {Nielsen}, {Petigura}, {de Rosa}, {Isaacson}, {Weiss}, {Sinukoff}, \& {Macintosh}}]{Hirsch2021}
{Hirsch}, L.~A., {Rosenthal}, L., {Fulton}, B.~J., {et~al.} 2021, \aj, 161, 134, \dodoi{10.3847/1538-3881/abd639}

\bibitem[{{Hobson} {et~al.}(2021){Hobson}, {Brahm}, {Jord{\'a}n}, {Espinoza}, {Kossakowski}, {Henning}, {Rojas}, {Schlecker}, {Sarkis}, {Trifonov}, {Thorngren}, {Binnenfeld}, {Shahaf}, {Zucker}, {Ricker}, {Latham}, {Seager}, {Winn}, {Jenkins}, {Addison}, {Bouchy}, {Bowler}, {Briegal}, {Bryant}, {Collins}, {Daylan}, {Grieves}, {Horner}, {Huang}, {Kane}, {Kielkopf}, {McLean}, {Mengel}, {Nielsen}, {Okumura}, {Jones}, {Plavchan}, {Shporer}, {Smith}, {Tilbrook}, {Tinney}, {Twicken}, {Udry}, {Unger}, {West}, {Wittenmyer}, {Wohler}, {Torres}, \& {Wright}}]{Hobson2021}
{Hobson}, M.~J., {Brahm}, R., {Jord{\'a}n}, A., {et~al.} 2021, \aj, 161, 235, \dodoi{10.3847/1538-3881/abeaa1}

\bibitem[{{Hogg} {et~al.}(2010){Hogg}, {Myers}, \& {Bovy}}]{Hogg2010}
{Hogg}, D.~W., {Myers}, A.~D., \& {Bovy}, J. 2010, \apj, 725, 2166, \dodoi{10.1088/0004-637X/725/2/2166}

\bibitem[{{Horne}(1986)}]{Horne1986}
{Horne}, K. 1986, \pasp, 98, 609, \dodoi{10.1086/131801}

\bibitem[{{Howard} {et~al.}(2010){Howard}, {Johnson}, {Marcy}, {Fischer}, {Wright}, {Bernat}, {Henry}, {Peek}, {Isaacson}, {Apps}, {Endl}, {Cochran}, {Valenti}, {Anderson}, \& {Piskunov}}]{Howard2010}
{Howard}, A.~W., {Johnson}, J.~A., {Marcy}, G.~W., {et~al.} 2010, \apj, 721, 1467, \dodoi{10.1088/0004-637X/721/2/1467}

\bibitem[{{Howell} {et~al.}(2021){Howell}, {Matson}, {Ciardi}, {Everett}, {Livingston}, {Scott}, {Horch}, \& {Winn}}]{Howell2021}
{Howell}, S.~B., {Matson}, R.~A., {Ciardi}, D.~R., {et~al.} 2021, \aj, 161, 164, \dodoi{10.3847/1538-3881/abdec6}

\bibitem[{{Hsu} {et~al.}(2024){Hsu}, {Wang}, {Xuan}, {Ruffio}, {Morris}, {Echeverri}, {Xin}, {Liberman}, {Finnerty}, {Horstman}, {Sappey}, {Doppmann}, {Mawet}, {Jovanovic}, {Fitzgerald}, {Delorme}, {Wallace}, {Baker}, {Bartos}, {Blake}, {Calvin}, {Cetre}, {L{\'o}pez}, {Pezzato}, {Schofield}, {Skemer}, \& {Wang}}]{Hsu2024}
{Hsu}, C.-C., {Wang}, J.~J., {Xuan}, J.~W., {et~al.} 2024, \apj, 971, 9, \dodoi{10.3847/1538-4357/ad58d3}

\bibitem[{{Huber}(2017)}]{huber2017}
{Huber}, D. 2017, {Isoclassify: V1.2}, v1.2,  Zenodo, \dodoi{10.5281/zenodo.573372}

\bibitem[{{Huber} {et~al.}(2013){Huber}, {Carter}, {Barbieri}, {Miglio}, {Deck}, {Fabrycky}, {Montet}, {Buchhave}, {Chaplin}, {Hekker}, {Montalb{\'a}n}, {Sanchis-Ojeda}, {Basu}, {Bedding}, {Campante}, {Christensen-Dalsgaard}, {Elsworth}, {Stello}, {Arentoft}, {Ford}, {Gilliland}, {Handberg}, {Howard}, {Isaacson}, {Johnson}, {Karoff}, {Kawaler}, {Kjeldsen}, {Latham}, {Lund}, {Lundkvist}, {Marcy}, {Metcalfe}, {Silva Aguirre}, \& {Winn}}]{Huber2013}
{Huber}, D., {Carter}, J.~A., {Barbieri}, M., {et~al.} 2013, Science, 342, 331, \dodoi{10.1126/science.1242066}

\bibitem[{{Jang-Condell} {et~al.}(2015){Jang-Condell}, {Chen}, {Nesvold}, {Kuchner}, {Mittal}, {Puravankara}, {Watson}, \& {Lisse}}]{Jang-Condell2015}
{Jang-Condell}, H., {Chen}, C., {Nesvold}, E., {et~al.} 2015, in American Astronomical Society Meeting Abstracts, Vol. 225, American Astronomical Society Meeting Abstracts \#225, 330.04

\bibitem[{Jovanovic {et~al.}(2025)Jovanovic, Echeverri, Delorme, Finnerty, Schofield, Wang, Xin, Xuan, Wallace, Mawet, Sanghi, Baker, Bartos, Bond, Calvin, Cetre, Doppmann, Fitzgerald, Fucik, Gao, Ge, Guthery, Horstman, Hsu, Liberman, Leifer, Lilley, Lopez, Marin, Martin, Mennesson, Morris, Nash, Pezzato, Porter, Roberts, Ruane, Ruffio, Sappey, Serabyn, Shen, Skemer, Wang, Wetherell, Wizinowich, Salama, Chambouleyron, Jensen-Clem, \& Beichman}]{Jovanovic2025}
Jovanovic, N., Echeverri, D., Delorme, J.-R., {et~al.} 2025, Journal of Astronomical Telescopes, Instruments, and Systems, 11, 015005, \dodoi{10.1117/1.JATIS.11.1.015005}

\bibitem[{{Justesen} \& {Albrecht}(2019{\natexlab{a}})}]{Justesen2019}
{Justesen}, A.~B., \& {Albrecht}, S. 2019{\natexlab{a}}, \aap, 625, A59, \dodoi{10.1051/0004-6361/201834368}

\bibitem[{{Justesen} \& {Albrecht}(2019{\natexlab{b}})}]{tauboo2019}
---. 2019{\natexlab{b}}, \aap, 625, A59, \dodoi{10.1051/0004-6361/201834368}

\bibitem[{{Kane} {et~al.}(2015){Kane}, {Barclay}, {Hartmann}, {Hatzes}, {Jensen}, {Ciardi}, {Huber}, {Wright}, \& {Quintana}}]{Kane2015}
{Kane}, S.~R., {Barclay}, T., {Hartmann}, M., {et~al.} 2015, \apj, 815, 32, \dodoi{10.1088/0004-637X/815/1/32}

\bibitem[{{Katz} {et~al.}(2023){Katz}, {Sartoretti}, {Guerrier}, {Panuzzo}, {Seabroke}, {Th{\'e}venin}, {Cropper}, {Benson}, {Blomme}, {Haigron}, {Marchal}, {Smith}, {Baker}, {Chemin}, {Damerdji}, {David}, {Dolding}, {Fr{\'e}mat}, {Gosset}, {Jan{\ss}en}, {Jasniewicz}, {Lobel}, {Plum}, {Samaras}, {Snaith}, {Soubiran}, {Vanel}, {Zwitter}, {Antoja}, {Arenou}, {Babusiaux}, {Brouillet}, {Caffau}, {Di Matteo}, {Fabre}, {Fabricius}, {Fragkoudi}, {Haywood}, {Huckle}, {Hottier}, {Lasne}, {Leclerc}, {Mastrobuono-Battisti}, {Royer}, {Teyssier}, {Zorec}, {Crifo}, {Jean-Antoine Piccolo}, {Turon}, \& {Viala}}]{Katz2023}
{Katz}, D., {Sartoretti}, P., {Guerrier}, A., {et~al.} 2023, \aap, 674, A5, \dodoi{10.1051/0004-6361/202244220}

\bibitem[{{Kervella} {et~al.}(2019){Kervella}, {Arenou}, {Mignard}, \& {Th{\'e}venin}}]{Kervella2019}
{Kervella}, P., {Arenou}, F., {Mignard}, F., \& {Th{\'e}venin}, F. 2019, \aap, 623, A72, \dodoi{10.1051/0004-6361/201834371}

\bibitem[{{Khandelwal} {et~al.}(2023){Khandelwal}, {Sharma}, {Chakraborty}, {Chaturvedi}, {Ulmer-Moll}, {Ciardi}, {Boyle}, {Baliwal}, {Bieryla}, {Latham}, {Prasad}, {Nayak}, {Lendl}, \& {Mordasini}}]{Khandelwal2023}
{Khandelwal}, A., {Sharma}, R., {Chakraborty}, A., {et~al.} 2023, \aap, 672, L7, \dodoi{10.1051/0004-6361/202245608}

\bibitem[{{Kraus} {et~al.}(2016){Kraus}, {Ireland}, {Huber}, {Mann}, \& {Dupuy}}]{Kraus2016}
{Kraus}, A.~L., {Ireland}, M.~J., {Huber}, D., {Mann}, A.~W., \& {Dupuy}, T.~J. 2016, \aj, 152, 8, \dodoi{10.3847/0004-6256/152/1/8}

\bibitem[{{Lester} {et~al.}(2021){Lester}, {Matson}, {Howell}, {Furlan}, {Gnilka}, {Scott}, {Ciardi}, {Everett}, {Hartman}, \& {Hirsch}}]{Lester2021}
{Lester}, K.~V., {Matson}, R.~A., {Howell}, S.~B., {et~al.} 2021, \aj, 162, 75, \dodoi{10.3847/1538-3881/ac0d06}

\bibitem[{{Lester} {et~al.}(2023){Lester}, {Howell}, {Matson}, {Furlan}, {Gnilka}, {Littlefield}, {Ciardi}, {Everett}, {Fajardo-Acosta}, \& {Clark}}]{Lester2023}
{Lester}, K.~V., {Howell}, S.~B., {Matson}, R.~A., {et~al.} 2023, \aj, 166, 166, \dodoi{10.3847/1538-3881/acf563}

\bibitem[{{Lindegren} {et~al.}(2018){Lindegren}, {Hern{\'a}ndez}, {Bombrun}, {Klioner}, {Bastian}, {Ramos-Lerate}, {de Torres}, {Steidelm{\"u}ller}, {Stephenson}, {Hobbs}, {Lammers}, {Biermann}, {Geyer}, {Hilger}, {Michalik}, {Stampa}, {McMillan}, {Casta{\~n}eda}, {Clotet}, {Comoretto}, {Davidson}, {Fabricius}, {Gracia}, {Hambly}, {Hutton}, {Mora}, {Portell}, {van Leeuwen}, {Abbas}, {Abreu}, {Altmann}, {Andrei}, {Anglada}, {Balaguer-N{\'u}{\~n}ez}, {Barache}, {Becciani}, {Bertone}, {Bianchi}, {Bouquillon}, {Bourda}, {Br{\"u}semeister}, {Bucciarelli}, {Busonero}, {Buzzi}, {Cancelliere}, {Carlucci}, {Charlot}, {Cheek}, {Crosta}, {Crowley}, {de Bruijne}, {de Felice}, {Drimmel}, {Esquej}, {Fienga}, {Fraile}, {Gai}, {Garralda}, {Gonz{\'a}lez-Vidal}, {Guerra}, {Hauser}, {Hofmann}, {Holl}, {Jordan}, {Lattanzi}, {Lenhardt}, {Liao}, {Licata}, {Lister}, {L{\"o}ffler}, {Marchant}, {Martin-Fleitas}, {Messineo}, {Mignard}, {Morbidelli}, {Poggio}, {Riva}, {Rowell}, {Salguero}, {Sarasso}, {Sciacca}, {Siddiqui}, {Smart},
  {Spagna}, {Steele}, {Taris}, {Torra}, {van Elteren}, {van Reeven}, \& {Vecchiato}}]{Lindegren2018}
{Lindegren}, L., {Hern{\'a}ndez}, J., {Bombrun}, A., {et~al.} 2018, \aap, 616, A2, \dodoi{10.1051/0004-6361/201832727}

\bibitem[{{Lissauer} {et~al.}(2011){Lissauer}, {Ragozzine}, {Fabrycky}, {Steffen}, {Ford}, {Jenkins}, {Shporer}, {Holman}, {Rowe}, {Quintana}, {Batalha}, {Borucki}, {Bryson}, {Caldwell}, {Carter}, {Ciardi}, {Dunham}, {Fortney}, {Gautier}, {Howell}, {Koch}, {Latham}, {Marcy}, {Morehead}, \& {Sasselov}}]{Lissauer2011}
{Lissauer}, J.~J., {Ragozzine}, D., {Fabrycky}, D.~C., {et~al.} 2011, \apjs, 197, 8, \dodoi{10.1088/0067-0049/197/1/8}

\bibitem[{L{\'o}pez {et~al.}(2020)L{\'o}pez, Hoffman, Doppmann, Fitzgerald, Johnson, Kassis, Lanclos, Lyke, Martin, McLean, Sohn, \& Weiss}]{lopez_Characterization_2020}
L{\'o}pez, R.~A., Hoffman, E.~B., Doppmann, G., {et~al.} 2020, in Ground-Based and {{Airborne Instrumentation}} for {{Astronomy VIII}}, Vol. 11447 ({SPIE}), 1436--1450, \dodoi{10.1117/12.2563075}

\bibitem[{{Lubin} {et~al.}(2022){Lubin}, {Van Zandt}, {Holcomb}, {Weiss}, {Petigura}, {Robertson}, {Akana Murphy}, {Scarsdale}, {Batygin}, {Polanski}, {Batalha}, {Crossfield}, {Dressing}, {Fulton}, {Howard}, {Huber}, {Isaacson}, {Kane}, {Roy}, {Beard}, {Blunt}, {Chontos}, {Dai}, {Dalba}, {Gary}, {Giacalone}, {Hill}, {Mayo}, {Mo{\v{c}}nik}, {Kosiarek}, {Rice}, {Rubenzahl}, {Latham}, {Seager}, {Winn}, \& {Gary}}]{Lubin2022}
{Lubin}, J., {Van Zandt}, J., {Holcomb}, R., {et~al.} 2022, \aj, 163, 101, \dodoi{10.3847/1538-3881/ac3d38}

\bibitem[{{Lubow} \& {Martin}(2016)}]{Lubow2016}
{Lubow}, S.~H., \& {Martin}, R.~G. 2016, \apj, 817, 30, \dodoi{10.3847/0004-637X/817/1/30}

\bibitem[{{Lund} {et~al.}(2017){Lund}, {Rodriguez}, {Zhou}, {Gaudi}, {Stassun}, {Johnson}, {Bieryla}, {Oelkers}, {Stevens}, {Collins}, {Penev}, {Quinn}, {Latham}, {Villanueva}, {Eastman}, {Kielkopf}, {Oberst}, {Jensen}, {Cohen}, {Joner}, {Stephens}, {Relles}, {Corfini}, {Gregorio}, {Zambelli}, {Esquerdo}, {Calkins}, {Berlind}, {Ciardi}, {Dressing}, {Patel}, {Gagnon}, {Gonzales}, {Beatty}, {Siverd}, {Labadie-Bartz}, {Kuhn}, {Col{\'o}n}, {James}, {Pepper}, {Fulton}, {McLeod}, {Stockdale}, {Calchi Novati}, {DePoy}, {Gould}, {Marshall}, {Trueblood}, {Trueblood}, {Johnson}, {Wright}, {McCrady}, {Wittenmyer}, {Johnson}, {Sergi}, {Wilson}, \& {Sliski}}]{Lund2017}
{Lund}, M.~B., {Rodriguez}, J.~E., {Zhou}, G., {et~al.} 2017, \aj, 154, 194, \dodoi{10.3847/1538-3881/aa8f95}

\bibitem[{{Maciejewski} {et~al.}(2024){Maciejewski}, {Niedzielski}, {Gozdziewski}, {Wolszczan}, {Villaver}, {Fernandez}, {Adamow}, \& {Sierzputowska}}]{Maciejewski2024}
{Maciejewski}, G., {Niedzielski}, A., {Gozdziewski}, K., {et~al.} 2024, arXiv e-prints, arXiv:2407.11706, \dodoi{10.48550/arXiv.2407.11706}

\bibitem[{{Manara} {et~al.}(2019){Manara}, {Tazzari}, {Long}, {Herczeg}, {Lodato}, {Rota}, {Cazzoletti}, {van der Plas}, {Pinilla}, {Dipierro}, {Edwards}, {Harsono}, {Johnstone}, {Liu}, {Menard}, {Nisini}, {Ragusa}, {Boehler}, \& {Cabrit}}]{Manara2019}
{Manara}, C.~F., {Tazzari}, M., {Long}, F., {et~al.} 2019, \aap, 628, A95, \dodoi{10.1051/0004-6361/201935964}

\bibitem[{Martin {et~al.}(2018)Martin, Fitzgerald, McLean, Doppmann, Kassis, Aliado, Canfield, Johnson, Kress, Lanclos, Magnone, Sohn, Wang, \& Weiss}]{martin_overview_2018}
Martin, E.~C., Fitzgerald, M.~P., McLean, I.~S., {et~al.} 2018, \procspie, 10702, 107020A, \dodoi{10.1117/12.2312266}

\bibitem[{{Masuda} {et~al.}(2020){Masuda}, {Winn}, \& {Kawahara}}]{Masuda2020}
{Masuda}, K., {Winn}, J.~N., \& {Kawahara}, H. 2020, \aj, 159, 38, \dodoi{10.3847/1538-3881/ab5c1d}

\bibitem[{{Matson} {et~al.}(2025){Matson}, {Gore}, {Howell}, {Ciardi}, {Christiansen}, {Clark}, {Crossfield}, {Fajardo-Acosta}, {Fernandes}, {Furlan}, {Gilbert}, {Gonzales}, {Lester}, {Lund}, {Matthews}, {Polanski}, {Schlieder}, \& {Ziegler}}]{Matson2025}
{Matson}, R.~A., {Gore}, R., {Howell}, S.~B., {et~al.} 2025, \aj, 169, 76, \dodoi{10.3847/1538-3881/ad9923}

\bibitem[{{Mayor} {et~al.}(2003){Mayor}, {Pepe}, {Queloz}, {Bouchy}, {Rupprecht}, {Lo Curto}, {Avila}, {Benz}, {Bertaux}, {Bonfils}, {Dall}, {Dekker}, {Delabre}, {Eckert}, {Fleury}, {Gilliotte}, {Gojak}, {Guzman}, {Kohler}, {Lizon}, {Longinotti}, {Lovis}, {Megevand}, {Pasquini}, {Reyes}, {Sivan}, {Sosnowska}, {Soto}, {Udry}, {van Kesteren}, {Weber}, \& {Weilenmann}}]{Mayor2003}
{Mayor}, M., {Pepe}, F., {Queloz}, D., {et~al.} 2003, The Messenger, 114, 20

\bibitem[{{Millholland} {et~al.}(2021){Millholland}, {He}, {Ford}, {Ragozzine}, {Fabrycky}, \& {Winn}}]{Millholland2021}
{Millholland}, S.~C., {He}, M.~Y., {Ford}, E.~B., {et~al.} 2021, \aj, 162, 166, \dodoi{10.3847/1538-3881/ac0f7a}

\bibitem[{{Morgan} {et~al.}(2024){Morgan}, {Bowler}, {Tran}, {Petigura}, {Nagpal}, \& {Blunt}}]{Morgan2024}
{Morgan}, M., {Bowler}, B.~P., {Tran}, Q.~H., {et~al.} 2024, \aj, 167, 48, \dodoi{10.3847/1538-3881/ad0728}

\bibitem[{{Ogilvie}(2000)}]{Ogilvie2000}
{Ogilvie}, G.~I. 2000, \mnras, 317, 607, \dodoi{10.1046/j.1365-8711.2000.03654.x}

\bibitem[{{Perdelwitz} {et~al.}(2024){Perdelwitz}, {Trifonov}, {Teklu}, {Sreenivas}, \& {Tal-Or}}]{Perdelwitz2024}
{Perdelwitz}, V., {Trifonov}, T., {Teklu}, J.~T., {Sreenivas}, K.~R., \& {Tal-Or}, L. 2024, \aap, 683, A125, \dodoi{10.1051/0004-6361/202348263}

\bibitem[{{Petigura} {et~al.}(2013){Petigura}, {Howard}, \& {Marcy}}]{Petigura2013}
{Petigura}, E.~A., {Howard}, A.~W., \& {Marcy}, G.~W. 2013, Proceedings of the National Academy of Science, 110, 19273, \dodoi{10.1073/pnas.1319909110}

\bibitem[{{Polanski} {et~al.}(2024){Polanski}, {Lubin}, {Beard}, {Akana Murphy}, {Rubenzahl}, {Hill}, {Crossfield}, {Chontos}, {Robertson}, {Isaacson}, {Kane}, {Ciardi}, {Batalha}, {Dressing}, {Fulton}, {Howard}, {Huber}, {Petigura}, {Weiss}, {Angelo}, {Behmard}, {Blunt}, {Brinkman}, {Dai}, {Dalba}, {Fetherolf}, {Giacalone}, {Hirsch}, {Holcomb}, {Kosiarek}, {Mayo}, {MacDougall}, {Mo{\v{c}}nik}, {Pidhorodetska}, {Rice}, {Rosenthal}, {Scarsdale}, {Turtelboom}, {Tyler}, {Van Zandt}, {Yee}, {Coria}, {Dulz}, {Hartman}, {Householder}, {Lange}, {Langford}, {Louden}, {Siegel}, {Gilbert}, {Gonzales}, {Schlieder}, {Boyle}, {Christiansen}, {Clark}, {Fernandes}, {Lund}, {Savel}, {Gill}, {Beichman}, {Matson}, {Matthews}, {Furlan}, {Howell}, {Scott}, {Everett}, {Livingston}, {Ershova}, {Cheryasov}, {Safonov}, {Lillo-Box}, {Barrado}, \& {Morales-Calder{\'o}n}}]{Polanski2024}
{Polanski}, A.~S., {Lubin}, J., {Beard}, C., {et~al.} 2024, \apjs, 272, 32, \dodoi{10.3847/1538-4365/ad4484}

\bibitem[{{Pueyo}(2016)}]{Pueyo2016}
{Pueyo}, L. 2016, \apj, 824, 117, \dodoi{10.3847/0004-637X/824/2/117}

\bibitem[{{Quintana} {et~al.}(2007){Quintana}, {Adams}, {Lissauer}, \& {Chambers}}]{Quintana2007}
{Quintana}, E.~V., {Adams}, F.~C., {Lissauer}, J.~J., \& {Chambers}, J.~E. 2007, \apj, 660, 807, \dodoi{10.1086/512542}

\bibitem[{{Raghavan} {et~al.}(2010){Raghavan}, {McAlister}, {Henry}, {Latham}, {Marcy}, {Mason}, {Gies}, {White}, \& {ten Brummelaar}}]{Raghavan2010}
{Raghavan}, D., {McAlister}, H.~A., {Henry}, T.~J., {et~al.} 2010, \apjs, 190, 1, \dodoi{10.1088/0067-0049/190/1/1}

\bibitem[{{Rodriguez} {et~al.}(2015){Rodriguez}, {Duch{\^e}ne}, {Tom}, {Kennedy}, {Matthews}, {Greaves}, \& {Butner}}]{Rodriguez2015}
{Rodriguez}, D.~R., {Duch{\^e}ne}, G., {Tom}, H., {et~al.} 2015, \mnras, 449, 3160, \dodoi{10.1093/mnras/stv483}

\bibitem[{{Ros{\'a}rio} {et~al.}(2024){Ros{\'a}rio}, {Demangeon}, {Barros}, {Gandolfi}, {Egger}, {Serrano}, {Osborn}, {Beck}, {Benz}, {Flor{\'e}n}, {Guterman}, {Wilson}, {Alibert}, {Fossati}, {Hooton}, {Delrez}, {Santos}, {Sousa}, {Bonfanti}, {Salmon}, {Adibekyan}, {Nigioni}, {Venturini}, {Alonso}, {Anglada}, {Asquier}, {B{\'a}rczy}, {Barrado Navascues}, {Barrag{\'a}n}, {Baumjohann}, {Beck}, {Billot}, {Biondi}, {Bonfils}, {Borsato}, {Brandeker}, {Broeg}, {Cessa}, {Charnoz}, {Collier Cameron}, {Csizmadia}, {Cubillos}, {Davies}, {Deleuil}, {Deline}, {Demory}, {Ehrenreich}, {Erikson}, {Esposito}, {Fortier}, {Fridlund}, {Gillon}, {G{\"u}del}, {G{\"u}nther}, {Helling}, {Hoyer}, {Isaak}, {Kiss}, {Lam}, {Laskar}, {Lecavelier des Etangs}, {Lendl}, {Luntzer}, {Magrin}, {Maxted}, {Mordasini}, {Nascimbeni}, {Olofsson}, {Osborne}, {Ottensamer}, {Pagano}, {Pall{\'e}}, {Peter}, {Piotto}, {Pollacco}, {Queloz}, {Ragazzoni}, {Rando}, {Rauer}, {Ribas}, {Scandariato}, {S{\'e}gransan}, {Simon}, {Smith}, {Stalport}, {Szab{\'o}},
  {Thomas}, {Udry}, {Van Eylen}, {Van Grootel}, {Villaver}, {Walter}, \& {Walton}}]{ros2024}
{Ros{\'a}rio}, N.~M., {Demangeon}, O.~D.~S., {Barros}, S.~C.~C., {et~al.} 2024, \aap, 686, A282, \dodoi{10.1051/0004-6361/202347759}

\bibitem[{{Rota} {et~al.}(2022){Rota}, {Manara}, {Miotello}, {Lodato}, {Facchini}, {Koutoulaki}, {Herczeg}, {Long}, {Tazzari}, {Cabrit}, {Harsono}, {M{\'e}nard}, {Pinilla}, {van der Plas}, {Ragusa}, \& {Yen}}]{Rota2022}
{Rota}, A.~A., {Manara}, C.~F., {Miotello}, A., {et~al.} 2022, \aap, 662, A121, \dodoi{10.1051/0004-6361/202141035}

\bibitem[{Service {et~al.}(2016)Service, Lu, Campbell, Sitarski, Ghez, \& Anderson}]{Service2016}
Service, M., Lu, J.~R., Campbell, R., {et~al.} 2016, Publications of the Astronomical Society of the Pacific, 128, 095004, \dodoi{10.1088/1538-3873/128/967/095004}

\bibitem[{{Silsbee} \& {Rafikov}(2021)}]{Silsbee2021}
{Silsbee}, K., \& {Rafikov}, R.~R. 2021, \aap, 652, A104, \dodoi{10.1051/0004-6361/202141139}

\bibitem[{{Sing} {et~al.}(2009){Sing}, {D{\'e}sert}, {Lecavelier Des Etangs}, {Ballester}, {Vidal-Madjar}, {Parmentier}, {Hebrard}, \& {Henry}}]{Sing2009}
{Sing}, D.~K., {D{\'e}sert}, J.~M., {Lecavelier Des Etangs}, A., {et~al.} 2009, \aap, 505, 891, \dodoi{10.1051/0004-6361/200912776}

\bibitem[{{Soummer} {et~al.}(2012){Soummer}, {Pueyo}, \& {Larkin}}]{Soummer2012}
{Soummer}, R., {Pueyo}, L., \& {Larkin}, J. 2012, \apjl, 755, L28, \dodoi{10.1088/2041-8205/755/2/L28}

\bibitem[{Speagle(2020)}]{speagle_DYNESTY_2020}
Speagle, J.~S. 2020, Monthly Notices of the Royal Astronomical Society, 493, 3132, \dodoi{10.1093/mnras/staa278}

\bibitem[{{Sulis} {et~al.}(2024){Sulis}, {Crossfield}, {Santerne}, {Saillenfest}, {Sousa}, {Mary}, {Aguichine}, {Deleuil}, {Delgado Mena}, {Mathur}, {Polanski}, {Adibekyan}, {Boisse}, {Costes}, {Cretignier}, {Heidari}, {Lebarb{\'e}}, {Forveille}, {Hara}, {Meunier}, {Santos}, {Balcarcel-Salazar}, {Cort{\'e}s-Zuleta}, {Dalal}, {Gorjian}, {Halverson}, {Howard}, {Kosiarek}, {Lopez}, {Martin}, {Mousis}, {Rajkumar}, {Str{\o}m}, {Udry}, {Venot}, \& {Willett}}]{Sulis2024}
{Sulis}, S., {Crossfield}, I.~J.~M., {Santerne}, A., {et~al.} 2024, \aap, 688, A14, \dodoi{10.1051/0004-6361/202449559}

\bibitem[{{Talens} {et~al.}(2018){Talens}, {Justesen}, {Albrecht}, {McCormac}, {Van Eylen}, {Otten}, {Murgas}, {Palle}, {Pollacco}, {Stuik}, {Spronck}, {Lesage}, {Grundahl}, {Fredslund Andersen}, {Antoci}, \& {Snellen}}]{Talens2018}
{Talens}, G.~J.~J., {Justesen}, A.~B., {Albrecht}, S., {et~al.} 2018, \aap, 612, A57, \dodoi{10.1051/0004-6361/201731512}

\bibitem[{{Tayar} {et~al.}(2022){Tayar}, {Claytor}, {Huber}, \& {van Saders}}]{Tayar2022}
{Tayar}, J., {Claytor}, Z.~R., {Huber}, D., \& {van Saders}, J. 2022, \apj, 927, 31, \dodoi{10.3847/1538-4357/ac4bbc}

\bibitem[{{Teske} {et~al.}(2021){Teske}, {Wang}, {Wolfgang}, {Gan}, {Plotnykov}, {Armstrong}, {Butler}, {Cale}, {Crane}, {Howard}, {Jensen}, {Law}, {Shectman}, {Plavchan}, {Valencia}, {Vanderburg}, {Ricker}, {Vanderspek}, {Latham}, {Seager}, {Winn}, {Jenkins}, {Adibekyan}, {Barrado}, {Barros}, {Benkhaldoun}, {Brown}, {Bryant}, {Burt}, {Caldwell}, {Charbonneau}, {Cloutier}, {Collins}, {Collins}, {Colon}, {Conti}, {Demangeon}, {Eastman}, {Elmufti}, {Feng}, {Flowers}, {Guerrero}, {Hojjatpanah}, {Irwin}, {Isopi}, {Lillo-Box}, {Mallia}, {Massey}, {Mori}, {Mullally}, {Narita}, {Nishiumi}, {Osborn}, {Paegert}, {de Leon}, {Quinn}, {Reefe}, {Schwarz}, {Shporer}, {Soubkiou}, {Sousa}, {Stockdale}, {Str{\o}m}, {Tan}, {Tang}, {Tenenbaum}, {Wheatley}, {Wittrock}, {Yahalomi}, \& {Zohrabi}}]{Teske2021}
{Teske}, J., {Wang}, S.~X., {Wolfgang}, A., {et~al.} 2021, \apjs, 256, 33, \dodoi{10.3847/1538-4365/ac0f0a}

\bibitem[{{Tokovinin} \& {Moe}(2020)}]{Tokovinin2020}
{Tokovinin}, A., \& {Moe}, M. 2020, \mnras, 491, 5158, \dodoi{10.1093/mnras/stz3299}

\bibitem[{{Trotta}(2008)}]{Trotta2008}
{Trotta}, R. 2008, Contemporary Physics, 49, 71, \dodoi{10.1080/00107510802066753}

\bibitem[{{Vogt} {et~al.}(1994){Vogt}, {Allen}, {Bigelow}, {Bresee}, {Brown}, {Cantrall}, {Conrad}, {Couture}, {Delaney}, {Epps}, {Hilyard}, {Hilyard}, {Horn}, {Jern}, {Kanto}, {Keane}, {Kibrick}, {Lewis}, {Osborne}, {Pardeilhan}, {Pfister}, {Ricketts}, {Robinson}, {Stover}, {Tucker}, {Ward}, \& {Wei}}]{Vogt1994}
{Vogt}, S.~S., {Allen}, S.~L., {Bigelow}, B.~C., {et~al.} 1994, Society of Photo-Optical Instrumentation Engineers (SPIE) Conference Series, Vol. 2198, {HIRES: the high-resolution echelle spectrometer on the Keck 10-m Telescope}, 362, \dodoi{10.1117/12.176725}

\bibitem[{{Wang} {et~al.}(2015){Wang}, {Fischer}, {Xie}, \& {Ciardi}}]{Wang2015}
{Wang}, J., {Fischer}, D.~A., {Xie}, J.-W., \& {Ciardi}, D.~R. 2015, \apj, 813, 130, \dodoi{10.1088/0004-637X/813/2/130}

\bibitem[{Wang {et~al.}(2021)Wang, Ruffio, Morris, Delorme, Jovanovic, Pezzato, Echeverri, Finnerty, Hood, Zanazzi, Bryan, Bond, Cetre, Martin, Mawet, Skemer, Baker, Xuan, Wallace, Wang, Bartos, Blake, Boden, Buzard, Calvin, Chun, Doppmann, Dupuy, Duch{\^e}ne, Feng, Fitzgerald, Fortney, Freedman, Knutson, Konopacky, Lilley, Liu, Lopez, Lupu, Marley, Meshkat, Miles, {Millar-Blanchaer}, Ragland, Roy, Ruane, Sappey, Schofield, Weiss, Wetherell, Wizinowich, \& Ygouf}]{wang_Detection_2021}
Wang, J.~J., Ruffio, J.-B., Morris, E., {et~al.} 2021, The Astronomical Journal, 162, 148, \dodoi{10.3847/1538-3881/ac1349}

\bibitem[{{Winters} {et~al.}(2022){Winters}, {Cloutier}, {Medina}, {Irwin}, {Charbonneau}, \& {LTT 1445A planet mass paper co-authors}}]{Winters2022}
{Winters}, J., {Cloutier}, R., {Medina}, A., {et~al.} 2022, in Bulletin of the American Astronomical Society, Vol.~54, 102.417

\bibitem[{{Winters} {et~al.}(2019){Winters}, {Medina}, {Irwin}, {Charbonneau}, {Astudillo-Defru}, {Horch}, {Eastman}, {Vrijmoet}, {Henry}, {Diamond-Lowe}, {Winston}, {Barclay}, {Bonfils}, {Ricker}, {Vanderspek}, {Latham}, {Seager}, {Winn}, {Jenkins}, {Udry}, {Twicken}, {Teske}, {Tenenbaum}, {Pepe}, {Murgas}, {Muirhead}, {Mink}, {Lovis}, {Levine}, {L{\'e}pine}, {Jao}, {Henze}, {Fur{\'e}sz}, {Forveille}, {Figueira}, {Esquerdo}, {Dressing}, {D{\'\i}az}, {Delfosse}, {Burke}, {Bouchy}, {Berlind}, \& {Almenara}}]{Winters2019}
{Winters}, J.~G., {Medina}, A.~A., {Irwin}, J.~M., {et~al.} 2019, \aj, 158, 152, \dodoi{10.3847/1538-3881/ab364d}

\bibitem[{{Xuan} \& {Wyatt}(2020)}]{Xuan2020}
{Xuan}, J.~W., \& {Wyatt}, M.~C. 2020, \mnras, 497, 2096, \dodoi{10.1093/mnras/staa2033}

\bibitem[{Xuan {et~al.}(2022)Xuan, Wang, Ruffio, Knutson, Mawet, Mollière, Kolecki, Vigan, Mukherjee, Wallack, Wang, Baker, Bartos, Blake, Bond, Bryan, Calvin, Cetre, Chun, Delorme, Doppmann, Echeverri, Finnerty, Fitzgerald, Horstman, Inglis, Jovanovic, López, Martin, Morris, Pezzato, Ragland, Ren, Ruane, Sappey, Schofield, Skemer, Venenciano, Wallace, \& Wizinowich}]{Xuan2022}
Xuan, J.~W., Wang, J., Ruffio, J.-B., {et~al.} 2022, The Astrophysical Journal, 937, 54, \dodoi{10.3847/1538-4357/ac8673}

\bibitem[{Xuan {et~al.}(2024)Xuan, Wang, Finnerty, Horstman, Grimm, Peck, Nielsen, Knutson, Mawet, Isaacson, Howard, Liu, Walker, Phillips, Blake, Ruffio, Zhang, Inglis, Wallack, Sanghi, Gonzales, Dai, Baker, Bartos, Bond, Bryan, Calvin, Cetre, Delorme, Doppmann, Echeverri, Fitzgerald, Jovanovic, Liberman, López, Martin, Morris, Pezzato, Ruane, Sappey, Schofield, Skemer, Venenciano, Wallace, Wang, Wizinowich, Xin, Agrawal, Ó, Hsu, \& Phillips}]{Xuan2024}
Xuan, J.~W., Wang, J., Finnerty, L., {et~al.} 2024, The Astrophysical Journal, 962, 10, \dodoi{10.3847/1538-4357/ad1243}

\bibitem[{{Xuan} {et~al.}(2024){Xuan}, {Hsu}, {Finnerty}, {Wang}, {Ruffio}, {Zhang}, {Knutson}, {Mawet}, {Mamajek}, {Inglis}, {Wallack}, {Bryan}, {Blake}, {Molli{\`e}re}, {Hejazi}, {Baker}, {Bartos}, {Calvin}, {Cetre}, {Delorme}, {Doppmann}, {Echeverri}, {Fitzgerald}, {Jovanovic}, {Liberman}, {L{\'o}pez}, {Morris}, {Pezzato}, {Sappey}, {Schofield}, {Skemer}, {Wallace}, {Wang}, {Agrawal}, \& {Horstman}}]{Xuan2024_Survey}
{Xuan}, J.~W., {Hsu}, C.-C., {Finnerty}, L., {et~al.} 2024, The Astrophysical Journal, 970, 71, \dodoi{10.3847/1538-4357/ad4796}

\bibitem[{Yee {et~al.}(2018)Yee, Petigura, Fulton, Knutson, Batygin, Bakos, Hartman, Hirsch, Howard, Isaacson, Kosiarek, Sinukoff, \& Weiss}]{Yee2018}
Yee, S.~W., Petigura, E.~A., Fulton, B.~J., {et~al.} 2018, \aj, 155, 255, \dodoi{10.3847/1538-3881/aabfec}

\bibitem[{{Zanazzi} \& {Lai}(2018{\natexlab{a}})}]{Zanazzi2018}
{Zanazzi}, J.~J., \& {Lai}, D. 2018{\natexlab{a}}, \mnras, 477, 5207, \dodoi{10.1093/mnras/sty951}

\bibitem[{{Zanazzi} \& {Lai}(2018{\natexlab{b}})}]{zanazzi2018b}
---. 2018{\natexlab{b}}, \mnras, 478, 835, \dodoi{10.1093/mnras/sty1075}

\bibitem[{{Zhang} {et~al.}(2021){Zhang}, {Weiss}, {Huber}, {Blunt}, {Chontos}, {Fulton}, {Grunblatt}, {Howard}, {Isaacson}, {Kosiarek}, {Petigura}, {Rosenthal}, \& {Rubenzahl}}]{Zhang2021}
{Zhang}, J., {Weiss}, L.~M., {Huber}, D., {et~al.} 2021, \aj, 162, 89, \dodoi{10.3847/1538-3881/ac0634}

\bibitem[{{Zhang} {et~al.}(2024{\natexlab{a}}){Zhang}, {Weiss}, {Huber}, {Jensen}, {Brandt}, {Collins}, {Conti}, {Isaacson}, {Lewin}, {Marino}, {Massey}, {Murgas}, {Palle}, {Radford}, {Relles}, {Srdoc}, {Stockdale}, {Tan}, \& {Wang}}]{Zhang2024a}
---. 2024{\natexlab{a}}, \aj, 167, 89, \dodoi{10.3847/1538-3881/ad1189}

\bibitem[{{Zhang} {et~al.}(2024c){Zhang}, {Huber}, {Weiss}, {Xuan}, {Burt}, {Dai}, {Saunders}, {Petigura}, {Rubenzahl}, {Winn}, {Wang}, {Van Zandt}, {Brodheim}, {Claytor}, {Crossfield}, {Deich}, {Fulton}, {Gibson}, {Halverson}, {Hill}, {Holden}, {Householder}, {Howard}, {Isaacson}, {Kaye}, {Lanclos}, {Laher}, {Lubin}, {Payne}, {Roy}, {Schwab}, {Shaum}, {Walawender}, {Wishnow}, \& {Yeh}}]{Zhang2024c}
{Zhang}, J., {Huber}, D., {Weiss}, L.~M., {et~al.} 2024c, \aj, 168, 295, \dodoi{10.3847/1538-3881/ad86c4}

\bibitem[{{Zhang} {et~al.}(2025){Zhang}, {Weiss}, {Huber}, {Xuan}, {Bottom}, {Fulton}, {Isaacson}, {MacDougall}, \& {Saunders}}]{Zhang2024d}
{Zhang}, J., {Weiss}, L.~M., {Huber}, D., {et~al.} 2025, \aj, 169, 200, \dodoi{10.3847/1538-3881/ada60a}

\bibitem[{{Zhang} {et~al.}(2024{\natexlab{b}}){Zhang}, {Xuan}, {Mawet}, {Wang}, {Hsu}, {Ruffio}, {Knutson}, {Inglis}, {Blake}, {Chachan}, {Horstman}, {Baker}, {Bartos}, {Calvin}, {Cetre}, {Delorme}, {Doppmann}, {Echeverri}, {Finnerty}, {Fitzgerald}, {Jovanovic}, {Liberman}, {L{\'o}pez}, {Morris}, {Pezzato}, {Sappey}, {Schofield}, {Skemer}, {Wallace}, {Wang}, \& {Do {\'O}}}]{YZhang2024}
{Zhang}, Y., {Xuan}, J.~W., {Mawet}, D., {et~al.} 2024{\natexlab{b}}, \aj, 168, 131, \dodoi{10.3847/1538-3881/ad6609}

\bibitem[{{Zhang} {et~al.}(2022){Zhang}, {Bowler}, {Dupuy}, {Brandt}, {Brandt}, {Cochran}, {Endl}, {MacQueen}, {Kratter}, {Isaacson}, {Franson}, {Kraus}, {Morley}, \& {Zhou}}]{Zhangzj2022}
{Zhang}, Z., {Bowler}, B.~P., {Dupuy}, T.~J., {et~al.} 2022, arXiv e-prints, arXiv:2210.07252.
\newblock \doarXiv{2210.07252}

\bibitem[{{Zhou} {et~al.}(2022){Zhou}, {Wirth}, {Huang}, {Venner}, {Franson}, {Quinn}, {Bouma}, {Kraus}, {Mann}, {Newton}, {Dragomir}, {Heitzmann}, {Lowson}, {Douglas}, {Battley}, {Gillen}, {Triaud}, {Latham}, {Howell}, {Hartman}, {Tofflemire}, {Wittenmyer}, {Bowler}, {Horner}, {Kane}, {Kielkopf}, {Plavchan}, {Wright}, {Addison}, {Mengel}, {Okumura}, {Ricker}, {Vanderspek}, {Seager}, {Jenkins}, {Winn}, {Daylan}, {Fausnaugh}, \& {Kunimoto}}]{zhou2022}
{Zhou}, G., {Wirth}, C.~P., {Huang}, C.~X., {et~al.} 2022, \aj, 163, 289, \dodoi{10.3847/1538-3881/ac69e3}

\bibitem[{{Ziegler} {et~al.}(2020){Ziegler}, {Tokovinin}, {Brice{\~n}o}, {Mang}, {Law}, \& {Mann}}]{Ziegler2020}
{Ziegler}, C., {Tokovinin}, A., {Brice{\~n}o}, C., {et~al.} 2020, \aj, 159, 19, \dodoi{10.3847/1538-3881/ab55e9}

\bibitem[{{Ziegler} {et~al.}(2021){Ziegler}, {Tokovinin}, {Latiolais}, {Brice{\~n}o}, {Law}, \& {Mann}}]{Ziegler2021}
{Ziegler}, C., {Tokovinin}, A., {Latiolais}, M., {et~al.} 2021, \aj, 162, 192, \dodoi{10.3847/1538-3881/ac17f6}

\end{thebibliography}
\bibliographystyle{aasjournal}


\end{CJK*}
\end{document}